\documentclass[longbibliography,nofootinbib,tightenlines,superscriptaddress]{revtex4-1}
\usepackage{amsmath,amssymb,graphicx}
\graphicspath{{figs/}}
\usepackage{epsf,verbatim}
\usepackage{hyperref}
\usepackage{color}
\usepackage{slashed}
\usepackage{comment}
\usepackage{mdwlist, paralist}
\usepackage{rotating}
\usepackage{bm}
\usepackage{multirow}
\usepackage{fancyhdr}
\usepackage{empheq}
\usepackage[super,negative]{nth}

\newcommand{\beq}{\begin{equation}}
\newcommand{\eeq}{\end{equation}}
\newcommand{\bal}{\begin{align}}
\newcommand{\eal}{\end{align}}
\newcommand{\bit}{\begin{itemize}}
\newcommand{\eit}{\end{itemize}}
\newcommand{\ben}{\begin{enumerate}}
\newcommand{\een}{\end{enumerate}}

\renewcommand{\eqref}[1]{Eq.~(\ref{#1})}

\newcommand{\secref}[1]{Sec.~\ref{#1}}

\newcommand{\appref}[1]{Appendix~\ref{#1}}
\newcommand{\figref}[1]{Fig.~\ref{#1}}

\newcommand{\tabref}[1]{Table~\ref{#1}}

\newcommand{\met}{\slashed{E}_\text{T}}

\renewcommand{\t}{\tilde}

\newcommand{\fb}{\,{\rm fb}^{-1}}

\newcommand{\gev}{{\ \rm GeV}}
\newcommand{\tev}{{\ \rm TeV}}

\newcommand{\pt}{p_\text{T}}

\begin{document}

\title{The Leptoquark Hunter's Guide: Pair Production}

\author{Bastian Diaz}
\thanks{bastian.diaz@alumnos.usm.cl}
\affiliation{Departamento de Fisica and Centro Cientfico-Tecnologico de Valparaiso, Universidad Tecnica Federico Santa Maria, Casilla 110-V, Valparaiso, Chile}

\author{Martin Schmaltz}
\thanks{schmaltz@bu.edu}
\affiliation{Physics Department, Boston University, Boston, MA 02215, USA}

\author{Yi-Ming Zhong}
\thanks{ymzhong@bu.edu}
\affiliation{Physics Department, Boston University, Boston, MA 02215, USA}

\begin{abstract}
\vspace{.5in}
\hspace{2.2in} {\bf ABSTRACT}
\vspace{.35in}

Leptoquarks occur in many new physics scenarios and could be the next big discovery at the LHC.  The purpose of this paper is to point out that a model-independent search strategy covering all possible leptoquarks is possible and has not yet been fully exploited. To be systematic we organize the possible leptoquark final states according to a leptoquark matrix with entries corresponding to nine experimentally distinguishable leptoquark decays: any of \{light-jet, $b$-jet, top\} with any of \{neutrino, $e$/$\mu$, $\tau$\}. The 9 possibilities can be explored in a largely model-independent fashion with pair-production of leptoquarks at the LHC. We review the status of experimental searches for the 9 components of the leptoquark matrix, pointing out which 3 have not been adequately covered. We plead that experimenters publish bounds on leptoquark cross sections as functions of mass for as wide a range of leptoquark masses as possible. Such bounds are essential for reliable recasts to general leptoquark models. To demonstrate the utility of the leptoquark matrix approach we collect and summarize searches with the same final states as leptoquark pair production and use them to derive bounds on a complete set of Minimal Leptoquark models which span all possible flavor and gauge representations for scalar and vector leptoquarks.
\end{abstract}

\maketitle

\section{Introduction and overview}
\label{intro}

\begin{figure}[!htbp]
   \centering
   \includegraphics[width=0.22\textwidth]{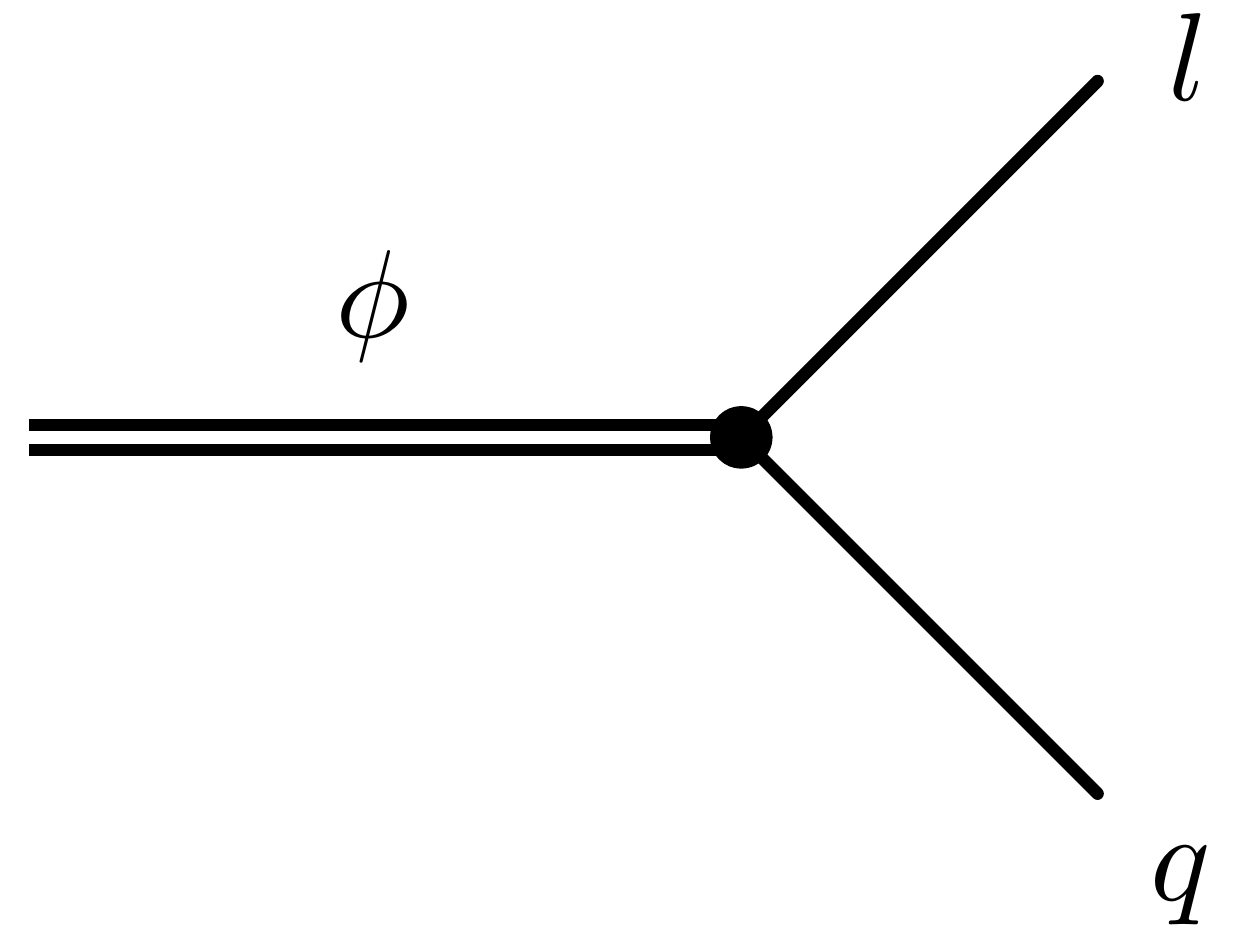}\hskip1.2in
   \includegraphics[width=0.22\textwidth]{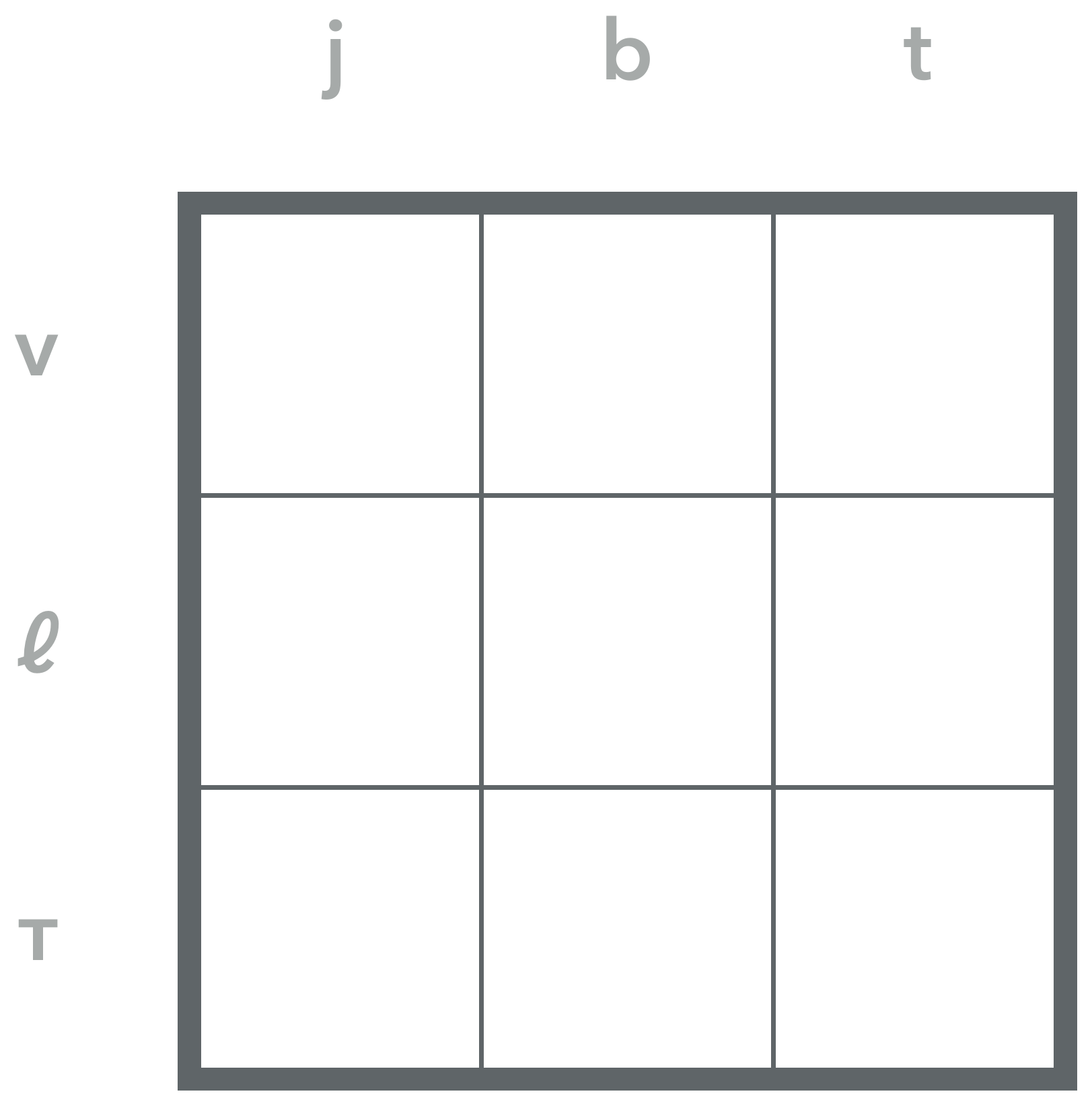}
    \caption{(\emph{left}) LQ coupling to quark $q$ and lepton $l$ (see ~\appref{app:notation} for notations used in the paper). (\emph{right}) LQ matrix. Each entry of the matrix represents one of the experimentally distinguishable leptoquark decays.  Rows label leptonic decay products, columns label hadronic decay products.
A more precise definition of the LQ matrix is given in the text. For a  summary of existing LHC searches corresponding to each matrix element see \secref{sec:matrix}.}   
\label{fig:leptoquark}
\end{figure}

A leptoquark (LQ) is a particle with a coupling which allows it to decay to a quark (or anti-quark) and a lepton, \figref{fig:leptoquark}. It carries color and electric charge, and possibly also weak charges. Because of their color, leptoquarks can be pair-produced as LQ  anti-LQ pairs with large QCD cross sections. Searches based on this production mode with subsequent decay,
$p p \rightarrow \phi \bar \phi \rightarrow (l q)(\bar l \bar q)$, are the subject of this paper (for single LQ production see~\cite{inprogress}). Here $\phi$ and $\bar \phi$
are the leptoquark and its antiparticle, $q$ can be any Standard Model (SM) quark or antiquark, $l$ is any lepton, and $\bar q$ and $\bar l$ are the corresponding antiparticles. We use parenthesis to indicate which final state particles reconstruct resonances.
Our goal is to provide a simple organizing principle which makes it straightforward to systematically search for all possible leptoquarks. The idea is that we identify a minimum set of independent final states which must be searched for. These final 
states can be arranged into a $3 \times 3$ matrix which we call the ``leptoquark final state matrix",
or simply ``LQ matrix",  \figref{fig:leptoquark}. For each final state, what is needed from experiment is an upper bound on
the cross section times branching fraction as a function of LQ mass. Lower bounds on the mass of an arbitrary
leptoquark can then be obtained by calculating the theoretical cross section times branching fraction into the final
states in the LQ matrix and comparing to the experimental cross section bounds. 

This paper consists of two complementary parts. In Part 1, we introduce the LQ matrix and show that bounds on the cross
sections into each of the 9 final states of the matrix are both efficient and sufficient for searching for all possible LQs.
In Part 2 we demonstrate the utility of this approach. We collect the best currently available cross section bounds from
LHC searches organized by which element of the LQ matrix they cover and use them to put bounds on a
complete set of ``Minimal Leptoquark" (MLQ) models. The MLQs include scalar and vector LQs with all possible flavor, charge,
and isospin quantum numbers.      
 
We begin with a few words on the theoretical motivation for leptoquarks. LQs are predicted in many physics
beyond the SM scenarios. Vector LQs may be related to unification, squarks automatically become scalar LQs in
supersymmetry (SUSY) with $R$-parity violation (RPV), and LQs arise in models in which quarks and leptons are composites of the
same underlying dynamics. Recently, additional motivation for LQs
arrived from a number of anomalies observed in $B$ meson decays~\cite{Aaij:2014ora,Aaij:2017vbb} which could
be explained in models with leptoquarks of TeV scale masses. Leptoquarks with couplings to muons might also be relevant to the persistent anomaly in the muon anomalous magnetic moment~\cite{Bennett:2006fi}.
On the other hand, there are no convincing theoretical arguments for preferring a particular pattern of leptoquark flavors or
quantum numbers, and the experimental hints, however intriguing, may still change or disappear.
This motivates a systematic approach to search for a minimal but sufficient set of final states which can
discover LQs with arbitrary decays to quarks and leptons.    

What are the possible LQ final states? By definition, a leptoquark can decay to any of the six SM quarks with any of the three charged leptons or three neutrinos. The quarks and leptons can be particle or antiparticle and left- or right-handed (we allow for right-handed neutrinos), corresponding to 
hundreds of different possibilities. However, many of these are covered by the same searches.
For example, left- and right-handed final state particles are generally not distinguishable.
Light quarks and antiquarks $u, d, s, c, \bar u, \bar d, \bar s, \bar c$ from leptoquark decays all show up as light-jets.%
\footnote{$c$-tagging is not yet efficient enough to improve leptoquark searches. We show this with an example in \secref{sec:nujbt}.}
Heavy quarks, $t$ and $b$ can be distinguished. For leptons, we have three distinct charged leptons but neutrino flavor is not observable in leptoquark decays. Thus we arrive at a matrix of nine distinct final states from the decay of a single LQ,
three ``flavors" of quarks $j, b, t$ times three ``flavors" of leptons%
\footnote{Note that it is customary in LHC searches to refer to electrons or muons as light leptons or  ``$\ell$" because search strategies for them are similar; we reflect this custom by assigning $e/\mu$ to a single row, $\ell$.
Of course, one could instead work with a $4\times 3$ matrix.}\ %
$\nu$, $\ell$ ($=e/\mu$), $\tau$, see \figref{fig:leptoquark}b.

To obtain the final states of pair production we must consider the decay of the LQ and its antiparticle. For a LQ with a unique coupling to a quark-lepton bilinear $q l$, the decay of the LQ anti-LQ pair is also unique to the ``symmetric" final state $(lq)(\bar l \bar q)$. Enumerating all possible such symmetric final states is trivial, they are determined by adding the corresponding antiparticles to the final states of a single LQ. Thus we can use the same $3\times 3$ LQ matrix to classify the distinct symmetric final states.  

However, when LQs have multiple decay channels then LQ pair production also produces ``asymmetric" final states which are not
covered by the classification in terms of a single $3\times 3$ matrix. We now argue that the symmetric final states are all we
need.       
Consider for simplicity the case where a LQ couples to two quark-lepton pairs, $lq$ and $l'q'$.
We obtain the pair production final states: 
\begin{eqnarray}
&{\rm symmetric:} &\quad (lq)(\bar l \bar q) + (l'q')(\bar l' \bar q')\nonumber \\
& {\rm asymmetric:} &\quad (lq)(\bar l' \bar q') + (l'q')(\bar l \bar q)\nonumber
\end{eqnarray}
The two symmetric ones are contained in the LQ matrix classification, the asymmetric ones are not.
But here is the point: all leptoquarks which produce asymmetric final states necessarily also produce symmetric final states.
Therefore the symmetric final states are sufficient to search for all possible leptoquarks. Moreover, one might expect
the couplings of leptoquarks to different quark-lepton pairs to not all be the same. Then the branching fraction to the quark-lepton
pair with the largest coupling $lq$ will dominate the decays, and the branching fraction to the symmetric state $(lq)(\bar l \bar q)$ will be largest.
To sum up, symmetric final states from pair production of LQs can be represented by the LQ matrix. They are
sufficient for searching for all possible leptoquarks. When LQ couplings to quark-lepton pairs are at least somewhat
hierarchical, a symmetric final state will usually also yield the most sensitive search.

An interesting special case arises when a symmetry relates couplings
of the leptoquark to different quark-lepton pairs. In fact, isospin, $SU(2)_{weak}$, is such a symmetry. Consider for
concreteness an $SU(2)_{weak}$ singlet scalar leptoquark which couples to the $SU(2)_{weak}$ doublets $q_L$ and $l_L$, both
of the first generation.
$SU(2)_{weak}$ enforces that the coupling of the LQ is to the combination $d\nu - u e^-$ so that pair production gives
rise to the four possible final states $(\nu d)(\bar \nu \bar d) + (e^- u)(e^+ \bar u) + (\nu d)(e^+ \bar u) + (e^- u)(\nu \bar d)$,
each with 25\% branching fraction. One could search for this leptoquark with the symmetric LQ matrix final states
$(e^- j)(e^+ j)$ and $j j \met$ or with non-matrix final states $(e^+ j) (j \met)$ and $(e^- j)(j\met)$.
However even in this case where no single symmetric final state dominates the ``easiest" and likely most
sensitive search is the symmetric search for $(e^- j)(e^+ j)$.
We will discuss many more examples with multiplicities of final states from $SU(2)_{weak}$ in \secref{sec:matrix}.

So far, we have argued that in order to search for all possible LQs, it is sufficient for experiments to cover 9
symmetric final states which are in one-to-one correspondence with elements of the LQ matrix. What information
do we need from each of these 9 searches in order to be able to put bounds on arbitrary leptoquarks? 

In order to maximize sensitivity, LHC leptoquark searches adjust their cuts as a function of the mass of LQs that
are being searched for. In addition, signal efficiencies depend on the LQ mass. Therefore, the natural
result of a leptoquark search is a ``Brazil plot" which shows the experimentally observed upper bound on the allowed signal
cross section times branching fraction as a function of LQ mass (along with the range of expected bounds).
This plot of the upper bound on $\sigma \times Br$ for each final state in the LQ matrix is precisely the result
which is needed from experiment. With the 9 plots from each of the LQ matrix searches anyone can
then easily determine a lower bound on the mass of their favorite leptoquark by comparing their
theoretical $\sigma \times Br$  as a function of mass to the bound. This is true for LQs of 
arbitrary spin, charge, or couplings.

This last claim requires further examination. In order to determine the bound on $\sigma \times Br$
experimenters need to divide the observed bound on the signal cross section for events in a signal region
by the efficiency of signal events to pass the signal region cuts. In general, this signal efficiency
can depend on many properties of the signal under consideration, and one may worry that
model-independent bounds on $\sigma \times Br$ cannot be obtained.
Fortunately, the situation for symmetric LQ searches is nice. 
To fairly good precision (better than 10 \%) the signal efficiency only depends on the mass
of the leptoquark. The basic reason for this model independence is easy to understand:
current LQ bounds already require masses well above 500 GeV and center
of mass energies in excess of 1 TeV. Thus leptoquarks are produced with only moderate boosts.
Furthermore they two-body-decay directly to SM particles, yielding highly energetic widely separated SM particles
with approximately isotropic distributions. Thus signal efficiencies for standard kinematic cuts are high and
only very weakly dependent on any details of the LQ model other than the LQ mass.
Two concerns which could have invalidated this argument
are dependence on LQ spin and width, we discuss them in turn.

\begin{figure}[!htbp]
   \centering
   \includegraphics[width=0.48\textwidth]{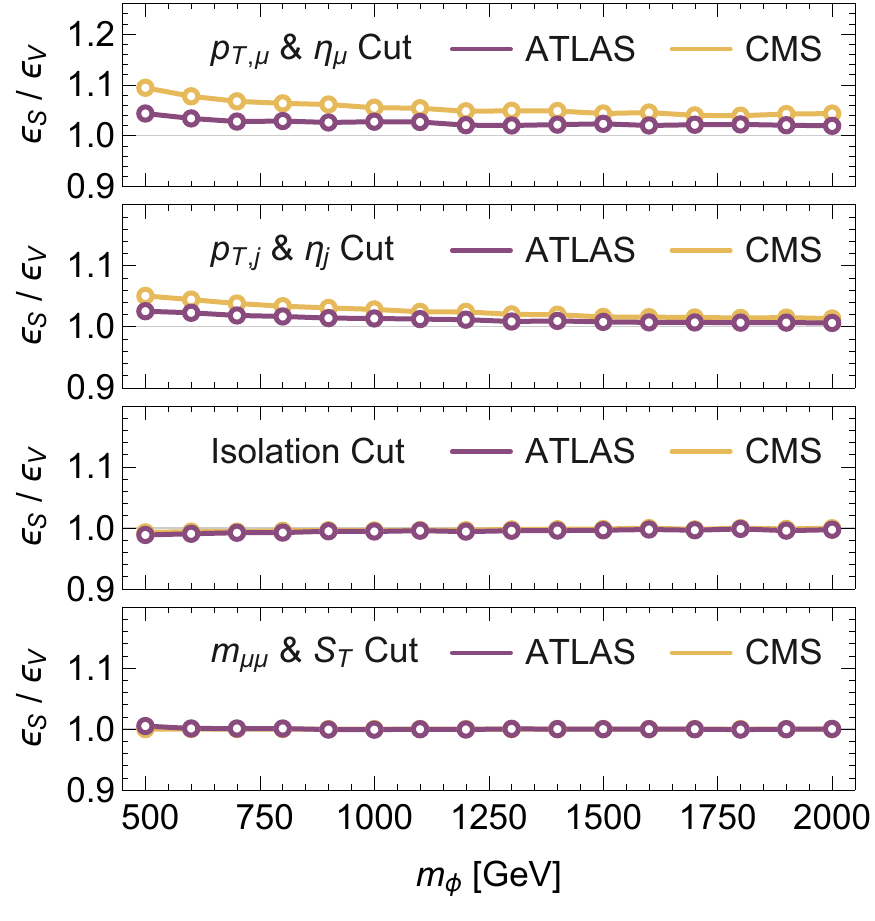} 
   \caption{Ratios of signal efficiencies $\epsilon_{S}/\epsilon_{V}$ for  pair-produced scalar and vector LQs with various masses. The LQ decays are to muons and jets. The panels show the relative efficiencies of a series of  selection cuts from recent ATLAS~\cite{Aaboud:2016qeg} and CMS ~\cite{CMS-PAS-EXO-16-007} analyses. Details of the cuts and plots of phase space distributions
are in~\appref{app:efficiencies}.}
   \label{fig:cutsofsandv}
\end{figure}

{\it Spin:} Leptoquarks can be spin 0 or 1, i.e., scalars or vectors. In either case the cross section
is dominated by gluon-gluon interactions. Scalars are produced in a $p$-wave, implying that the cross section is more
suppressed near the threshold, and typical scalar LQ events have more energy than typical vector
events. Thus we expect the efficiencies of $p_T$ cuts to be slightly higher for scalars. This is a small effect
because even LQs decaying at rest would give their decay products sufficient momentum to pass basic $p_T$ cuts.
More important are angular distributions resulting from LQ spin. Because of the large LQ mass, decay
products are well-isolated and approximately back-to-back. However the pseudo-rapidity, $\eta$, distribution of vector
LQs is more forward than for the scalars, causing the efficiency of especially lepton $\eta$ cuts to be lower in the vector case.  
But this effect is also small.
To demonstrate this we show the ratio of the two signal efficiencies, $\epsilon_S/\epsilon_V$, as a function of
mass for the analysis cuts used in two typical ATLAS~\cite{Aaboud:2016qeg} and CMS~\cite{CMS-PAS-EXO-16-007}
analyses in \figref{fig:cutsofsandv} (details are shown in \appref{app:efficiencies}).
This analysis is for LQs decaying to muons and light
quarks and with signal simulated at the parton level. 
One sees that for LQ masses in the range of current bounds (800 GeV -- 2 TeV)  the difference between scalar and vector
efficiencies is always smaller than 10\%.

We conclude that searches with cuts optimized for scalar LQ detection are also very sensitive to vectors and 
the differences in efficiencies are small enough that we can ignore them. An analysis aiming for higher precision
could account for the small difference in efficiencies  between vectors and scalars by determining separate
cross section bounds for the two cases.

{\it Width:} For leptoquarks with very large couplings ($\gg 1$) to quarks and leptons the width becomes large and a significant
number of events may have one or both LQs off-shell. In that case signal efficiencies can become much more model-dependent.
However large couplings to quarks and leptons of the first and second generation for TeV scale leptoquarks are ruled from
single leptoquark production, $e^+ e^-$ collisions, and precision electroweak constraints.  Large couplings to third generation
particles remain a possibility which could introduce some model dependence. We ignore this special possibility and assume that LQ widths are smaller than 10\% so that on-shell LQ production dominates.

To summarize our results so far, we have argued that in order to search for arbitrary LQs it is sufficient to
consider 9 final states corresponding to the LQ matrix. Experiments should provide the upper bound on the
cross section times branching fraction as a function of LQ mass for each of the different final states (Brazil plot).
The signal efficiencies needed to produce the cross section bounds are largely model independent and can
be determined from Monte Carlo (MC) for any convenient leptoquark implementation (scalar or vector with arbitrary couplings).  
Armed with cross section bounds for each LQ matrix final state one can determine lower mass bounds on arbitrary LQs
by computing the cross section times branching fractions and comparing to the bounds.

\begin{figure}[!htbp]
   \centering
   \includegraphics[width=0.45\textwidth]{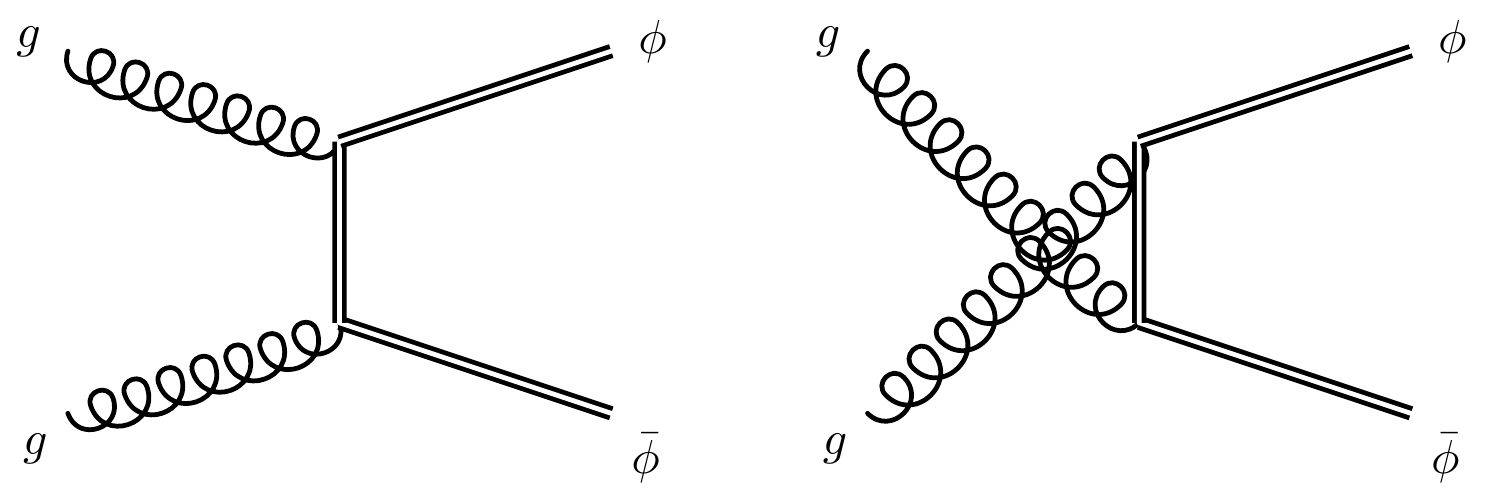}~
   \includegraphics[width=0.45\textwidth]{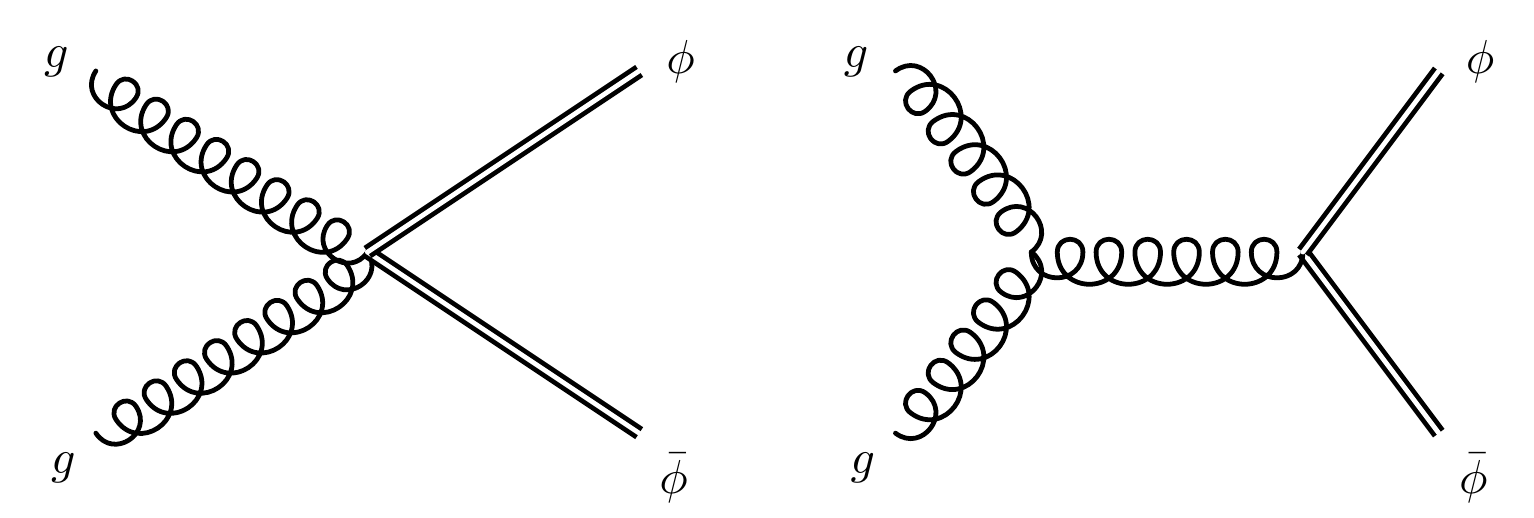}\\
   \includegraphics[width=0.45\textwidth]{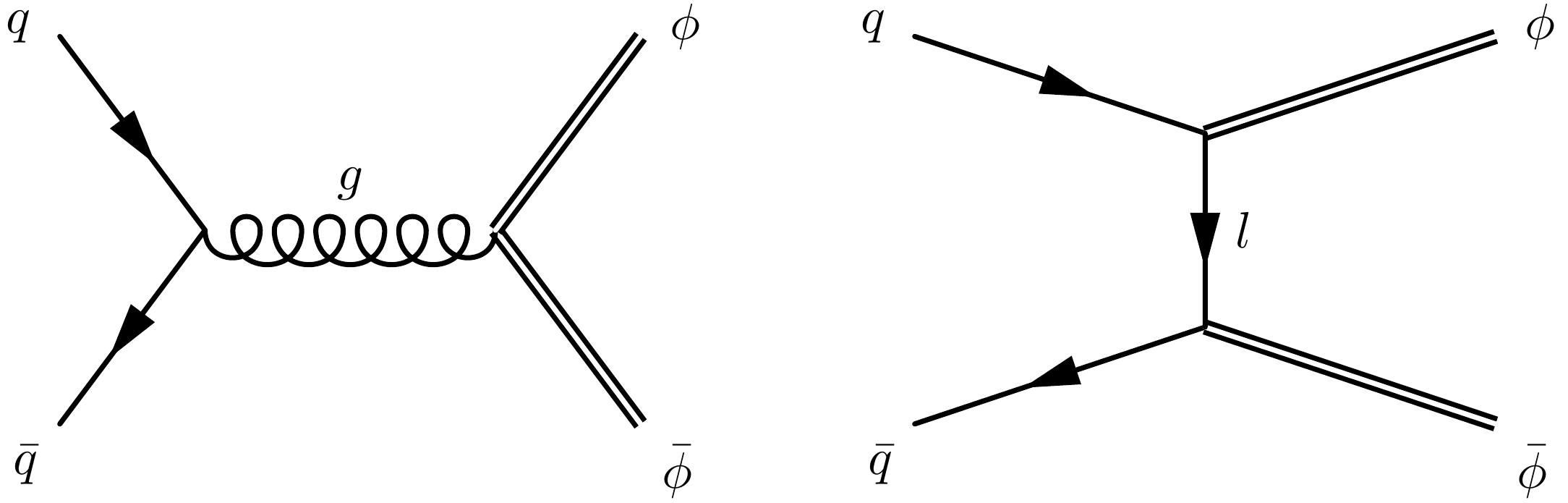} 
   \caption{Diagrams for leptoquark pair production: gluon-initiated (\emph{upper row}), quark-initiated (\emph{lower-left}), and diagram proportional to the square of the leptoquark-lepton-quark coupling, $\lambda^2$ (\emph{lower-right}).}
   \label{fig:production}
\end{figure}

The leading Feynman diagrams contributing to the LQ pair-production cross section are shown in~\figref{fig:production}.
Gluon-initiated diagrams dominate for all but the largest LQ masses. The diagram proportional to the square of the leptoquark-lepton-quark coupling $\lambda$ is model dependent but it is never important for pair production. This is because once the coupling
is large enough to make a difference for pair production it also contributes to single LQ production at the same order in $\lambda$ and single production will yield a stronger limit~\cite{Raj:2016aky, inprogress}. Therefore we will ignore this diagram in the following.

\begin{figure}[!htbp]
   \centering
   \includegraphics[width=0.48\textwidth]{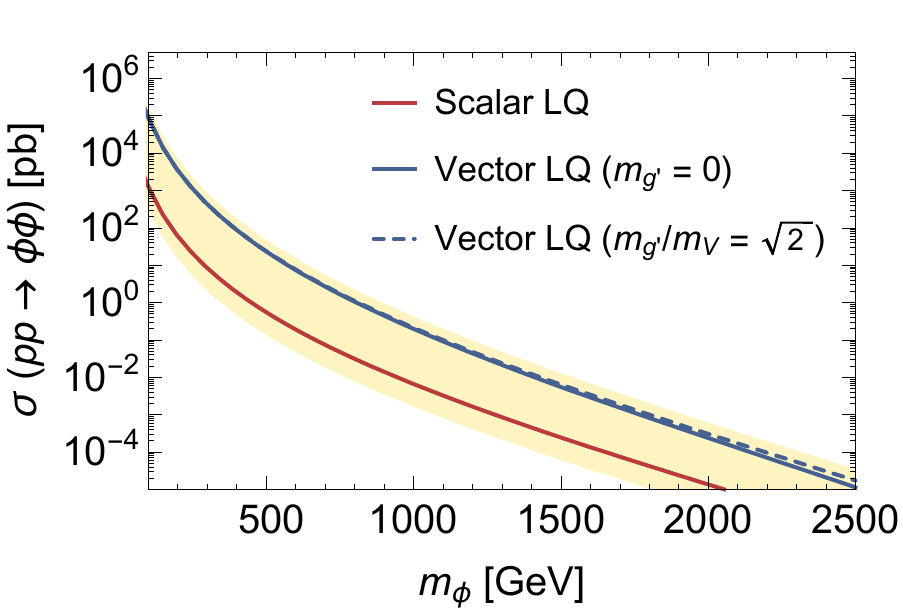} 
   \caption{Pair-production cross sections of LQs. Predictions are leading order with $K$-factors from~\cite{Blumlein:1998ym} for scalar $S$ and vector $V$ leptoquarks ($K$-factors for pair-production cross sections at hadron colliders are given in~\cite{Blumlein:1996qp, Kramer:1997hh, Blumlein:1998ym, Kramer:2004df, Mandal:2015lca}). For details of the vector model see~\appref{app:model}. The cream band is explained in~\secref{sec:flavor}.}
   \label{fig:crossec}
\end{figure}

LQ cross sections times branching fractions into the dominant final states are large because LQs carry color and are
strongly produced.
Choosing for example a leptoquark mass of 1 TeV, hundreds of scalar leptoquarks or
thousands of vector leptoquarks (would) have been produced
with the $\sim$ 36 fb$^{-1}$ data of 13 TeV  accumulated in 2016 at the LHC, see~\figref{fig:crossec}.%
\footnote{Leptoquarks carry color, charge, baryon and lepton numbers. Hence they cannot be hidden with invisible or fully hadronic decays.}
The scalar LQ cross section is model independent when the last diagram in~\figref{fig:production} is negligible.
The vector case has a subtlety related to unitarity of theories with massive vectors. A consistent theory
must have additional states beyond the LQ with masses not too far above the LQ mass. For computing the vector cross section
in the figure we used a minimal model implementation described in in ~\appref{app:model}. The model has a massive gluon partner,  $g'$, which contributes to LQ pair production. Varying the mass of the gluon partner across its theoretically allowed range we
obtain the two limiting cross sections labeled $m_{g'}/m_{V}=0,\sqrt{2}$ in~\figref{fig:crossec}.

\begin{figure}[!htbp]
   \centering
      \includegraphics[width=0.48\textwidth]{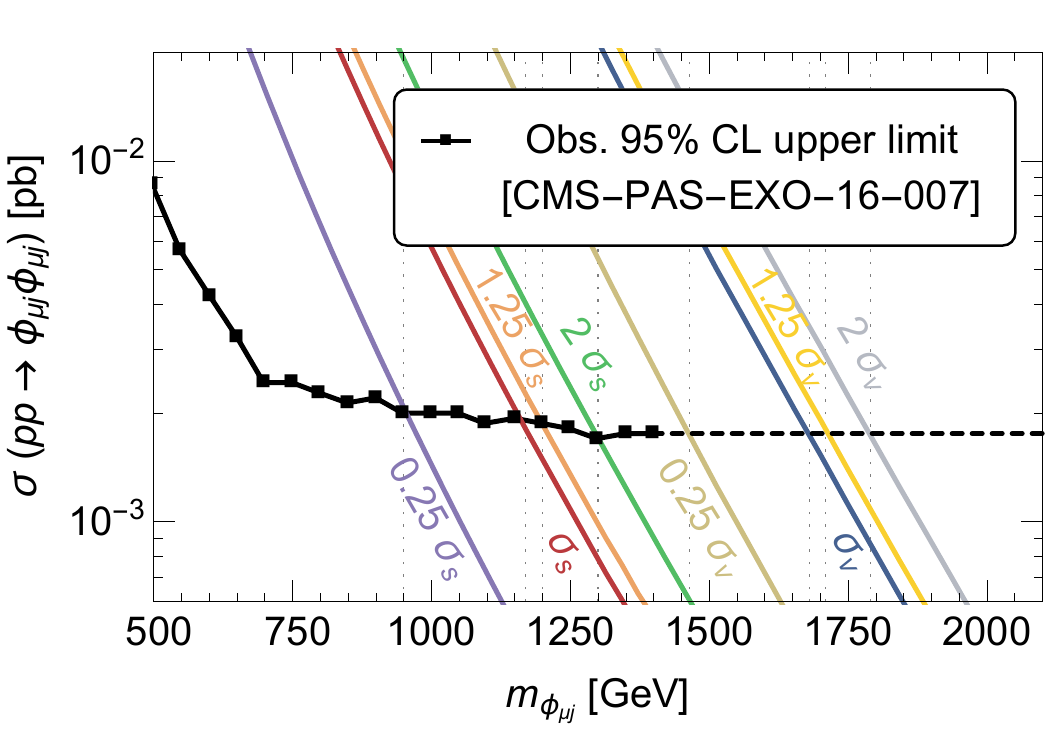}~  \includegraphics[width=0.48\textwidth]{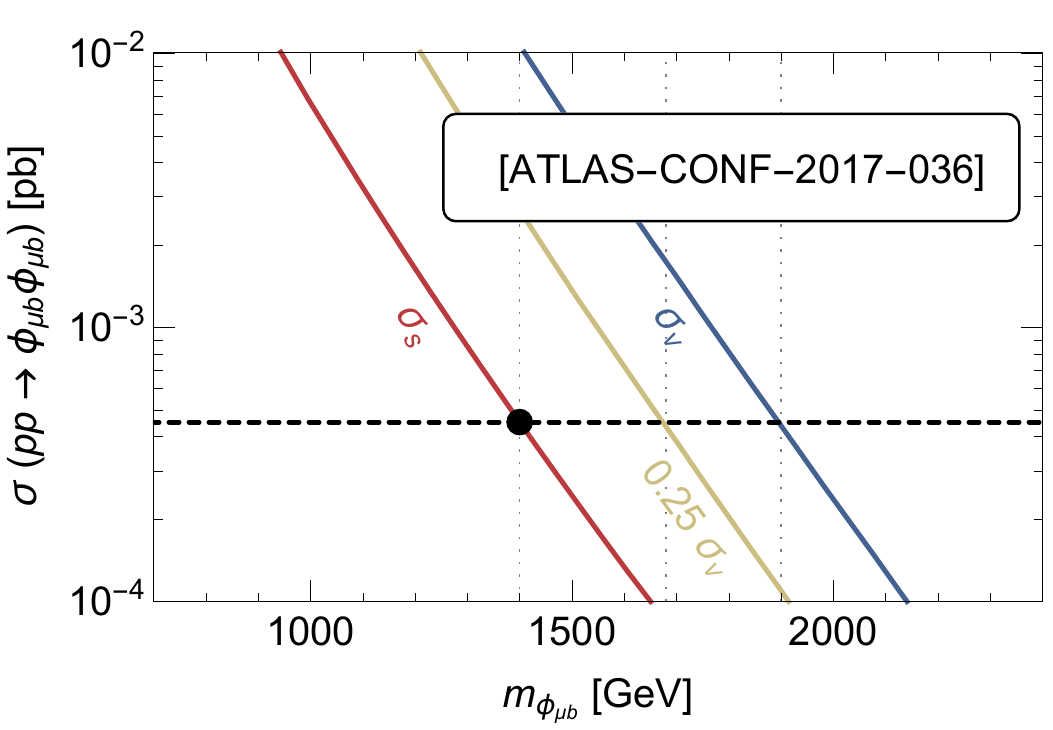} 
    \caption{(\emph{left plot}) Bounds on scalar and vector leptoquarks which decay to ($\mu j$)($\mu j$) final states. The observed 95\% CL upper limit on the production cross section times branching fraction (black solid line) is taken from a search for leptoquarks in this final state at CMS~\cite{CMS-PAS-EXO-16-007}. The search stopped at $1.4$ TeV. In order to obtain a plausible estimate for the mass limit on vector LQs we extended the cross section limit beyond 1.4 TeV by assuming that it becomes mass-independent (black dashed line). The theoretical cross sections for different MLQs which can decay to ($\mu j$)($\mu j$) final states are shown as colored lines and labeled with corresponding cross sections. The intersects (vertical dotted lines) determine the lower limits on the MLQ masses. (\emph{right plot}) Bounds on scalar and vector MLQs which decay into ($\mu b$)($\mu b$) final states. Since only a 95\% CL on the stop/leptoquark mass (black dot) is provided in~\cite{ATLAS-CONF-2017-036} we cannot derive rigorous mass bounds on other LQs. To obtain an estimate, we assume that the experimental upper limit on the cross section is flat across the LQ mass range and agrees with the value for the scalar fiducial cross section, $\sigma_S$, at the stop/leptoquark mass limit (black dashed horizontal line). Cross sections for other MLQs that can decay into ($\mu b$)($\mu b$) final states are shown as colored lines and labeled with their cross sections. The intersects (vertical dotted lines) determine our estimates for lower limits on the MLQ masses.}
   \label{fig:mujandmub}
\end{figure}

We now show as an example how to obtain lower limits on LQ masses using the LQ matrix final
state $(\mu^+ j)(\mu^- j)$, which is obtained from pair-production and decay of a LQ
decaying to muons and jets. A nice analysis for this final state has been presented by CMS~\cite{CMS-PAS-EXO-16-007}. CMS
provided a 95\% confidence level (CL) upper bound on the cross section times branching fraction to this final state
which we represent with the solid black curve in the left panel of \figref{fig:mujandmub}.
Also shown as colorful lines are the
theoretical cross sections times branching fractions for a number of Minimal LQ models which we define
in the next Section. Four of the models have scalar LQs with pair production cross sections times branching fraction
ranging from $0.25 \sigma_S$ to $2 \sigma_S$. Here $\sigma_S$ is the ``fiducial" scalar cross section shown in
\figref{fig:crossec}. The factors $0.25, 1, 1.25, 2$ correspond to multiplicity factors and branching fractions which 
will also be explained in~\secref{sec:flavor}. We just use them as examples to make the point that there
are many different LQ models which can contribute to this final state. We can easily obtain lower bounds on the
different MLQs' masses by using the plot of the experimental cross section bound.
For example, the red leptoquark labeled $\sigma_S$ is excluded up to 1.17 TeV.
This leptoquark was the designated target of the CMS search and the limit 1.17 TeV can also be found
in~\cite{CMS-PAS-EXO-16-007}. However none of the other MLQs were considered explicitly by CMS, and had the
collaboration decided to only publish the mass bound a simple determination of mass bounds for all
LQs would have been impossible. 

Unfortunately, while CMS did provide the very useful plot of the 95\% CL cross section limit they did not extend the
mass reach of their plot far enough in the leptoquark mass to allow a reliable extrapolation
to vector leptoquarks which have much larger cross sections.
In absence of a cross section limit provided by experimenters we are forced to guess. In this 
case it seems reasonable to guess that the cross section limit flattens out for large LQ masses because signal efficiencies
from very heavy LQs approach saturation. Having made this guess (the dotted black line extension of the experimental
limit) we can then put lower bounds on the masses of the four vector MLQ models. However, we emphasize that
these mass limits are estimates based on our guess for how the sensitivity of the CMS search might continue to
higher masses.

The situation depicted in the right panel of~\figref{fig:mujandmub} is less satisfactory. This plot
corresponds to the LQ matrix search for $( \mu^- b)(\mu^+ \bar b )$. ATLAS performed this search as
a SUSY search for RPV stop decays and obtained an impressive lower mass bound of 1.4 TeV. This RPV squark is identical
in production and decay to a particular scalar leptoquark, and therefore this mass bound can be directly applied to
the scalar leptoquark model.
Given this bound and our knowledge of the theoretical cross section times branching
fraction for this leptoquark we can infer that ATLAS found a 95\% CL upper bound
of $4.5\times 10^{-4}$ pb on the signal cross section for a LQ with mass 1.4 TeV (indicated with the black dot in~\figref{fig:mujandmub}). Unfortunately,
that is all the information we have. From this one data point it is not possible to infer bounds on the masses of any other
LQ models with differing cross sections or branching fractions. The best we can do is make the assumption that 
perhaps the 95\% CL limit on the cross section is approximately mass independent (as it appears to be for the CMS search
in the left panel for larger LQ masses). Using our guess for the cross section bound (horizontal dashed line) we can now
provide estimates for the mass bounds of other MLQ models by reading off the intersects of the cross section
predictions with the extrapolated cross section bound.

In the remainder of the paper we present results for LHC searches for all 9 of the LQ matrix elements. We found fully
satisfactory cross section limit plots in six cases,  somewhat less satisfactory mass limits for specific 
leptoquark models in one case and no dedicated leptoquark (or equivalent RPV squark searches)
in two cases. In all cases, we perform a recast of the bounds to obtain mass limits in our MLQ models. For the
cases where experimenters provided full 95\% CL bounds on the cross section our MLQ model limits
can be regarded with confidence. In the other cases we regard them as estimates for what the limits could have been.

We close this Section with the LQ matrix with elements filled with references to experiments which performed
searches corresponding to the LQ matrix final states. Green indicates that at least one experiment published
the 95\% CL upper bound on the pair production cross section as a function of LQ mass.

\begin{figure}[t]
   \centering
   \includegraphics[width=0.64\textwidth]{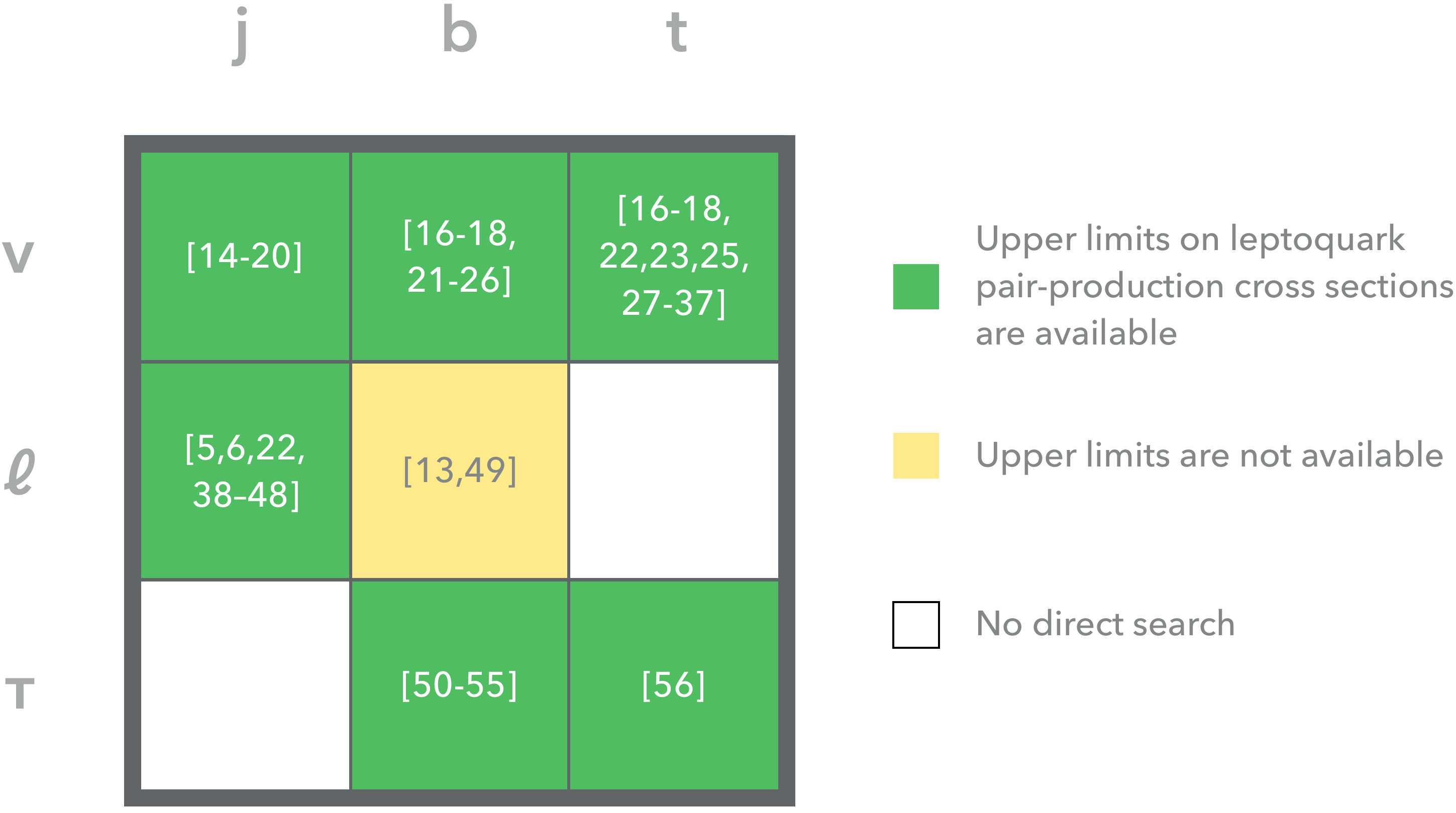} 
   \caption{Summary of the current status of LQ searches at LHC after Runs 1 and 2.  We include LQ searches and SUSY searches which have identical final states and decay topologies to LQ searches. Final states without any direct search are left white. The numbers give the references relevant to the various final states:
\hyperref[sec:nujbt]{$({\nu j})$}~\cite{Chatrchyan:2013mys, Khachatryan:2015vra, CMS-PAS-SUS-16-033,CMS-PAS-SUS-16-036,Sirunyan:2017cwe,Aad:2015gna, ATLAS-CONF-2017-022}, 
\hyperref[sec:nujbt]{$({\nu b})$} ~\cite{Aad:2013ija, Aad:2015caa, Aad:2015pfx, Aaboud:2016nwl, Sirunyan:2016jpr,Sirunyan:2017cwe,ATLAS-CONF-2017-38,CMS-PAS-SUS-16-033,CMS-PAS-SUS-16-036}, 
\hyperref[sec:nujbt]{$({\nu t})$}~\cite{Chatrchyan:2013xna, Aad:2014kra, Aad:2015caa, Aad:2015pfx,  ATLAS-CONF-2016-077, Sirunyan:2016jpr, Khachatryan:2016pxa, Khachatryan:2016oia, Khachatryan:2016pup, ATLAS-CONF-2017-020,Sirunyan:2017cwe,ATLAS-CONF-2017-024,CMS-PAS-SUS-17-001,CMS-PAS-SUS-16-049,CMS-PAS-SUS-16-033,CMS-PAS-SUS-16-036, Sirunyan:2017wif}, 
\hyperref[sec:emuj]{$({ej})$} and/or
\hyperref[sec:emuj]{$({\mu j})$}~\cite{Khachatryan:2010mq, Khachatryan:2010mp, Aad:2011ch, Aad:2011uv, Chatrchyan:2011ar, ATLAS:2012aq, Chatrchyan:2012vza, CMS-PAS-EXO-12-041, CMS-PAS-EXO-12-042, Aad:2015caa, Khachatryan:2015vaa, Aaboud:2016qeg,CMS-PAS-EXO-16-007, CMS-PAS-EXO-16-043},
\hyperref[sec:emub]{$({eb})$} and 
\hyperref[sec:emub]{$({ \mu b})$}~\cite{ATLAS-CONF-2015-015,ATLAS-CONF-2017-036},
\hyperref[sec:taub]{$({\tau b})$}~\cite{Chatrchyan:2012sv,Chatrchyan:2012st,ATLAS:2013oea,Khachatryan:2014ura,Khachatryan:2016jqo,Sirunyan:2017yrk}, and 
\hyperref[sec:taut]{$({\tau t})$}~\cite{Khachatryan:2015bsa}.}
   \label{fig:keymatrix}
\end{figure}

\section{Minimal leptoquark models}
\label{sec:flavor}

Here we define a set of MLQ models. The idea behind those models is simple. Each
model has one leptoquark in a single $SU(3)_{color}\times SU(2)_{weak}\times U(1)_Y$ representation coupling to only one
SM lepton-quark fermion bilinear. Therefore to define the MLQ models we just need to identify all possible lepton-quark
bilinears and their transformation properties under the SM gauge group. For each MLQ model we then
obtain the lower bound on the mass of the MLQ by comparing the predicted cross section times branching fraction
into LQ final states with the upper bound from experiment. To do so we need to determine the cross section times branching
fraction into the possible symmetric finals states for each MLQ. It will be convenient to parameterize the answer
as $\mu \times \sigma_S$ or $\mu \times \sigma_V$ for scalar or vector MLQs. Here $\sigma_S$ and $\sigma_V$ are the fiducial
cross sections shown in~\figref{fig:crossec} and $\mu$ is a number defined as
\beq
\mu \equiv \frac{\sigma(p p \rightarrow S \bar S)}{\sigma_S} Br(S \bar S \rightarrow {\rm final}\ {\rm state})\,, \quad
\mu \equiv \frac{\sigma(p p \rightarrow V \bar V) }{\sigma_V} Br(V \bar V \rightarrow {\rm final}\ {\rm state})\, .
\eeq
$\mu$ may be greater than 1 when multiple LQs are produced and less when branching fractions into the
final state are nontrivial.

To define the MLQ models, we identify the lepton-quark bilinears in the SM that we can couple to LQs.  Ignoring the three
generations for the moment, these are given by arbitrary quark or antiquark fields multiplying arbitrary lepton fields.
Enumerating all the possibilities is easiest when all SM fermions are interpolated by left chirality fields.%
\footnote{To pass to this notation right-chirality fields are replaced by their (left-chirality) charge-conjugates
$\psi_R \rightarrow \psi^c_L \equiv i \sigma_2 \psi_R^*$. For more on the $QUDLEN$ notation see~\appref{app:notation}.}
One generation of fermion fields is $Q,U^c,D^c,L,E^c,N^c$. Note that we have included possible singlet right-handed neutrinos. Scalar $S$ and vector $V$ leptoquarks couple to quark-lepton bilinears as
\beq
 S ({\bm Q \bm L}) \quad~\text{and} \quad  V^\mu ({\bm Q}^\dagger \bar \sigma_\mu \bm{L}) \ ,
\label{eq:bilinears}
\eeq 
respectively. Here $\bar \sigma_\mu$ is vector of $2\times 2$ matrices which contracts the vector index of the LQ with the fermion spinors. This matrix appears in all vector LQ couplings to fermions. To avoid notational clutter we suppress it in subsequent formulas. $\bm Q$ and $ \bm L$ can be any of the quark and
lepton fields, i.e., ${\bm Q} \in \{Q,U^c,D^c\}$ and ${\bm L}\in \{L,E^c,N^c\}$. For both scalars and vectors there are $3\times3$ different possibilities (before considering generation number). 

Adding in the three generations, each of the above described scalar or vector LQs we can couple to one of 3 generations
of quarks and 3 generations of leptons for an additional multiplicity of 9. Thus in total there are
$3 (Q,U^c,D^c) \times 3 (L,E^c,N^c) \times 3 ({\rm quark\ gen.}) \times 3 ({\rm lepton\ gen.}) \times 2 (S~{\rm or\ } V) = 162$
different MLQ models. Note that we are defining the MLQ models in the mass eigenstate basis for the quarks and charged leptons. In doing so, we ignore the small mismatch between mass eigenstate bases for the doublet up- and down-type quark fields, this is equivalent to ignoring leptoquark decays which are proportional to $\lambda$ times small CKM matrix elements. The basis for neutrinos is irrelevant because the different neutrino flavors are indistinguishable final states.

For example, we can have a scalar leptoquark coupling as $\lambda S (Q_1 E^c_1)$ where the subscripts
are generation indices. This is a scalar which is an $SU(2)_{weak}$ doublet. One of its components couples to up quarks and $e^+$ and the other couples to down quarks and $e^+$. Thus this MLQ model actually contains two mass-degenerate scalar leptoquarks which means that the total cross section for leptoquark production is twice as large as $\sigma_S$. Each of the two components of $S$ couple to only one fermion bilinear. Therefore both decay with 100\% branching fraction, one decays to $(e^+ u) $, the other to $(e^+ d)$. In terms 
of experimentally distinguishable final states both LQs decay to $(e^+ j)$. Thus the total cross section times branching fraction to the final state
$(e^+ j)(e^- j)$ in this model is two times the scalar fiducial cross section in~\figref{fig:keymatrix}, and the $\mu$-factor for this model is $\mu = 2$. We show the pair production and decay $\mu$-factors of all MLQs in Table~\ref{tab:multipliers}.
Consider as another example the leptoquark coupling to $U^{c\dagger}_3 E^c_1$. This is a vector leptoquark with a unique coupling and decay to top quarks and $e^+$. It is an $SU(2)_{weak}$ singlet and therefore has no multiplicity. The cross section times branching to the final state
$(e^- \bar t)(e^+ t)$ is then equal to the fiducial cross section $\sigma_V$,
and the $\mu$-factor is 1. Note that the final state of this leptoquark is distinguishable
from the final state of the scalar leptoquark coupling to $U^{c}_3 E^c_1$ which decays to $(e^+ \bar t)$ giving the pair production final state
 $(e^- t)(e^+ \bar t)$ which has the same final particles but different resonance pairs. Its $\mu$-factor is also 1.

To get MLQs with non-trivial branching fractions we must consider scalar leptoquarks coupling to $QL$ and vectors coupling to $Q^\dagger L$. In each case, there are two possibilities, the leptoquark can be isospin singlet or triplet.

Turning first to the singlet case, the scalar couples to $\epsilon_{\alpha \beta} Q_\alpha L_\beta$ while
$\delta_{\alpha \beta} Q_\alpha^\dagger L_\beta$ is the vector coupling.%
\footnote{Here the  $ \epsilon = \begin{pmatrix} 
      0 & 1 \\
      -1 & 0 \\
   \end{pmatrix}$ and $\delta = \begin{pmatrix} 
      1 & 0 \\
      0 & 1 \\
   \end{pmatrix}$.}
Consider as an example the singlet LQ coupling to third generation quarks and first generation leptons. It couples to the fermion bilinear $(t e^- - b \nu)$. Thus there are 4 distinct final states
\beq
S: \frac14\left[(e^- t)( e^+ \bar t) +  b \bar b \met +   (e^- t) \bar b \met +  ( e^+ \bar t) b \met\right]  \, ,
\eeq
where the $1/4$ indicates that each final state has branching fraction 25\%.
This is an example in which isospin predicts symmetric final states $(e^- t)( e^+ \bar t)$ and $b \bar b \met$
as well as asymmetric final states $(e^- t) \bar b \met$, $(e^+ \bar t) b \met$.
Again, we focus only on the symmetric case because it does not have neutrinos and is usually more
sensitive than the asymmetric one even though that has multiplicity of 2.
In the literature this case is sometimes called $\beta=1/2$ because the LQ
has two different decays with branching fractions $Br\equiv\beta=1/2$.
 
An isospin singlet vector coupling to third generation quarks and first generation leptons couples to $(\bar t \nu  + \bar b e^-)$ and also gives rise to
25\% branching fractions
\beq
V: \frac14 \left[( e^- \bar b)( e^+ b) +  t \bar t \met +  (e^- \bar b) t \met +  (e^+ b ) \bar t \met\right] \, .
\eeq
In both the scalar and the vector cases the branching fractions to any single final state are reduced to 25\%. There is no multiplicity factor
from the cross section, thus both have $\mu=0.25$.

An isospin triplet MLQ model contains three LQs which each couple to a different combination of fermions.
For example, choosing the MLQ with couplings to third generation quarks and first generation leptons, the couplings are to
$\left(t \nu, (t e^- + b \nu)/\sqrt{2}, b e^-\right)$ for the scalar case and $\left(\bar t e^-, (\bar t \nu - \bar b e^-)/\sqrt{2}, \bar b \nu\right)$ for the vector case. Since each of the three LQs can be pair produced with the fiducial cross section one obtains the final states
\begin{align}
S: &~t \bar t \met + (e^- b)(e^+ \bar b ) + \frac14\left[(e^- t)(e^+ \bar t )+b \bar b \met + (e^- t)\bar b \met + (e^+ \bar t ) b \met\right],\\
V: &~(e^- \bar t)(e^+ t)+ \bar b b \met + \frac14\left[(e^- \bar b )(e^+ b)+t \bar t \met + (e^- \bar b )t \met + (e^+ b) \bar t \met\right].
\end{align}
Here the most promising search channels are the final states $ (e^- b)(e^+ \bar b ) $ and $(e^- \bar t)(e^+ t)$ which occur with $\mu$-factors 1. 

Finally, for isospin triplet MLQs coupling to quarks of the first or second generation the final state quarks produce indistinguishable light-jets. Thus here the final states are
\begin{eqnarray}
S~\text{or}~V: \frac54\left[(e^- j)(e^+ j) + jj\met \right] + \frac14\left[(e^- j)j \met + (e^+ j) j \met\right].
\end{eqnarray}
yielding $\mu$-factors of $1.25$ for both of the symmetric final states $(e^- j)(e^+ j)$ and $jj\met$.

We summarize all possible $\mu$-factors in \tabref{tab:multipliers}. For each final state, i.e. each row in the table, one can read off the $\mu$-factor for different MLQ models contributing to this final state. 
Alternatively, given a particular MLQ model defined by the MLQs' coupling to SM fermion bilinears $[{\bm L \bm Q}]$, the entries in the corresponding column show the $\mu$-factors for each attainable LQ matrix final state. For example, for $S$ coupling to $ [{Q_3 L_2}~\rm{triplet}]$, the allowed final states are $({\nu b})$, $({\nu t})$, $({\mu b})$, and $({\mu t})$
with corresponding $\mu$-factors $0.25, 1, 1, 0.25$. 
Thus the cross sections for $p p \to ({\nu t}) ({\nu \bar t})$ and $p p \to ({\mu b}) ({\mu b})$ are 4 times larger
than $p p \to ({\nu b}) ({\nu b})$ or $p p \to ({\mu t}) ({\mu \bar t})$ and hence better suited for putting bounds on the mass of this MLQ.

\begin{table}[!htbp]
   \caption{(\emph{Upper table}) $\mu$-factors for production and decay
$\sigma(p p \to S \bar S \to ({\bm l \bm q}) ({ \bar {\bm l}  \bar {\bm q}}))$ in different scalar MLQ models. Rows correspond to different final states whereas columns correspond to different MLQ models. For example, the first row of the first column corresponds to pair production of a scalar LQ which decays to $(\nu j)$ via a coupling in which $Q$ and $L$ are contracted into an isospin triplet. 
For any final state $(\bm q \bm l)(\bm q \bm l)$, the $\mu$-factor is the factor which multiplies the fiducial cross section $\sigma_S$ in \figref{fig:crossec} to account for multiplicities and branching fractions.
MLQs which couple to $SU(2)_{weak}$-singlet fermions always have $\mu=1$. A ``$\times$" entry indicates that that MLQ model does not contribute to the final state. (\emph{Lower table}) $\mu$-factors for the vector leptoquark $V$. Comparing to the scalar case, differences exist only in the first two columns. Note that the $\mu$-factors for scalar MLQs decaying to $({\nu t}), ({eb}),({\mu b}),({\tau b})$ and vector MLQs $({\nu b}), ({e t}), ({\mu t}), ({\tau t})$ are equal to 1 irrespective of MLQ model.}
   \label{tab:multipliers}
   \centering
   \vspace{0.1in}
   \begin{tabular}{@{} ccccccccccc @{}} 
      \hline
   $pp\to S \bar S \to ({\bm {l q}})({\bm {\bar l\bar q}})$         & $QL$ triplet &  $QL$ singlet & $U^c L$ & $D^c L$ & $QE^c$ & $U^cE^c$ & $D^cE^c$ & $QN^c$ & $U^cN^c$ & $D^cN^c$ \\
               \hline
$({\nu j})$ & 1.25 & 0.25 & 1 & 1 & $\times$ & $\times$ & $\times$ & 2 & 1 & 1 \\
$({\nu b})$ & 0.25 & 0.25 & $\times$ & 1 & $\times$ & $\times$ & $\times$ & 1 & $\times$ & 1 \\
$({\nu t})$ & 1 & $\times$ & 1 & $\times$ & $\times$ & $\times$ & $\times$ & 1 & 1 & $\times$ \\
$({ej})$, $({\mu j})$, $({\tau j})$ & 1.25 & 0.25 & 1 & 1 & 2 & 1 & 1 & $\times$ & $\times$ & $\times$\\
$({eb})$, $({\mu b})$, $({\tau b})$ & 1 & $\times$ & $\times$ & 1 & 1 & $\times$ & 1 & $\times$ & $\times$ & $\times$\\
$({et})$, $({\mu t})$, $({\tau t})$ & 0.25 & 0.25 & 1 & $\times$ & 1 & 1 & $\times$ & $\times$ & $\times$ & $\times$\\
      \hline
      \hline
      $pp\to V \bar V \to ({\bm {lq}})({\bm {\bar l\bar q}})$           & $Q^\dag L$ triplet &  $Q^ \dag L$ singlet & ${U^c}^\dag L$ & ${D^c}^\dag L$ & $Q^\dag E^c$ & ${U^c}^\dag E^c$ & ${D^c}^\dag E^c$ & $Q^\dag N^c$ & ${U^c}^\dag N^c$ & ${D^c}^\dag N^c$ \\
               \hline
$({\nu j})$ & 1.25 & 0.25 & 1 & 1 & $\times$ & $\times$ & $\times$ & 2 & 1 & 1 \\
$({\nu b})$ & 1 & $\times$ & $\times$ & 1 & $\times$ & $\times$ & $\times$ & 1 & $\times$ & 1 \\
$({\nu t})$ & 0.25 & 0.25 & 1 & $\times$ & $\times$ & $\times$ & $\times$ & 1 & 1 & $\times$ \\
$({ej})$, $({\mu j})$, $({\tau j})$ & 1.25 & 0.25 & 1 & 1 & 2 & 1 & 1 & $\times$ & $\times$ & $\times$\\
$({eb})$, $({\mu b})$, $({\tau b})$ & 0.25 & 0.25 & $\times$ & 1 & 1 & $\times$ & 1 & $\times$ & $\times$ & $\times$\\
$({et})$, $({\mu t})$, $({\tau t})$ & 1 & $\times$ & 1 & $\times$ & 1 & 1 & $\times$ & $\times$ & $\times$ & $\times$\\
      \hline
   \end{tabular}
\end{table}

To place bounds on all MLQ models with the same final state, it is important for the experimental collaborations to search for sufficiently high and low LQ masses. The lower masses are required to set bounds on MLQs with small cross sections whereas the highest masses are relevant for MLQs with large cross sections. The range in cross sections of MLQ models is bounded by 
$0.25\times \sigma_S$ and $2\times \sigma_V$. We illustrate this range as the cream band in~\figref{fig:crossec}. 
Other models can in principle have even larger or smaller $\mu$ factors because of larger LQ multiplicities or smaller branching fractions to the final state under consideration. Thus a search mass window that is as large as possible is desirable.

\section{Leptoquark searches organized by the LQ matrix}
\label{sec:matrix}

LQs may be classified by their decay products and organized into the LQ matrix. The symmetric final states from pair-production where the LQ and
its antiparticle decay in the same way give the strongest bounds. We therefore focus on the symmetric final states which can also be organized into the LQ matrix.
The current status of searches for each final state is summarized in \figref{fig:keymatrix}. The color of each entry of the matrix indicates whether a cross section bound as a function of mass has been made available; the numbers indicate relevant references.%
\footnote{Searches in asymmetric final states  $(\ell j )(\nu j)$ are given in~\cite{Aad:2011ch, Aad:2011uv, ATLAS:2012aq, Khachatryan:2015vaa}. Searches in asymmetric final states $(\tau t) (\nu b)$ and $(\tau b) (\nu t)$ are given in~\cite{Gripaios:2010hv}. } 
In the following subsections we review LQ searches organized by entries of the LQ matrix.

\subsection{${\nu j}$, ${\nu b}$ and ${\nu t}$}
\label{sec:nujbt}
\begin{figure}[!htbp]
   \centering
   \includegraphics[width=0.7\textwidth]{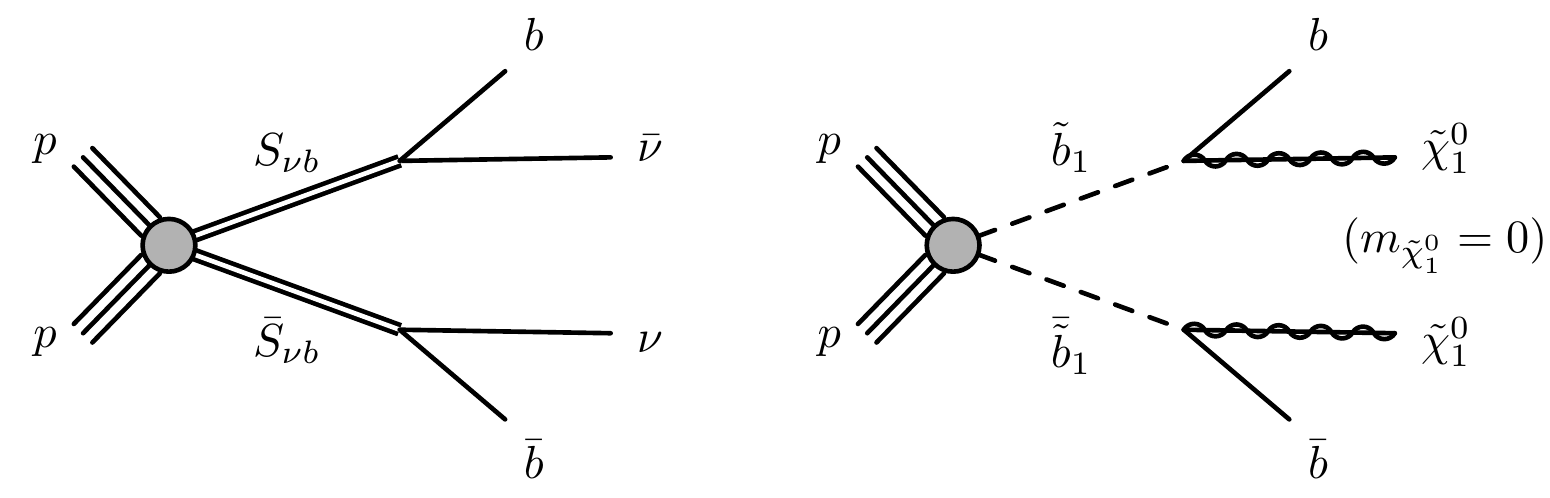} 
   \caption{Diagrams for the process $pp\to S_{\nu b} \bar{S}_{\nu b}\to (b \bar{\nu})(\bar b\nu)$ (left) and the corresponding SUSY process $pp\to \t b_1 \bar{\t b}_1 \to (b \tilde \chi_1^0)( \bar b \t \chi_1^0)$ (right). For massless neutralinos the final-states and kinematic distributions of the two processes are identical. Similar identifications can be made for other  $S_{\nu \bm q}$ and squark searches. Here the ``1" subscripts are the usual SUSY notation for lightest $b$-squark and lightest neutralino.}
   \label{fig:feyn1}
\end{figure}

Searches for pair-produced scalar leptoquarks decaying into $(\nu j)(\nu j)$, $(\nu b)(\nu b)$, and $(\nu t)(\nu t)$ final states are identical to those for pair-produced squarks $\t q_1$~\cite{Chatrchyan:2013mys, Khachatryan:2015vra, CMS-PAS-SUS-16-033,CMS-PAS-SUS-16-036,Sirunyan:2017cwe,Aad:2015gna, ATLAS-CONF-2017-022} ($\t q_1 = \t u_1$, $\t d_1$, $\t s_1$, or $\t c_1$), $\t b_1$~\cite{Aad:2013ija, Aad:2015caa, Aad:2015pfx, Aaboud:2016nwl, Sirunyan:2016jpr,Sirunyan:2017cwe,ATLAS-CONF-2017-38,CMS-PAS-SUS-16-033,CMS-PAS-SUS-16-036}, and $\t t_1$~\cite{Chatrchyan:2013xna, Aad:2014kra, Aad:2015caa, Aad:2015pfx,  ATLAS-CONF-2016-077, Sirunyan:2016jpr, Khachatryan:2016pxa, Khachatryan:2016oia, Khachatryan:2016pup, ATLAS-CONF-2017-020,Sirunyan:2017cwe,ATLAS-CONF-2017-024,CMS-PAS-SUS-17-001,CMS-PAS-SUS-16-049,CMS-PAS-SUS-16-033,CMS-PAS-SUS-16-036,Sirunyan:2017wif}, namely the pair-produced particles share the same spin, charge, decay final states and kinematics, if the neutralino $\t \chi_1^0$ is taken to be massless. An example is shown in~\figref{fig:feyn1} with the leptoquark production and decay diagram on the left and the corresponding squark diagram on the right. Thus constraints on squarks can be directly applied to $S_{ \nu \bm q}$ and vise visa. Such a translation has already been performed by ATLAS~\cite{Aad:2015caa}. We obtain the strongest 95\% CL level limits by applying squark bounds from CMS Run 2 analyses, $m_{S_{\nu j}} \geq 1.05 \tev$~\cite{CMS-PAS-SUS-16-033,CMS-PAS-SUS-16-036},
$m_{S_{\nu b}}\geq 1.175 \tev$~\cite{CMS-PAS-SUS-16-036}, and $m_{S_{\nu t}} \geq 1.07 \tev$~\cite{CMS-PAS-SUS-16-036}. 

\begin{figure}[htbp]
   \centering
   \includegraphics[width=0.48\textwidth]{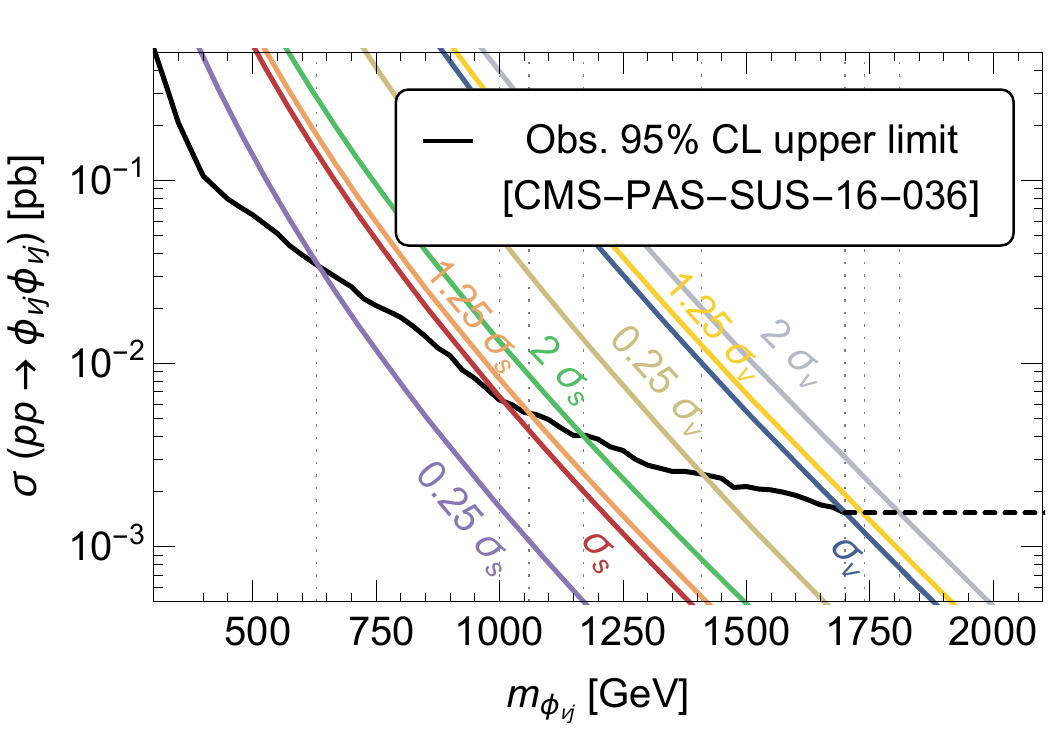} 
   \caption{Bounds on scalar and vector MLQs which decay into ($\nu j$)($\nu j$) final states. The observed 95\% CL upper limit on the production cross section (black solid line) is taken from light squark searches at CMS~\cite{CMS-PAS-SUS-16-036}. The search stopped at $1.7$ TeV, but we extrapolated the cross section limit by assuming that it becomes independent of LQ mass for large masses (black dashed line). The theoretical cross sections for different MLQs which can decay into ($\nu j$)($\nu j$) final states are shown as colored lines and labeled with their cross sections. The intersects (vertical dotted lines) correspond to lower limits on the MLQ masses.}
   \label{fig:nuj}
\end{figure}

The recast of the bounds to other MLQs is performed as follows. We digitize the temperature plot for the observed  95\% CL upper limit on the light squark pair production cross section from searches at CMS~\cite{CMS-PAS-SUS-16-036}, setting $m_{\t \chi^0_1}=0$. The digitized cross section limit is shown as  the  black solid curve in \figref{fig:nuj}. The colored lines in the same plot show the theoretical cross sections of different MLQs that can decay into $(\nu j)(\nu j)$ final states. The curves are labeled with the cross sections given in terms of the $\mu$-factor times the relevant fiducial cross section, $\sigma_S$ or $\sigma_V$, for scalar or vector LQs respectively. The intersects (dotted black lines) of the cross section curves with the experimental bound correspond to the lower mass bounds obtained in the different models. We obtain the bounds $630 \gev$, $1 \tev$, $1.1 \tev$, and $1.2 \tev$ for scalar leptoquarks with $\mu$-factors of 0.25, 1, 1.25, and 2, respectively, and $1.4$ TeV and $1.7$ TeV for vector leptoquarks with $\mu$-factors of 0.25 and 1, respectively. The CMS search stopped at 1.7 TeV and therefore we cannot obtain reliable experimental bounds on heavier leptoquark candidates. To get an estimate, we extrapolated the experimental bound on the cross sections as independent of LQ mass for masses larger than 1.7 TeV (shown as the black dotted line). This allows us to recast bounds on vector leptoquarks of 1.75 TeV and 1.8 TeV for $\mu$-factors of 1.25 and 2 respectively. Referring to the row labeled $S_{\nu j}$ and $V_{\nu j}$ in \tabref{tab:multipliers} we can then associate the mass bounds corresponding to different $\mu$-factors to the different MLQ models. These bounds are summarized in~\tabref{tab:nuj}, \tabref{tab:nub}, and \tabref{tab:nut}. Mass bounds in parenthesis are estimated bounds obtained using our extrapolated experimental cross section limits. Mass bounds without parenthesis are rigorous, they rely only on cross section bounds published by the experiments.

\begin{table}[!htbp]
\caption{(\emph{upper})~95\% CL lower limit on the mass of scalar MLQs $S_{\nu j}$ obtained from~\figref{fig:nuj}. The bounds apply to LQs coupling to the first or second generation quarks and neutrinos of any generation. (\emph{lower}) 95\% CL lower limit on the  mass of vector MLQs $V_{\nu j}$ which decay into ($\nu j$)($\nu j$) final states. For bounds greater than 1.7 TeV, we have assumed that the experimental upper limit on the cross section is flat as the mass increases. Since these bounds are based on our assumption they cannot be considered solid and we marked them with parentheses. Note that~\cite{CMS-PAS-SUS-16-036} obtain a mass bound of 1050 GeV while our plot~\figref{fig:nuj} shows an intersect at $\sim$ 1 TeV, this small discrepancy may be a reflection of the inaccuracy of our digitization of the experimental cross section bound from the provided temperature plot or a difference in assumed $K$-factor. }  
   \vspace{0.1in}
   \centering
   \begin{tabular}{@{}| c c| c |c| c| c| @{}} 
      \hline
           & & $U^c L, D^c L, U^cN^c, D^c N^c$&  $Q N^c$ & $Q L$ triplet & $Q L$ singlet \\
           \hline
     \multirow{2}{*}{ $S_{\nu j}$}   & $\sigma_\text{prod}  $ & $\times1$  & $\times2$ & $\times1.25$ & $\times 0.25$ \\     
      & $m_{S_{\nu j}}$  & $\geq 1 \tev$~\cite{CMS-PAS-SUS-16-036} & $\geq  1.2 \tev$ & $\geq 1.1 \gev$  & $\geq 630 \gev$  \\
       \hline
           \hline
           & & $U^{c\dag} L, D^{c\dag} L, U^{c\dag}N^c, D^{c\dag} N^c$&  $Q^\dag N^c$ & $Q^\dag L$ triplet & $Q^\dag L$ singlet \\
           \hline
     \multirow{2}{*}{ $V_{\nu j}$}   & $\sigma_\text{prod}  $ & $\times1$  & $\times2$ & $\times1.25$ & $\times 0.25$ \\     
      & $m_{V_{\nu j}}$  & $\geq 1.7  \tev$ & $(\geq  1.8 \tev)$ & $(\geq  1.75 \tev)$  & $\geq  1.4 \tev$  \\
       \hline  
   \end{tabular}
   
        \label{tab:nuj}
\end{table}

\begin{figure}[htbp]
   \centering
   \includegraphics[width=0.48\textwidth]{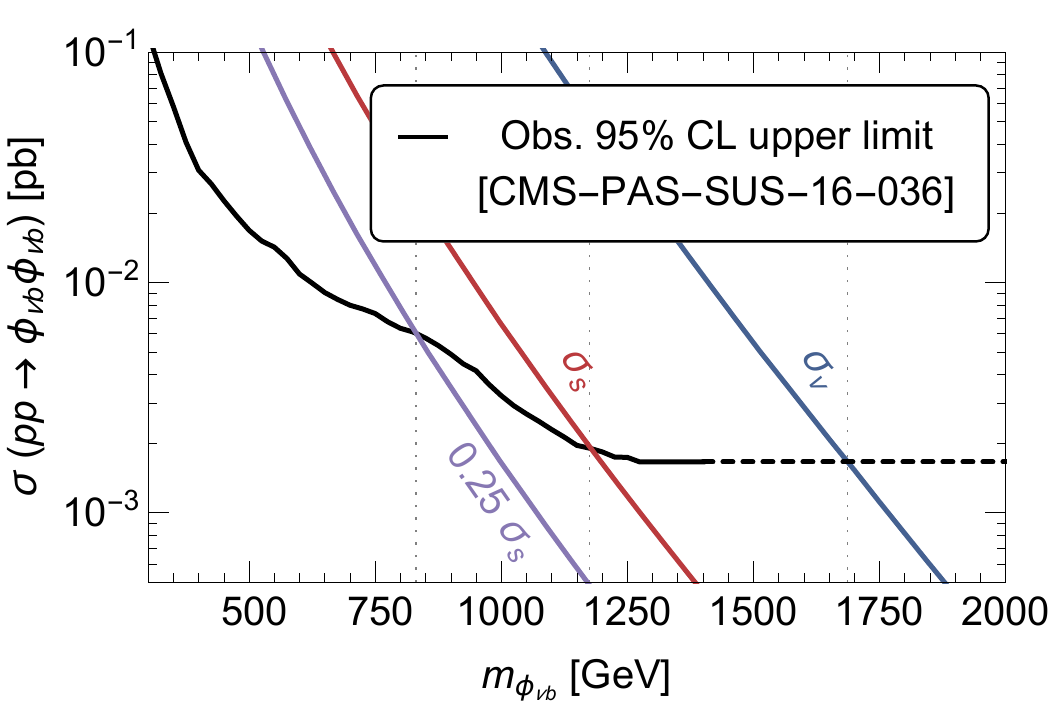} 
   \caption{Bounds on scalar and vector MLQs that decay into ($\nu b$)($\nu b$) final states. The observed 95\% CL upper limit on the production cross section (black solid line) is taken from sbottom searches at CMS~\cite{CMS-PAS-SUS-16-036} which stopped at 1.4 TeV. Our analysis parallels that of~\figref{fig:nuj}.}
   \label{fig:nub}
\end{figure}

\begin{table}[!htbp]
   \caption{95\% CL lower limits on the masses of MLQs which decay into ($\nu b$)($\nu b$) final states for the scalar LQs (\emph{left}) and vector MLQs (\emph{right}).}
      \vspace{0.1in}
   \begin{tabular}{@{} |c c |c |c| @{}} 
      \hline
           & & $D^c L, Q N^c, D^c N^c$  & $ (Q L$ triplet), $Q L$ singlet\\
           \hline
     \multirow{2}{*}{ $S_{\nu b}$}   & $\sigma_\text{prod}  $ & $\times1$  & $\times 0.25$  \\     
      & $m_{S_{\nu b}}$  & $\geq 1.18 \tev$~\cite{CMS-PAS-SUS-16-036}&$ \geq  830 \gev$   \\
       \hline
   \end{tabular}
   \quad
    \begin{tabular}{@{}| c c| c|  @{}} 
      \hline
           & & $D^{c\dag} L, Q^\dag N^c, D^{c\dag} N^c, (Q^\dag L$ triplet)\\
           \hline
     \multirow{2}{*}{ $V_{\nu b}$}   & $\sigma_\text{prod}  $ & $\times1$   \\     
      & $m_{V_{\nu b}}$  & $(\geq 1.7 \tev)$    \\
       \hline
   \end{tabular}
   \label{tab:nub}
\end{table}

\begin{figure}[htbp]
   \centering
   \includegraphics[width=0.48\textwidth]{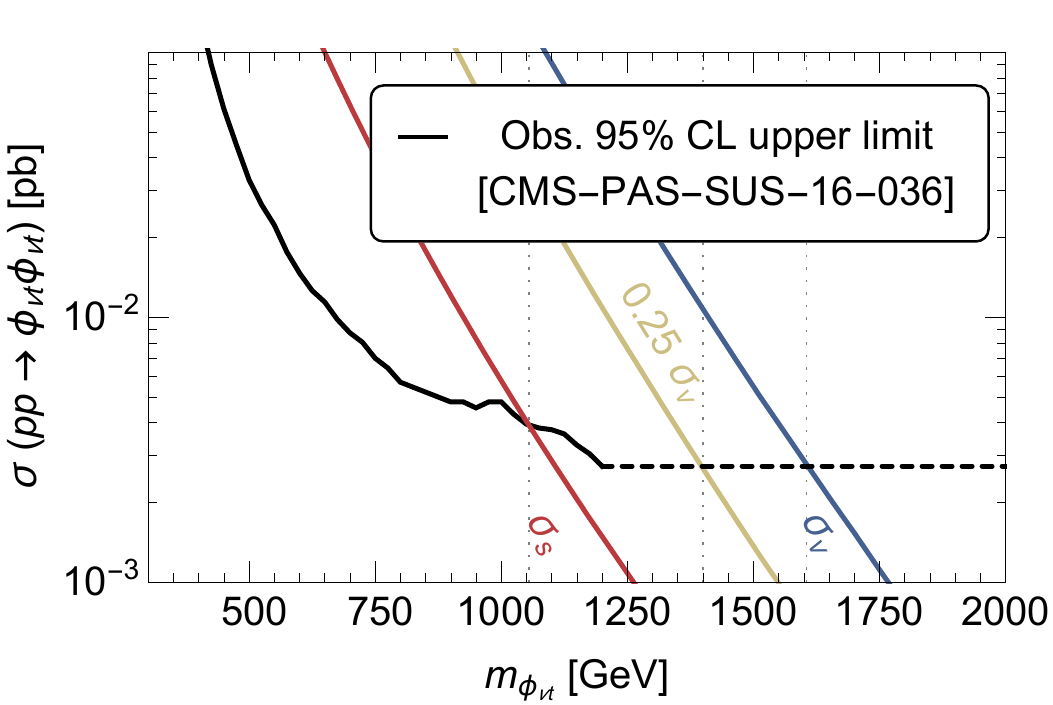} 
   \caption{Bounds on scalar and vector MLQs which decay into ($\nu t$)($\nu t$) final states. The observed 95\% CL upper limit on the production cross section (black solid line) is taken from stop searches at CMS~\cite{CMS-PAS-SUS-16-036} which stopped at 1.2 TeV. Our analysis parallels that of~\figref{fig:nuj}.}
   \label{fig:nut}
\end{figure}

\begin{table}[!htbp] 
   \caption{ 95\% CL lower limits on the masses of MLQs which decay into ($\nu t$)($\nu t$) final states for scalar LQs (\emph{left}) and vector LQs (\emph{right}).}
      \vspace{0.1in}
   \begin{tabular}{@{} |c c| c|  @{}} 
      \hline
           & &  $U^{c} L, Q N^c, U^{c} N^c\!, (Q L$ triplet)\\
           \hline
     \multirow{2}{*}{ $S_{\nu t}$}   & $\sigma_\text{prod}  $ & $\times1$   \\     
      & $m_{S_{\nu t}}$  & $\geq 1.07 \tev$~\cite{CMS-PAS-SUS-16-036} \\
       \hline
   \end{tabular}
   \quad
   \begin{tabular}{@{} |c c| c| c| @{}} 
      \hline
           & & \quad $U^{c\dag} L, Q^\dag N^c, U^{c\dag} N^c$\, &   $Q L$ triplet, $Q L$ singlet\\
           \hline
     \multirow{2}{*}{ $V_{\nu t}$}   & $\sigma_\text{prod}  $ & $\times1$  & $\times 0.25$ \\     
      & $m_{V_{\nu t}}$  & $(\geq 1.6 \tev)$   & $(\geq 1.4 \tev)$    \\
       \hline
   \end{tabular}
\label{tab:nut}
\end{table}

In addition ATLAS and CMS also looked for  asymmetric decays into $(ej)(\nu j)$ or $(\mu j)(\nu j)$ final states. These searches require leptoquark decays to both $(e j)$ and $( \nu j)$ which is true for the $SU(2)_{weak}$ singlet or triplet coupling to $QL$ or $Q^\dagger L$. 
The latest 95\% CL upper limit on the leptoquark mass from those channels is $m_{S_{\nu j}} \geq 850 \gev$~\cite{Khachatryan:2015vaa}. This is comparable to the bound from searches for the symmetric final state $(ej)(ej)$ in the case of a singlet leptoquark coupling to $[{Q_{1,2} L_{1,2,3}}~\rm{singlet}]$, but significantly weaker than the bound from the symmetric search for the triplet leptoquark coupling to  $[{Q_{1,2} L_{1,2,3}}~\rm{triplet}]$, see~\tabref{tab:nuj}. 

Also note that many searches for $(\nu j)(\nu j)$ final states and light squark searches did not veto $c$-jets. Therefore the limits also apply to $\phi_{\nu c}$. One might hope to improve the sensitivity of these searches to $\phi_{\nu c}$ with $c$-tagging. However, $c$-tagging efficiencies are currently too small for this to be effective. For example, Ref.~\cite{Aad:2015gna}  ($8 \tev$, 20 $\fb$) conducted a $\t c_1$ search with efficiencies: $c$-tag 19\%, $b$-misstagged-as-$c$ 13\%,  and $j$-misstagged-as-$c$ 0.5\%. They obtained the bound $m_{\t c_1(S_{\nu c})} \geq 540 \gev$ which is weaker than the bound from the flavor inclusive analysis with the same data set~\cite{Khachatryan:2015vra}, $m_{\t c_1(S_{\nu c})} \geq 580 \gev$. Efficiencies of $c$-tagging did not improve significantly in Run 2~\cite{CMS-PAS-BTV-16-001}.

\subsection{${e j}$ and ${\mu j}$}
\label{sec:emuj}

\begin{figure}[htbp]
   \centering
   \includegraphics[width=0.48\textwidth]{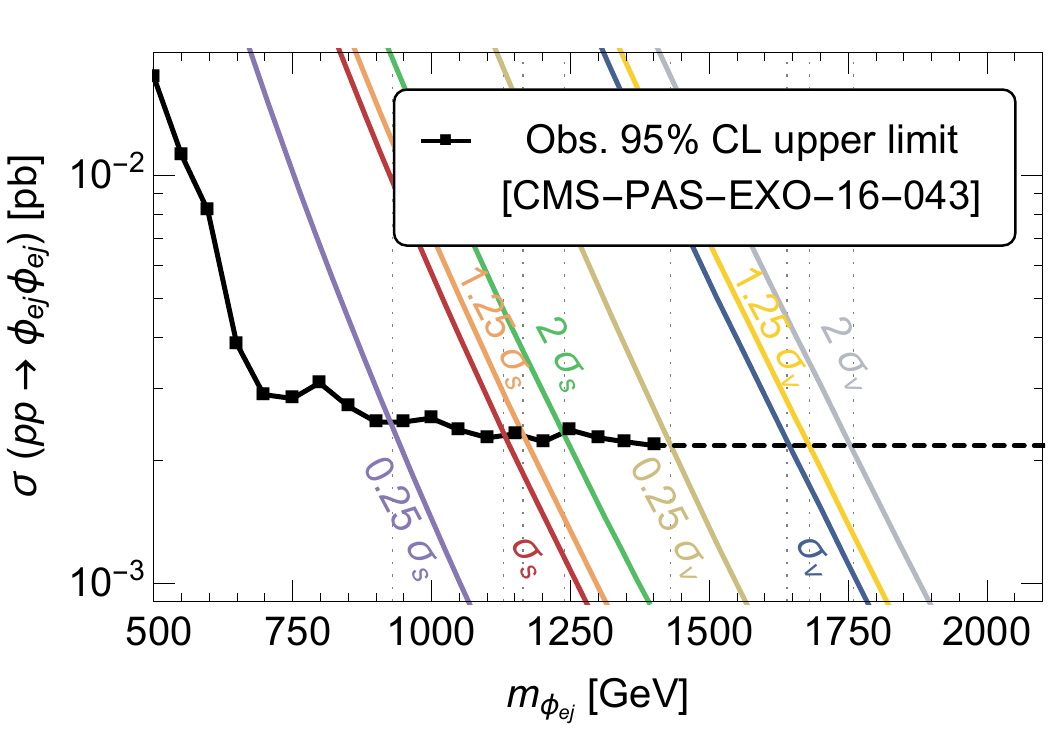} 
   \caption{Bounds on scalar and vector leptoquarks that decay into ($e j$)($e j$) final states. The observed 95\% CL upper limit on the production cross section (black solid line) is taken from a search for leptoquarks in the  ($e j$)($e j$) final state by CMS~\cite{CMS-PAS-EXO-16-043} which stopped at 1.4 TeV. See the caption of~\figref{fig:nuj} for details.}
   \label{fig:ej}
\end{figure}

Leptoquarks decaying into $(e j)(e j)$ and $(\mu j)(\mu j)$ have been a focus of LHC leptoquark searches with both Run 1 and Run 2 data~\cite{Khachatryan:2010mq, Khachatryan:2010mp, Aad:2011ch, Aad:2011uv, Chatrchyan:2011ar, ATLAS:2012aq, Chatrchyan:2012vza, CMS-PAS-EXO-12-041, CMS-PAS-EXO-12-042, Aad:2015caa, Khachatryan:2015vaa, Aaboud:2016qeg,CMS-PAS-EXO-16-007, CMS-PAS-EXO-16-043}. No significant signal has been discovered, and 95\% CL upper limits on leptoquark masses are: $m_{S_{ej}} \geq 1.13 \tev$~\cite{CMS-PAS-EXO-16-043} and $m_{S_{\mu j}}\geq 1.17 \gev$~\cite{CMS-PAS-EXO-16-007}. 
We summarize these bounds and our recasts from the cross section plots~\figref{fig:mujandmub} and~\figref{fig:ej} in~\tabref{tab:emuj}.
Note that there is a stronger limit on $m_{S_{ej}} \geq 1.73 \tev$ from single leptoquark production~\cite{Khachatryan:2015qda}. However this limit assumes a large coupling ($\lambda \ge 1$) between up quarks and the leptoquark. For smaller couplings the single leptoquark cross section becomes very small and no limit can be obtained.

\begin{table}[!htbp]
        \caption{ (\emph{upper table}) 95\% CL lower limits on the masses of the scalar MLQs $S_{e j}$  and $S_{\mu j}$. The mass bounds in the first column are from \cite{CMS-PAS-EXO-16-043} and \cite{CMS-PAS-EXO-16-007} while the mass bounds in the remaining columns are our recasts, which use our $\mu$-factors and the experimental bounds on the production cross section from \cite{CMS-PAS-EXO-16-043} and \cite{CMS-PAS-EXO-16-007} respectively. The SM fermions involved in the couplings are the first- or second-generation quarks and electrons or muons. (\emph{lower table}) 95\% CL recast lower limits on the masses of vector leptoquarks $V_{e j}$ and $V_{\mu j}$.}
           \vspace{0.1in}
   \centering
   \begin{tabular}{@{}| c c |c| c| c| c| @{}} 
      \hline
           & & $U^c L, D^c L, U^cE^c, D^cE^c$&  $QE^c$ & ($QL$ triplet) & ($QL$ singlet) \\
           \hline
     \multirow{2}{*}{ $S_{e j}$}   & $\sigma_\text{prod}  $ & $\times1$  & $\times2$ & $\times1.25$ & $\times 0.25$ \\     
      & $m_{S_{e j}}$  & $\geq 1.13 \tev$~\cite{CMS-PAS-EXO-16-043} & $\geq 1.2 \tev$ & $\geq 1.2 \tev$  & $\geq 930 \gev$  \\
           \hline
     \multirow{2}{*}{ $S_{\mu j}$}   & $\sigma_\text{prod}  $ & $\times1$  & $\times2$ & $\times1.25$ & $\times 0.25$ \\     
      & $m_{S_{\mu j}}$  & $\geq 1.17 \tev$~\cite{CMS-PAS-EXO-16-007} & $\geq 1.3 \tev$ & $\geq 1.2 \tev$  & $\geq 950 \gev$  \\
       \hline
        \hline
           & & $U^{c\dag} L, D^{c\dag} L, U^{c\dag}E^c, D^{c\dag}E^c$&  $Q^\dag E^c$ & $Q^\dag L$ triplet & $Q^\dag L$ singlet \\
           \hline
     \multirow{2}{*}{ $V_{e j}$}   & $\sigma_\text{prod}  $ & $\times1$  & $\times2$ & $\times1.25$ & $\times 0.25$ \\     
      & $m_{V_{e j}}$  & $(\geq 1.6 \tev)$ & $(\geq 1.8 \tev)$ & $(\geq 1.7 \tev)$  & $(\geq 1.45 \tev)$  \\
           \hline
     \multirow{2}{*}{ $V_{\mu j}$}   & $\sigma_\text{prod}  $ & $\times1$  & $\times2$ & $\times1.25$ & $\times 0.25$ \\     
      & $m_{V_{\mu j}}$  & $(\geq 1.7 \tev)$ & $(\geq 1.8 \tev)$ & $(\geq 1.7 \tev)$  & $(\geq 1.5\tev)$  \\
       \hline
   \end{tabular}
        \label{tab:emuj}
\end{table}

\subsection{${eb}$ and ${\mu b}$}
\label{sec:emub}

Many searches for $(e j)(e j)$ or $(\mu j)(\mu j)$ final states did not veto heavy flavors and their bounds apply to $(eb)(eb)$ or  $(\mu b)(\mu b)$ as well. However, one can improve the sensitivity of the searches with $b$-tagging.  

\begin{figure}[!htbp]
   \centering
   \includegraphics[width=0.7\textwidth]{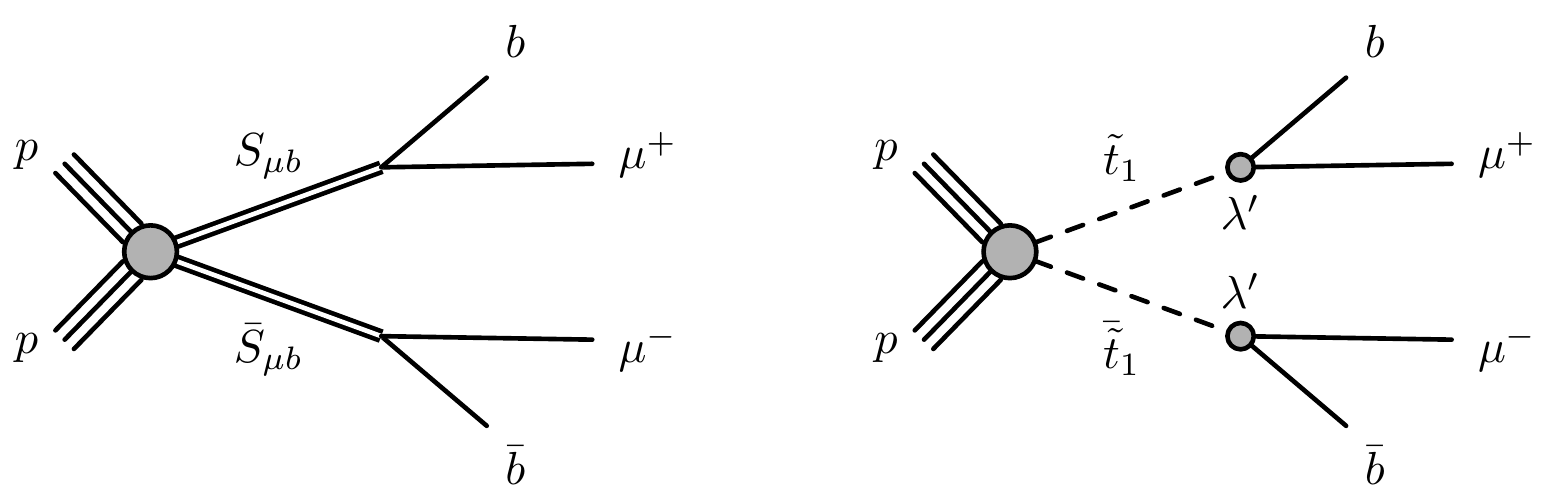} 
   \caption{Diagrams for $pp\to S_{\mu b} \bar{S}_{\mu b}\to (b \mu^+)(\bar b \mu^-)$ (\emph{left}) and $pp\to \t t_1 \bar{\t t}_1 \to (b \mu^+)( \bar b \mu^-)$ (\emph{right}). Final states and kinematic distributions of the two processes are identical. }
   \label{fig:feyn2}
\end{figure}

\begin{figure}[!htbp]
   \centering
   \includegraphics[width=0.48\textwidth]{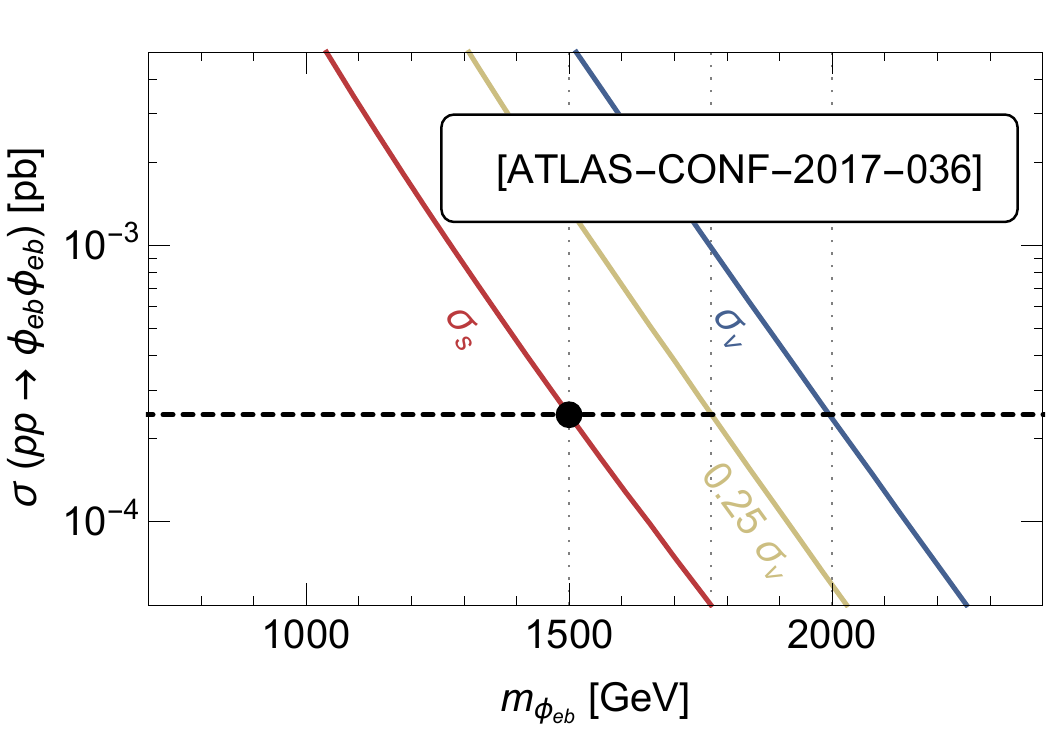} 
   \caption{Bounds on scalar and vector MLQs which decay into ($e b$)($e b$) final states. The 95\% CL on the stop/leptoquark mass (black dot) is provided in~\cite{ATLAS-CONF-2017-036}. The remainder of the analysis is analogous to the right plot of \figref{fig:mujandmub}.}
   \label{fig:eb}
\end{figure}

SUSY with RPV allows a superpotential coupling $\hat L \hat Q \hat D^c$ for the superfields. This superpotential coupling gives rise to a Yukawa coupling $\t d^{c\dagger} QL$ in which the $\t d^c$ squark couples like a leptoquark. The final-state signature and kinematic distributions of pair-produced squarks which decay via this Yukawa coupling are identical to production and decay of scalar leptoquarks (e.g., see \figref{fig:feyn2}).  Such a one-to-one correspondence to RPV squarks exists for all the leptoquarks coupling to $[LQ]$ or $[LD^c]$. 
ATLAS have conducted this type of RPV stop search with Run 1~\cite{ATLAS-CONF-2015-015} and Run 2~\cite{ATLAS-CONF-2017-036} data. The process searched for is $pp\to \t t_1 \t t_1$ where $\t t_1$ decays into the  final states $(e b)$ or $(\mu b)$
with at least one $b$-tag required in~\cite{ATLAS-CONF-2017-036}.  No excess was found for the stop masses searched. We take the strongest bounds from~\cite{ATLAS-CONF-2017-036} (13 TeV, $36\fb$) at the special points where the decays are 100\% to $(eb)$ or 100\% to $(\mu b)$. These can be directly reinterpreted as bounds on a scalar leptoquark decaying to $(\mu b)$ or $(e b)$.  ${S_{e b}}$ and ${S_{\mu b}}$ are excluded up to 1.5 TeV and 1.4 TeV with 95\% CL, respectively. Assuming that the bounds on the cross section times branching fraction to the signal regions do not depend on the LQ mass for even heavier LQs we can recast these bounds to vector MLQs (see the right panel in \figref{fig:mujandmub} for the cross section bound plot for $(\mu b)$ and \figref{fig:eb} for $(e b)$). We obtain the estimated bounds shown in parenthesis in~\tabref{tab:lb}.

\begin{table}[!htbp]
   
           \caption{(\emph{left}) 95\% CL lower limit on the masses of scalar LQs $S_{e b}$ and $S_{\mu b}$. Both occur only with $\mu$-factors equal to one so that no recasts are necessary. The limits are adopted from RPV stop searches from ATLAS~\cite{ATLAS-CONF-2017-036} (see~\figref{fig:eb} and~\figref{fig:mujandmub}).  (\emph{right}) Estimated 95\% CL recast lower limits on the masses of vector LQs $V_{e b}$ and $V_{\mu b}$.}
           \label{tab:lb} 
      \vspace{0.1in}
   \begin{tabular}{@{}| c c| c|  @{}} 
      \hline
           & &{$D^c L, D^c E^c, QE^c, (Q L$ triplet)} \\
           \hline
     \multirow{2}{*}{ $S_{e b}$}   & $\sigma_\text{prod}  $ & {$\times1$} \\     
      & $m_{S_{e b}}$  & {$\geq 1.5 \tev$}~\cite{ATLAS-CONF-2017-036}\\
\hline
     \multirow{2}{*}{ $S_{\mu b}$}   & $\sigma_\text{prod}  $ & {$\times1$}  \\     
      & $m_{S_{\mu b}}$  & {$\geq 1.4 \tev$}~\cite{ATLAS-CONF-2017-036}\\
       \hline
   \end{tabular}
   \quad 
      \begin{tabular}{@{} |c c| c| c|  @{}} 
             \hline
           & &  $D^{c\dag} L, D^{c\dag} E^c, Q^\dag E^c$ & $Q^\dag L$ triplet, $Q^\dag L$ singlet \\
           \hline
     \multirow{2}{*}{ $V_{e b}$}   & $\sigma_\text{prod}  $ & $\times1$  & $\times 0.25$ \\     
      & $m_{V_{e b}}$  & $(\geq 2.0 \tev)$ & $(\geq 1.8 \tev)$  \\

\hline
     \multirow{2}{*}{ $V_{\mu b}$}   & $\sigma_\text{prod}  $ & $\times1$ & $\times 0.25$ \\     
      & $m_{V_{\mu b}}$  & $(\geq 1.9 \tev)$ & $(\geq 1.7 \tev)$  \\
       \hline
   \end{tabular}

\end{table}

\subsection{${e t}$ and ${\mu t}$}
\label{sec:emut}

Decays of leptoquarks into $(et)(et)$ and $(\mu t)(\mu t)$ give rise to several different final states depending on the subsequent top decays. We may have fully hadronic decays, semi-leptonic  decays, and dileptonic decays:
\begin{subequations}
\label{eq:et}
\begin{empheq}[left={p p \to \phi_{\ell t} \bar \phi_{\ell t} \to }\empheqlbrace]{alignat=2}
   & (t_h \ell^+) (\bar t_h \ell^-)  &~ [45.7\%]\\
   &(t_l \ell^+) (\bar t_h \ell^-)+  (t_h \ell^+) (\bar t_l \ell^-)  &~  [43.8\%] \label{eq:mysimute}\\
   & (t_l \ell^+) (\bar t_l \ell^-)  &~  [10.5\%] 
\end{empheq}
\end{subequations}
where subscript $h$ or $l$ of $t$ indicates the top quark decays hadronically or leptonically, respectively. The first two cases of~\eqref{eq:et} are preferable searches because of their large branching fractions (numbers in the brackets). Subsequent $\tau$-decays can further increase the number of different final states. 

\begin{figure}[!htbp]
   \centering
   \includegraphics[width=0.38\textwidth]{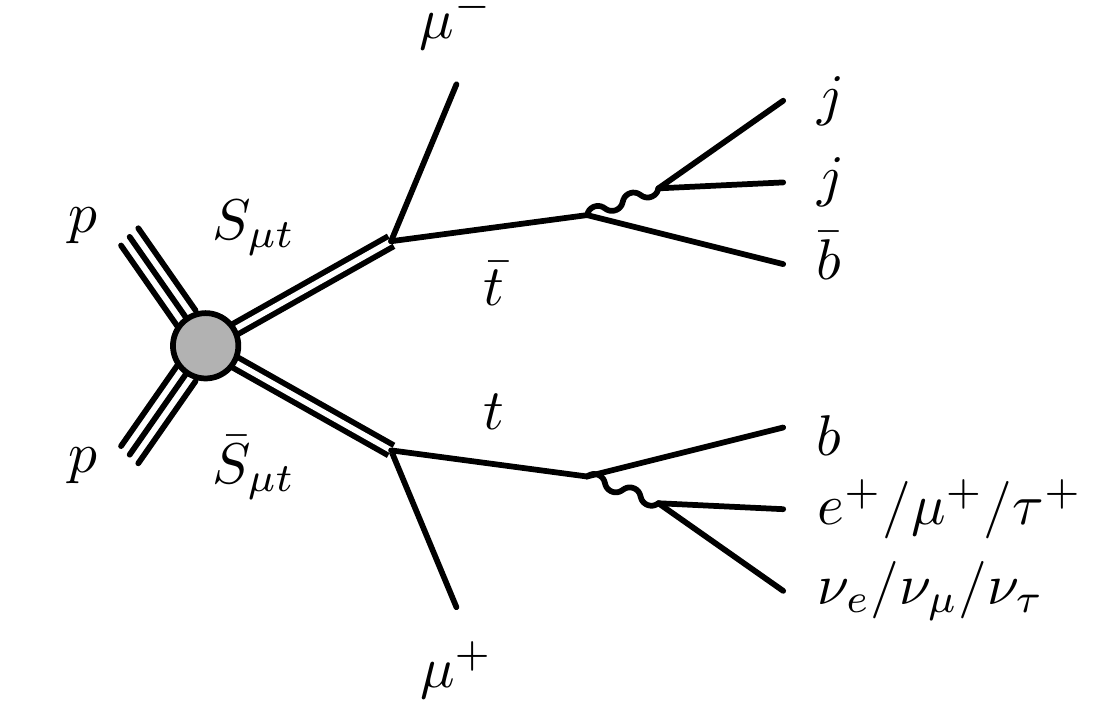} 
   \caption{Diagram for $pp\to S_{\mu t} \bar{S}_{\mu t}\to (t_l \mu^+) (\bar t_h \mu^-)$. Similar final states and kinematics occur in RPV SUSY models.}
   \label{fig:feyn_3}
\end{figure}

No dedicated searches for leptoquarks in any of these channels have been performed. However, the final states of $(et)(et)$ or $(\mu t)(\mu t)$ feature high jet-multiplicities, multiple electrons and/or muons, large missing energy, and multiple $b$ quarks.  These features are also looked for in many SUSY and Exotics searches. 
Hence we can recast existing SUSY searches to set lower limits on the masses of  $\phi_{et}$ or ${\phi_{\mu t}}$. 

Here we focus on the semi-leptonic case (\eqref{eq:mysimute} , see~\figref{fig:feyn_3}). CMS~\cite{CMS-PAS-SUS-16-041} presented a 13 TeV stop search with $35.9\,\fb$ of data with 50 distinct signal regions. The final states  can have multiple $b$-tags, triple leptons, and large missing energy. We generated signals with \texttt{MadGraph5 aMC@NLO 2.5.4}~\cite{Alwall:2014hca} at leading order parton level. We considered leptoquark masses ranging from 500 GeV to 1.6 TeV with 100 GeV steps, and applied the NLO $K$-factor from~\cite{Blumlein:1998ym}.  The signal can be triggered by two sets of dilepton triggers. The first set requires the loose isolation ($I_{\text{mini},\ell} < 0.4$) of two light leptons, with transverse momentum $p_\text{T}$ satisfying $p_{\text{T} \ell_1}, p_{\text{T} \mu_2 (e_2)} > 23, 8 (12) \gev$. The second set has no isolation requirement. It requires leptons with $p_{\text{T} \ell_{1,2}} > 8\gev$ and $H_\text{T}^{30}  \geq 300 \gev$.  The jets are clustered with the anti-$k_\text{T}$ algorithm with $R =0.4$. The reconstructed jets are required to have $p_\text{T} \geq 30\gev$ and pseudo-rapidity $|\eta| < 2.4$.  Reconstructed leptons are required to be well-isolated.   They need to satisfy $I_{\text{mini},\mu (e)} < 0.16\,(0.12)$ together with $p_{\text{T} \mu (e)}^\text{ratio} > 0.69\,(0.76)$ or $p_{\text{T} \mu (e)}^\text{rel} > 6.0 \,(7.2)\gev$. Besides isolation, the analysis requires the lepton pseudo-rapidity $|\eta_{\mu(e)}| <2.4\,(2.5)$ and the $p_\text{T}$ of the leading three leptons $ p_{\text{T} \ell_{1}}, p_{\text{T} \mu_2 (e_2)}, p_{\text{T} \ell_{3}} \geq 25, 10(15), 10 \gev$ if $H^{30}_\text{T} < 300 \gev$ or $ p_{\text{T} \mu_{1} (e_1)}, p_{\text{T} \mu_2 (e_2)}, p_{\text{T} \ell_{3}} \geq 10(15), 10(15), 10 \gev$ if $H^{30}_\text{T} > 300 \gev$. The identification efficiencies for muons (electrons) are taken to be 96\% (90\%). Kinematic variables used in the analysis are defined as:

\begin{table}[!htbp]
   \centering
   \begin{tabular}{@{} ll @{}} 
$H_\text{T}^{30}$ &hadronic activity, defined as the scalar sum of $p_\text{T}$ of all jets with $p_\text{T}>30$ GeV, \\
$I_\text{mini}$ & ratio between the amount of measured energy in a cone around the lepton, with radius $\Delta R_\ell$, and the lepton $p_\text{T}$\\
$\Delta R_\ell$ &  $=(0.2, 10 \gev/p_\text{T}, 0.05)$ for lepton ($p_\text{T} < 50 \gev$, $50\leq p_\text{T} \leq 200 \gev$, $\pt > 200 \gev$)\\
$p_\text{T}^\text{ratio}$ & ratio of the lepton $p_\text{T}$ and the $p_\text{T}$ of the jets around the lepton with $\Delta R_{j\ell} < 0.4$. If no jet is found near the lepton, \\
&   $p_\text{T}^\text{ratio} =1$.\\
$p_\text{T}^\text{rel}$ &  magnitude of the component of the lepton momentum perpendicular to the axis of the jet  that is near the lepton \\
& with $\Delta R_{j\ell} < 0.4$.  If no jet is found near the lepton,  $p_\text{T}^\text{rel} =0$.\\
$M_\text{T}^\text{min}$ &  minimum of the transverse mass of leptons $M_\text{T} \equiv [2 p_{\text{T},\ell} \met (1- \cos \Delta \phi_{\ell, \slashed{\vec{p}}_\text{T}})]^{1/2}$\\
$ \Delta \phi_{\ell, \slashed{\vec{p}}_\text{T}}$ &  azimuthal angle between a lepton and the missing transverse momentum
   \end{tabular}
\end{table}

After the trigger and jet and lepton reconstructions, we pass the events through the baseline selection and divide them into the 25 off-$Z$ signal regions (SR's) and super signal regions (SSR's). $b$-tagging is performed on jets with $p_\text{T}\geq 25\gev$ and $|\eta|< 2.4$. The $b$-tagging, $c$ miss-tagging, and light-jet miss-tagging rates are (70\%, 10\%, 1\%). We then compare the number of remaining signal events with the number of the observed events and expected background given in~\cite{CMS-PAS-SUS-16-041}. Signal regions SR5, SR6, SR9, SR10, and SR16a are most sensitive to leptoquarks.   Cuts for these regions, together with baseline cuts, are summarized in \tabref{tab:SSR}.  

\begin{table}[!htbp]

 \caption{Summary of baseline selection cuts and selection cuts  for the off-$Z$ signal regions (SR) that are most sensitive to $(e t)(e t)$ or $(\mu t)(\mu t)$~\cite{CMS-PAS-SUS-16-041}. We also list the observed and expected number of events for each SR. See text for more details.}
   \label{tab:SSR}
   \vspace{0.1in}
   \centering
   \begin{tabular}{@{} lccccc @{}} 
   \hline
   &\multicolumn{5}{c}{Baseline cuts}\\
   
    \hline
    $N_\text{sel. $\ell$}$ &\multicolumn{5}{c}{$\geq 3$} \\
    $N_\text{sel. jets}$ & \multicolumn{5}{c}{$\geq	2$} \\
$\met$ [GeV] &\multicolumn{5}{c}{$> 50$} \\
          $m_{\ell \ell}$ [GeV] &\multicolumn{5}{c}{$\geq 12$} \\
      \hline
             & SR5  & SR6 & SR9 & SR10 & SR16a \\
             
     \hline
     $b$-tag  & $1$ & $1$ & $2$ & $2$ & inclusive\\
     
     $H_\text{T}$ [GeV] & $60-400$ & $60-400$ & $60-400$ & $60-400$ & $\geq 60$   \\
     
     $\met$ [GeV]  & $50-150$ & $150-300$ & $50-150$ & $150-300$ & $\geq 300$\\
     
     $M_\text{T}^\text{min}$ [GeV] &  inclusive & inclusive & inclusive & inclusive & $<120$ \\ 
      \hline
 Expected & $202\pm4\pm 44$ & $25.6\pm 1.9\pm 4.6$ & $47.7 \pm 2.8 \pm 7.6$ & $5.3\pm 0.5\pm 0.6$ & $12.1\pm 1.5 \pm 1.9$\\
 
 Observed & 191 & 25 & 51 & 5 & 7\\
      \hline
   \end{tabular}
  \end{table}

Using the central value of the background estimation in \tabref{tab:SSR} and assuming Poisson statistics, the resulting 95\% CL lower limit on the mass of
$S_{\mu t}$ is 800 GeV from SR5 and SR9 and 700 GeV from SR6, SR10, and SR16a. The resulting constraint on the mass of $S_{et}$ 
is 900 GeV from SR16a, 800 GeV from SR10 and 700 GeV from SR5, SR6, SR9. A common feature of the most sensitive search regions
is that they allow relatively low hadronic activity. Note that our recast, using a simple cut and count search, is not optimized. Better sensitivity could be reached with an analysis that reconstructs at least one of the resonances of the signal. In~\tabref{tab:et}, we show the bounds on LQs  decaying into ($e t$)($e t$) or ($\mu t$)($\mu t$) which we obtained for other MLQ models. 

\begin{table}[!htbp]
 \caption{95\% CL lower limits on the mass of MLQs which decay into ($e t$)($e t$) or ($\mu t$)($\mu t$)  final states for the scalar leptoquarks (\emph{left}) and the vector leptoquarks (\emph{right}) from recasting the CMS SUSY search~\cite{CMS-PAS-SUS-16-041}.}
    \vspace{0.1in}
   \begin{tabular}{@{} |c c |c| c|  @{}} 
      \hline
           & & $U^c L, U^cE^c, QE^c$ & ($QL$ triplet),  ($QL$ singlet) \\
           \hline
     \multirow{2}{*}{ $S_{e t}$}   & $\sigma_\text{prod}  $ & $\times1$  & $\times 0.25$ \\     
      & $m_{S_{e t}}$  & $\geq 800 \gev$ & $\geq 600 \gev$  \\
           \hline
     \multirow{2}{*}{ $S_{\mu t}$}   & $\sigma_\text{prod}  $ & $\times1$ & $\times 0.25$ \\     
      & $m_{S_{\mu t}}$  & $\geq 800 \gev$ & $\geq 600 \gev$  \\
       \hline
   \end{tabular}
   \quad
     \begin{tabular}{@{}| c c| c|   @{}} 
       \hline
           & &          {$U^{c\dag} L, U^{c\dag} E^c, Q^\dag E^c, (Q^\dag L$ triplet)}  \\
                      \hline
     \multirow{2}{*}{ $V_{e t}$}   & $\sigma_\text{prod}  $ &  {$\times1$}  \\     
      & $m_{V_{e t}}$  &  {$\geq 1.4 \tev$}  \\

                      \hline
     \multirow{2}{*}{ $V_{\mu t}$}   & $\sigma_\text{prod}  $ &  {$\times1$}  \\     
      & $m_{V_{\mu t}}$  &  {$\geq 1.2 \tev$}  \\
      \hline
   \end{tabular}
           \label{tab:et}
\end{table}

\subsection{$\tau j$}
\label{sec:tauj}

In analogy with the $({et})(et)$ or $({\mu t})(\mu t)$ cases, the $\tau$'s in $({\tau j})(\tau j)$ final states can be either hadronic or leptonic. The branching fractions for $\phi_{\tau j} \bar \phi_{\tau j}$ decays are given by

\begin{subequations}
\label{eq:tauj}
\begin{empheq}[left={p p \to  \phi_{\tau j} \bar \phi_{\tau j} \to  }\empheqlbrace]{alignat=3}
 & (\tau_h j) (\tau_h j) &~{} [42.0\%] \\
  \label{eq:sltau}
 &(\tau^+_l j) (\tau_h j) +(\tau^-_l j) (\tau_h j) &~{} [45.6\%]\\
  & (\tau^+_l j) (\tau^-_l j) &~{} [12.4\%]
\end{empheq}
\end{subequations}
We are not aware of any dedicated searches for these final states. However, three types of existing searches can be relevant: (1)  leptoquarks which decay to $(\tau b)(\tau b)$ final states are intensively studied by ATLAS and CMS (see~\secref{sec:taub}). An analysis without imposing the $b$-tags would cover the $(\tau j)(\tau j)$ channel; (2) searches for right-handed $W_R$ and right-handed neutrinos $N$ using the process $p p \to W_R \to \tau N \to \tau (\tau q \bar{q})$ has been conducted by~\cite{Khachatryan:2016jqo,Sirunyan:2017yrk}. The final particles of this process are identical to $\phi_{\tau j} \bar \phi_{\tau j}$. To apply this search one would have to adjust the signal kinematics and re-analyze the data; (3) our LQ final state can be also be covered by SUSY or Exotics searches which look for large missing energy with high $p_\text{T}$-jets and zero, one, or two opposite-sign leptons.

Given the similarity of this final state with the well-covered $(\tau b)(\tau b)$ final state and given that SM backgrounds are small at large invariant masses we expect that a designated search will obtain bounds on leptoquark masses which are comparable
to those for the $(\tau b)(\tau b)$ case. We have not been able to obtain reliable bounds for any of the MLQs in this final state and therefore left the corresponding entries in~\tabref{tab:tauj} blank.

\begin{table}[!htbp]
 \caption{MLQ models which decay into the final state ($\tau j$)($\tau j$). We have not been able to find searches which can reliably and easily be recast to obtain mass limits for these models.}
    \vspace{0.1in}
   \centering
   \begin{tabular}{@{} |c c| c| c| c |c| @{}} 
      \hline
           & & $U^c L, D^c L, U^c E^c, D^c E^c$&  $Q E^c$ & $Q L$ triplet & $Q L$ singlet \\
           \hline
     \multirow{2}{*}{ $S_{\tau j}$}   & $\sigma_\text{prod}  $ & $\times1$  & $\times2$ & $\times1.25$ & $\times 0.25$ \\     
      & $m_{S_{\tau j}}$  & &  &   &  \\
           \hline
                 \hline
           & & $U^{c\dag} L, D^{c\dag} L, U^{c\dag}E^c, D^{c\dag}E^c$&  $Q^\dag E^c$ & $Q^\dag L$ triplet & $Q^\dag L$ singlet \\
           \hline
     \multirow{2}{*}{ $V_{\tau j}$}   & $\sigma_\text{prod}  $ & $\times1$  & $\times2$ & $\times1.25$ & $\times 0.25$ \\     
      & $m_{V_{\tau j}}$  & &  &   &  \\
      \hline
   \end{tabular}
   \label{tab:tauj}

\end{table}

\subsection{${\tau b}$}
\label{sec:taub}
LQ decays into the $({\tau b})(\tau b)$ final state have been well-covered by LHC searches for LQs coupling to third generation quarks and leptons with  Run 1 and Run 2 data~\cite{Chatrchyan:2012sv,Chatrchyan:2012st,ATLAS:2013oea,Khachatryan:2014ura,Khachatryan:2016jqo,Sirunyan:2017yrk}. The latest 95\% upper limit on the only scalar MLQ with this final state is $850\gev$~\cite{Sirunyan:2017yrk}. We recast the bound for the vector MLQs in \tabref{tab:taub}. 

\begin{figure}[!htbp]
   \centering
   \includegraphics[width=0.48\textwidth]{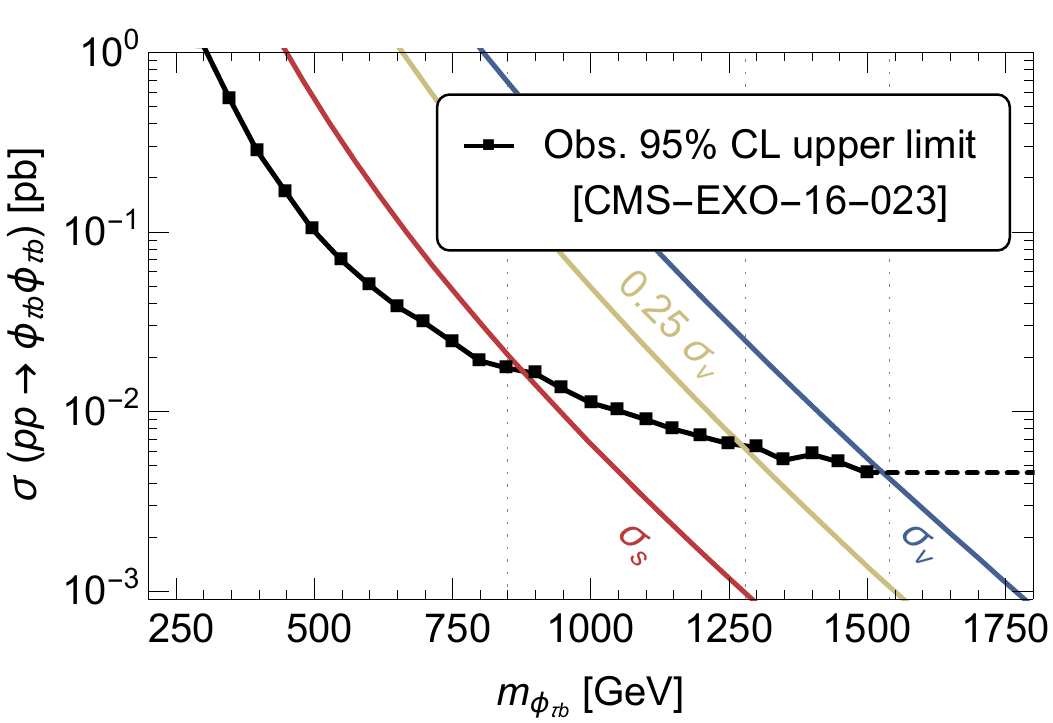} 
   \caption{Bounds on scalar and vector MLQs which decay into ($\tau b$)($\tau b$) final states. The observed 95\% CL upper limit on the production cross section times branching fraction (black solid line) is taken from searches for LQs in this final state at CMS~\cite{Sirunyan:2017yrk} that stopped at 1.5 TeV. The remainder of the analysis is analogous to~\figref{fig:nuj}.}
   \label{fig:taub}
\end{figure}

\begin{table}[!htbp]
     \caption{95\% CL lower limits on the mass of MLQs which decay into ($\tau b$)($\tau b$) final states for the scalar leptoquarks (\emph{left}) and the vector leptoquarks (\emph{right}).}
        \vspace{0.1in}
   \begin{tabular}{@{} |c c| c|  @{}} 
      \hline
           & &      {$D^c L, D^c E^c, Q E^c, (Q L$ triplet)}  \\
                      \hline
     \multirow{2}{*}{ $S_{\tau b}$}   & $\sigma_\text{prod}  $ &  {$\times1$}  \\     
      & $m_{S_{\tau b}}$  &  {$\geq 850 \gev$}~\cite{Sirunyan:2017yrk}  \\
         \hline
         \end{tabular}
         \quad
         \begin{tabular}{@{} |c c| c| c|  @{}}
         \hline
                    & & $D^{c\dag} L, D^{c\dag} E^c, Q^\dag E^c$ & $Q^\dag L$ triplet, $Q^\dag L$ singlet  \\
                    \hline
                         \multirow{2}{*}{ $V_{\tau b}$}   & $\sigma_\text{prod}  $ & $\times1$ & $\times 0.25$  \\     
      & $m_{V_{\tau b}}$  & $(\geq 1.55 \tev)$ & $\geq 1.3 \tev$ \\
                      \hline
   \end{tabular}

   \label{tab:taub}
\end{table}

\subsection{${\tau t}$}
\label{sec:taut}

$({\tau t})(\tau t)$ has the most complicated final states of the nine types of LQs in the LQ matrix because each of the four decay particles, $\tau^+$, $\tau^-$, $t$ or $\bar{t}$, can decay leptonically or hadronically. The multiplicity of possible final states, neglecting charge differences, is 10.

\beq
p p \to  \phi_{\tau t} \bar \phi_{\tau t} \to 
\left \{
  \begin{array}{cccccc}
  (t_h \tau_h) (t_h \tau_h) & [19.1\%] , & (t_h \tau_h) (t_h \tau_l) & [20.8\%] ,&  (t_h \tau_h) (t_l \tau_h) & [18.4\%], \\
    (t_h \tau_h) (t_l \tau_l) & [10.0\%],  &  (t_h \tau_l) (t_l \tau_h) & [10.0\%] ,&  (t_h \tau_l) (t_h \tau_l) & [5.7\%], \\
   (t_h \tau_l) (t_l \tau_l) & [5.4\%],& (t_l \tau_h) (t_l \tau_l) & [4.8\%], & (t_l \tau_h) (t_l \tau_h) & [4.4\%], \\
       (t_l \tau_l) (t_l \tau_l) & [1.3\%]. \\
  
  \end{array}
\right.
\eeq
Further complications can arise from $t_l\to \tau \bar \nu_\tau \bar b$ decays where $\tau$ can decay either hadronically or leptonically. 

Only one of the many possible final states has been covered. Ref.~\cite{Khachatryan:2015bsa} searched for $p p \to \phi_{\tau t} \bar \phi_{\tau t} \to (t_h \tau_h^-) (\bar t_l \tau_l^+)$ with $t_l$ decays to $e$ or $\mu$ at Run 1 ($ 8\tev, 20 \fb$).
The mass limit is $m_{S_{\tau t}} \geq 685 \gev$. Our recast bounds for other MLQs are shown in~\tabref{tab:taut}. Given that several other final states for $\phi_{\tau t}$ pair-production appear promising, we hope that searches in additional channels will also be performed.

\begin{figure}[!htbp]
   \centering
   \includegraphics[width=0.48\textwidth]{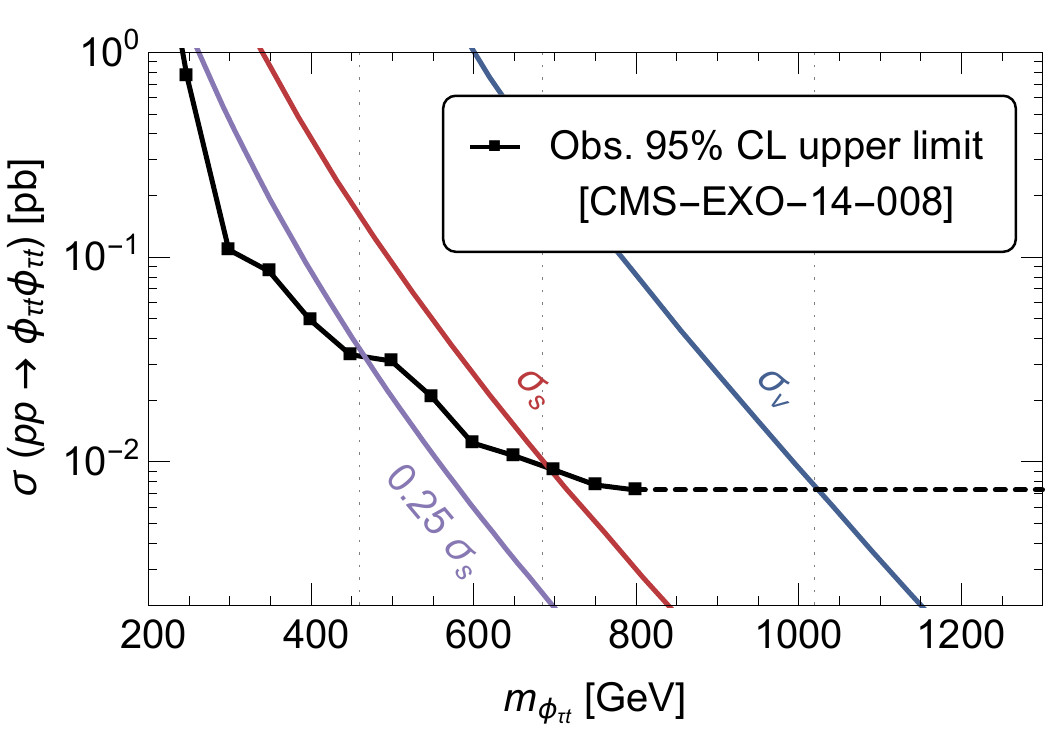} 
   \caption{Bounds on scalar and vector MLQs which decay into ($\tau t$)($\tau t$) final states. The observed 95\% CL upper limit on the production cross section times branching fraction (black solid line) is taken from LQ searches in this final state at CMS~\cite{Khachatryan:2015bsa} which stopped at 800 GeV. The remainder of the analysis is analogous to \figref{fig:nuj}.}
   \label{fig:taut}
\end{figure}

\begin{table}[!htbp]

 \caption{95\% CL lower limits on the mass for various MLQs which decay into ($\tau t$)($\tau t$) final states for scalar LQs (\emph{left}) and vector LQs (\emph{right}).}
   \vspace{0.1in}
   \begin{tabular}{@{} |c c| c| c|  @{}} 
      \hline
           & & $U^c L, U^cE^c, Q E^c$ & ($QL$ triplet), ($QL$ singlet) \\
           \hline
     \multirow{2}{*}{ $S_{\tau t}$}   & $\sigma_\text{prod}  $ & $\times1$  & $\times 0.25$ \\     
      & $m_{S_{\tau t}}$  & $\geq 685 \gev$~\cite{Khachatryan:2015bsa} & $\geq  460 \gev$  \\
              \hline
              \end{tabular}
              \quad
                \begin{tabular}{@{} |c c| c|   @{}}
                    \hline
           & &         {$U^{c\dag} L, U^{c\dag}E^c, Q^{\dag} E^c, (Q^\dag L$ triplet) }  \\
                      \hline
     \multirow{2}{*}{ $V_{\tau t}$}   & $\sigma_\text{prod}  $ &  {$\times1$}  \\     
      & $m_{V_{\tau t}}$  & {($\geq 1.0 \tev$)}  \\
         \hline
   \end{tabular}

   \label{tab:taut}

\end{table}

\section{Discussion and Summary}

In this paper, we have advocated that LQ pair production searches be organized by the leptoquark matrix. This naturally leads to 
9 distinct final states to be searched for. In order to make recasts to arbitrary leptoquark models possible it would be desirable if the experimental collaborations publish bounds on the cross section times branching fraction as a function of LQ mass for each of these final states. We have found strong bounds from existing searches for leptoquarks or equivalent searches for RPV squarks in 6 of the 9 cases. For $(\ell b)(\ell b)$ we have only been able to find a lower mass bound on a particular squark model which only allowed us to obtain recast estimates for other models in this channel. For $(\ell t)(\ell t)$ we were able to obtain bounds by recasting a general purpose cut-and-count SUSY search. These bounds could certainly be improved with a more dedicated search. Finally for $(\tau j)(\tau j)$ were not able to find a search which allowed us to obtain reliable strong bounds.

We also note that even in the cases where cross section bounds as a function of mass have been published it would be useful if the search range could be extended as far as possible in LQ mass. This is important because the theoretical cross section times branching fraction for LQs to particular final states can vary by orders of magnitude. At a minimum, the mass range covered should include the range required to set bounds on the MLQ states which can have cross sections times branching fractions ranging from $0.25\times \sigma_S$ to $2\times \sigma_V$.

We have focused on pair production of LQs with prompt decays. For leptoquarks with large couplings single (and off-shell) production can become important and potentially yield stronger lower bounds on LQ masses. This is especially true when the large
coupling is to first generation quarks with their large parton distribution functions.
We hope to extend the ``Leptoquark Hunter's Guide" to single and off-shell production in a
future publication~\cite{inprogress}. The LQ matrix is also convenient for organizing the possible single
LQ production final states. 

Another generalization that we have not covered here is the possibility of displaced vertices from LQ decays with long life times
or even LQs that are stable on detector time scales. In the former case exotic new searches using displaced vertices or kinks
in tracks would be interesting, in the latter case the LQs would leave strongly ionizing tracks similar to $R$-hadrons in
SUSY.  For LQs with TeV scale masses, moderate boosts, and decay width $\Gamma = m \lambda^2/(16\pi)$ the vertex
displacement is $d\sim 10^{-15}/\lambda^2~\text{cm}$. LQs with couplings larger than $\lambda \sim 10^{-6.5}$ have
prompt decays with vertex displacements smaller than typical LHC detector impact parameter resolution $\sim 100~\mu\text{m}$.
Couplings smaller than $10^{-9}$ lead to ``stable" LQs with decays outside the detector.

Is there a theoretically preferred range for the size of couplings? LQs are unlike any particles we have observed so far
and therefore we can only make guesses. If they are vectors one should perhaps expect couplings
similar to SM gauge couplings, $0.1$ -- $1$. If they are scalars one might expect
couplings similar to SM Yukawa couplings, $10^{-6}-1$. For LQs related to the recent
hints for new physics from $B$ meson decays the interesting range of couplings is $10^{-2} \leq \lambda \leq 1$.
All these would imply prompt decays but smaller couplings are certainly possible.

Before closing we would like to address a common misconception about leptoquark ``generations".
It is often said that stringent bounds on flavor violation require the couplings of leptoquarks to be limited
to first, second, or third generation leptoquarks. This is statement is both incorrect and uses misleading language.
What is correct is that large couplings to more than one generation of quarks lead to large quark flavor violation
(e.g. $K-\bar K$ mixing), and couplings to more than one generation of leptons lead to lepton flavor violation
(e.g. $\mu \to e \gamma$). However the generation number of the quark and the lepton do not have to be the
same to avoid flavor violation. For example, a leptoquark coupling only to right-handed strange quarks and electrons
(in the mass eigenstate basis for the fermions) exactly preserves all relevant $U(1)$ fermion number symmetries
(strangeness and electron-number). Thus flavor constraints do allow the more general form for leptoquark couplings
and decays that fill out the full LQ matrix. We discourage the use of the misleading terms ``first, second,
third generation LQs'' and instead suggest the use of more accurate language in which both the quark and lepton generation
are specified.

\subsection*{Acknowledgements}
We thank Gudrun Hiller, Simon Knapen,  Vojtech Pleskot and Ruth P\"ottgen for helpful discussions and comments on the manuscript. BD would like to thank Nicol\'as Neill, Sebasti\'an Tapia and Alfonso Zerwekh for discussions. MS would like to thank the IAS at HKUST for its hospitality during the completion of this project. BD was supported partially by Boston University HET visitor program and also partially supported by Conicyt Becas Chile. The work of MS and YZ work is supported by DOE grant DE-SC0015845. 

\appendix

\section{Notation}
\label{app:notation}

We apologize for our sometimes confusing notation. When referring to the SM fermion fields we use all left-handed 2-component spinors $Q, U^c, D^c, L, E^c, N^c$ to reflect 
the full $SU(3)_{color}\times SU(2)_{weak}\times U(1)_Y$ symmetry of the SM. The notation uses the fields $Q_i$ to denote the (left-handed) quark 
doublets with $i$ labeling the generation. For example, $Q_3=(t_L,b_L)$. $U^c_i$ and $D^c_i$ are the charge conjugates of the right-handed quark singlets, again $i$ labeling the generation. For example, $D^c_3=i\sigma_2 b_R^*$. $L_i$ are the lepton doublets and $E^c_i$ and $N^c_i$ are three generations of 
charged lepton and neutrino singlets, respectively.
\tabref{tab:notation} is a summary of the symbols  used in the paper.

\begin{table}[!htbp]
   \centering
      \caption{Summary of notation. 
   }
      \vspace{0.1in}
   \label{tab:notation}
   \begin{tabular}{c p{0.61\textwidth}} 
      \hline
       Symbol    & Denotes\\
      \hline
      \hline
      $S_{\bm{lq}}$      & Scalar leptoquark, we use the subscript to denote its leptonic and hadronic decay products. \\
      $V_{\bm{lq}}$ &  Vector leptoquark. \\
      $\phi_{\bm{lq}}$ & General leptoquark, either $S$ or $V$.\\
      \hline
      $Q, U^c, D^c, L, E^c$ & SM fermion fields. Subscripts $\{1,2,3\}$ indicate flavor. 
      \\
      $N^c$ &  Dirac partners of the SM left-handed neutrino fields. \\
      $\bm L$ & $L, E^c, N^c$ \\
      $\bm Q$ & $Q, U^c, D^c$ \\
      $\hat Q, \hat U, \hat D, \hat L, \hat E$ & MSSM fermion superfields which contain the corresponding SM fermion fields.\\
      \hline
      $\nu$ & $\nu_e, \nu_\mu, \nu_\tau$\\
      $j$, $b$ & light-jet (originating from $u, d, s, c$ or gluons), $b$-jet\\
      $\ell$ & $e, \mu$\\
      $\bm l$ & $e, \mu, \tau, \nu$\\
      $\bm q$ & $j, b, t$\\
      \hline
   \end{tabular}
   \label{tab:notation}
\end{table}

We also list the leptoquark nomenclature for the MLQ models used in this paper (\nth{1} column), their SM gauge quantum numbers  (\nth{2} column), the names used in Ref.~\cite{Olive:2016xmw} (PDG, \nth{3} column), and in Ref.~\cite{Buchmuller:1986zs} (BRW, \nth{4} column). Note that in our MLQ models two leptoquarks with the same gauge quantum numbers but different flavor quantum numbers are considered distinct and therefore have different names. In the PDG and BRW notation, LQs with the same gauge quantum numbers are given the same name. That is why the map between our models and PDG and BRW is not one-to-one.

\begin{table}[!htbp]
   \caption{Alternative LQ names in the literature.} 
      \vspace{0.1in}
   \begin{tabular}{@{} llll @{}} 
      \hline
      Scalar LQ  & $(SU(3),SU(2))_Y$    & BRW & PDG\\
      \hline
      $QL$ triplet & $(3,3)_{-1/3}$  & $S_3$ & $S_1^\dag$\\
      $QL$ singlet & $(3,1)_{-1/3}$  & $S_1$ & $S_0^\dag$\\
      $U^c L$ & $(\bar 3,2)_{-7/6}$  & $R_2$ & $S_{1/2}^\dag$\\
      $D^c L$ & $(\bar 3,2)_{-1/6}$ & $\t{R}_2$ & $\t{S}_{1/2}^\dag$\\
      $Q E^c$ &$(3,2)_{7/6}$  & $R_2$ & ${S}_{1/2}^\dag$\\
      $U^c E^c$ & $(\bar 3,1)_{1/3}$ & $S_1$ &  $S_0^\dag$ \\
      $D^c E^c$ & $(\bar 3,1)_{4/3}$  & $\t{S}_1$ & $\t{S}^\dag_0$\\
      $Q N^c$ &$(3,2)_{1/6}$   & $\t{R}_2$ & --\\
      $U^c N^c$ &$(\bar 3,1)_{-2/3}$ & $\bar S_1$& -- \\
      $D^c N^c$ & $(\bar 3,1)_{1/3}$& $ S_1 $ & --\\
      \hline
   \end{tabular}
                 \quad\quad\quad
   \begin{tabular}{@{} llll @{}} 
      \hline
      Vector LQ     & $(SU(3),SU(2))_Y$   & BRW & PDG \\
      \hline
      $Q^\dag L$ triplet  & $(\bar 3,3)_{-2/3}$ & $U_3$ & $V_1^\dag$ \\
      $Q^\dag L$ singlet & $(\bar 3,1)_{-2/3}$  & $U_1$ & $V_0^\dag$\\
      $U^{c \dag} L$ & $( 3,2)_{1/6}$  & $\t{V}_2$ & $\t{V}_{1/2}^\dag$ \\
      $D^{c\dag} L$ & $(3,2)_{-5/6}$& ${V}_2$ & ${V}_{1/2}^\dag$\\
      $Q^\dag E^c$ &$(\bar 3,2)_{5/6}$ & $V_2$ & ${V}_{1/2}^\dag$ \\
      $U^{c\dag} E^c$ & $( 3,1)_{5/3}$& $\t{U}_1$&  $\t{V}_0^\dag$  \\
      $D^{c\dag} E^c$ & $( 3,1)_{2/3}$ & ${U}_1$ & ${V}^\dag_0$\\
      $Q^\dag N^c$&$(\bar 3,2)_{-1/6}$  & $\t{V}_2$& -- \\
      $U^{c\dag} N^c$  &$( 3,1)_{2/3}$& $U_1$ & -- \\
      $D^{c\dag} N^c$ & $( 3,1)_{-1/3}$ & $\bar U_1$& --\\
      \hline
   \end{tabular}
   \label{tab:name}
\end{table}

\section{Comparison between vector and scalar LQ searches}
\label{app:efficiencies}
\begin{figure}[!htbp]
   \centering
   \includegraphics[width=0.96\textwidth]{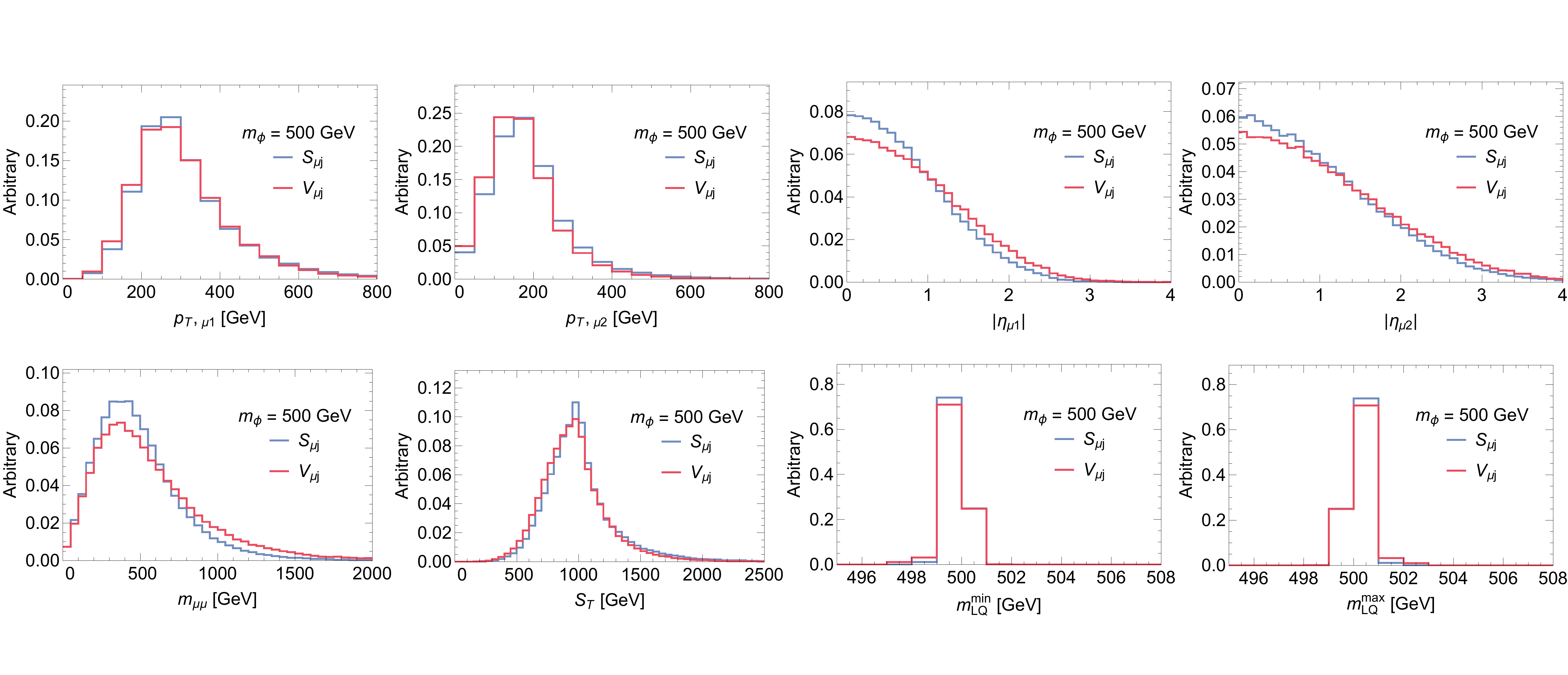}
   \caption{Shape variables for scalar and vector leptoquarks with masses $m_\phi = 500 \gev$ after passing the trigger cuts from ATLAS. Reconstructed leptoquark masses are artificially narrow because our parton level analysis does not include detector resolution. Resolution 
effects would also slightly smoothen the other plots and make vector and scalar distributions even more similar.}
   \label{fig:shapevarA}
\end{figure}

\begin{figure}[!htbp]
   \centering
    \includegraphics[width=0.96\textwidth]{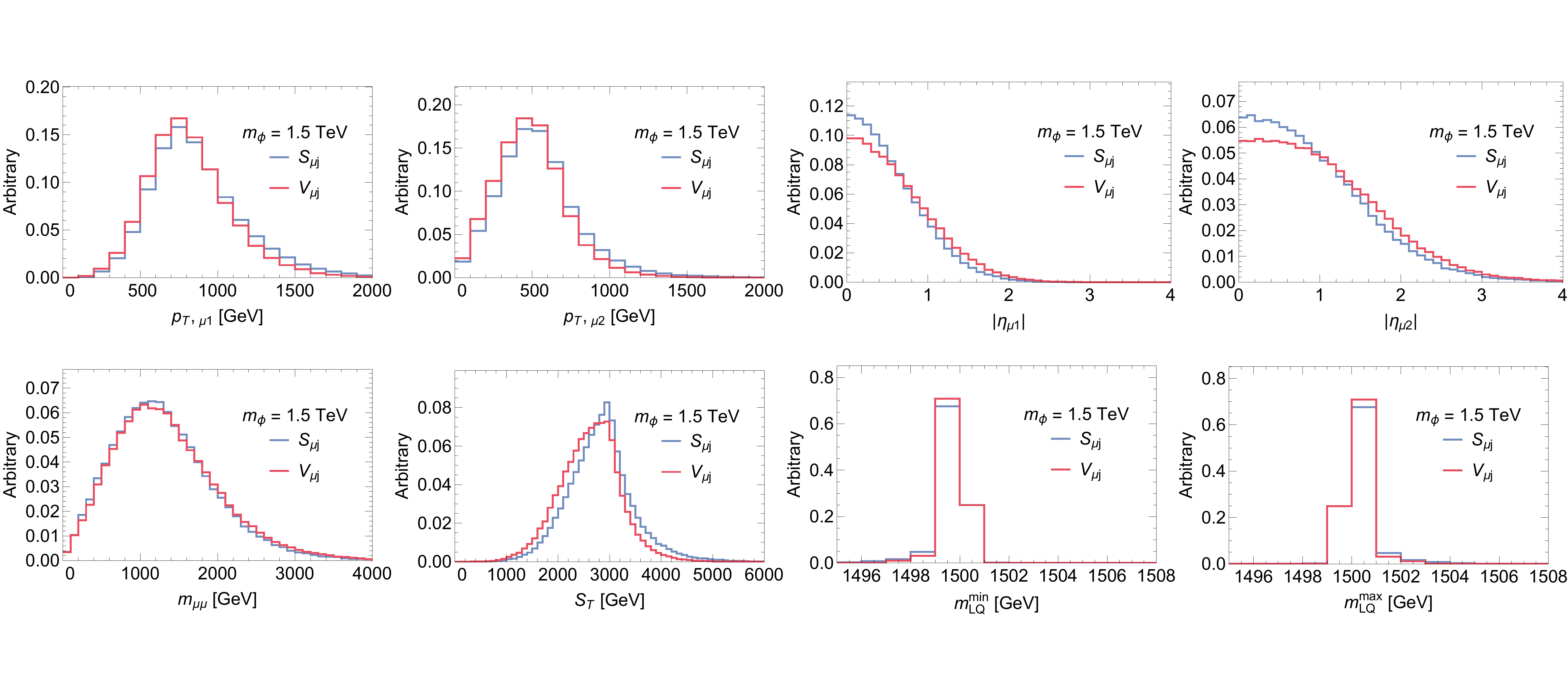}
   \caption{Same as \figref{fig:shapevarA} for leptoquarks with mass $m_\phi=1.5 \tev$.}
   \label{fig:shapevarB}
\end{figure}

The kinematics of pair-produced vector LQs is very similar to pair-produced scalar LQs. We generated the processes $pp\to S_{\mu j}S_{\mu j} \to (j \mu^+) (j \mu^-)$ and  $pp\to V_{\mu j}V_{\mu j} \to (j \mu^+) (j \mu^-)$ at leading order parton level using \texttt{MadGraph5 aMC@NLO 2.5.4} which retains polarization effects in production and decays. We then passed the events through the selection cuts for ATLAS'~\cite{Aaboud:2016qeg} and CMS'~\cite{CMS-PAS-EXO-16-007} analyses. The ratios of the relative efficiencies for various leptoquark masses are shown in~\figref{fig:cutsofsandv}. One sees that the efficiencies for scalar leptoquarks $S$ and vector leptoquarks $V$ are very similar. To better understand the origin of this similarity, we show in~\figref{fig:shapevarA} and~\figref{fig:shapevarB} the scalar and vector distributions for different leptoquark masses, and in the following we detail the variables used:

\begin{table}[!htbp]
   \centering
   \begin{tabular}{@{} ll @{}} 
$p_{\text{T}\mu_1}, p_{\text{T}\mu_2}$ & transverse momenta of the two leading leptons, \\
$m_{\mu\mu}$ & di-muon invariant mass, \\
$ S_\text{T}$ & scalar sum of the $p_\text{T}$ of the two leading jets and two leading muons, \\
$ m_{\text{LQ}}^{\min}$ & lower of the two balanced leptoquark-pair masses. The balanced leptoquark-pair is defined as the combination \\
& of leptons with quarks that yields a smaller mass differences between the two reconstructed leptoquarks, \\
$ m_{\text{LQ}}^{\max}$ & higher of the two balanced leptoquark-pair masses.
   \end{tabular}
\end{table}


Note that 
vector LQ production is slightly more forward than scalar LQ production.
This results in softer $p_\text{T}$ and larger $\eta$ for $VV$ compared to $SS$.
Consequently, the relative efficiency for $VV$ decay products to pass through the $p_\text{T}$ and $\eta$ cuts are lower
than those of $SS$ decay products. This can be seen from the first two panels of~\figref{fig:cutsofsandv}.
The difference is more pronounced for light leptoquarks where also overall efficiencies are smaller, but for $m_{\phi} \geq 500 \gev$
the difference in efficiencies is always below 10\%.

\begin{table}[!htbp]
   \caption{Summary of selection cuts for ATLAS~\cite{Aaboud:2016qeg} and CMS~\cite{CMS-PAS-EXO-16-007} searches for pair-produced $S_{\mu j}$.}
   \label{tab:atlasandcms}
      \vspace{0.1in}
   \centering
   \begin{tabular}{@{} lll @{}} 
      \hline
      	    & ATLAS & CMS\\
      \hline
      \hline
      Trigger      &   At least one muon with $|\eta|<2.5$, $p_{\text{T}} > 26 \gev$ & At least one muon with\\
      &  and isolated within $\Delta R = 10\gev/ p_{\text{T}}$ &  $p_{\text{T}} >45 \gev$, $|\eta|<2.1$\\
      & or $p_\text{T} > 50 \gev$ without isolation & \\
      \hline
      Muon $p_\text{T}$, $\eta$ & $p_{\text{T} \mu_{1,2}} > 40 \gev$,   & $p_{\text{T} \mu_{1,2}} > 50 \gev$,   \\
       & $|\eta_{\mu_{1,2}}| < 1.01$ or $1.10<|\eta_{\mu_{1,2}}| <2.5$ & $|\eta_{\mu_{1,2}}| < 2.1$ \\
             \hline
      Jet $p_\text{T}$, $\eta$ & $p_{\text{T} j_{1,2}} > 50 \gev$, & $p_{\text{T} j_{1,2}} > 50 \gev$,\\
       & $|\eta_{j_{1,2}}|<2.8$ & $ |\eta_{j_{1,2}}|<2.4$ \\
             \hline
      Isolation & $ \Delta R_{jj, \mu j} > 0.4$, & $ \Delta R_{jj} > 0.4$, $\Delta R_{\mu j, \mu\mu} >0.3$\\
        & isolated $\mu$'s with $\Delta R =10\gev/ p_{\text{T}} $  &\\
      \hline
      $m_{\mu\mu}$  & $> 130 \gev$ & $> 50 \gev$\\
      \hline
      $S_\text{T}$ & $>600 \gev$ & $ >300 \gev$ \\
      \hline
   \end{tabular}

\end{table}

\section{A vector leptoquark model}
\label{app:model}

Vector leptoquarks may either arise as strongly coupled states within a spectrum of composite resonances from
a strongly coupled theory or they may be fundamental, i.e., gauge bosons which obtain masses from spontaneous 
symmetry breaking. In fact, even composite spin-1 particles must have a spontaneously broken gauge invariance if 
the compositeness scale is parametrically larger than the mass of the spin-1 particles \cite{Weinberg:1980kq}.
In this Appendix we present a renormalizable model in which the vector leptoquarks are massive gauge bosons of a spontaneously
broken gauge theory.

Leptoquarks carry color and hypercharge, and therefore the gauge generators associated with the leptoquarks must not commute with color and hypercharge. The smallest gauge group which contains color and hypercharge generators and vector leptoquarks is
$SU(4)\times U(1)_{X} \supset SU(3) \times U(1)_Y$ where the $U(1)_X$ is necessary to obtain the correct hypercharges for
the SM fermions. The spectrum of gauge bosons of this minimal model contains the SM gauge bosons in addition to a complex
color triplet leptoquark $V$, and a massive singlet, a $Z'$. This $Z'$ is a phenomenological problem. The problem is that the $Z'$ cannot be much heavier than the leptoquark because their masses have a common origin. The $Z'$ couples strongly to SM fermions and therefore has a large production cross section as an $s$-channel resonance.  Resonance searches for the $Z'$ are much more sensitive probes of this model than leptoquark pair production. In other words, regions of parameter space with potentially discoverable leptoquarks are already ruled out by the non-observation of $Z'$ signals at the LHC.  This problem with ``light" $Z'$ resonances is generic to the most simple vector leptoquark models and makes viable models with TeV scale leptoquark masses non-trivial to construct.

In the following we offer a generic structure for leptoquark models which solves this problem. The trick is familiar from
``partial compositeness" models in which heavy gauge bosons arise as composite states from a ``strong sector". They obtain small mixing with weakly coupled gauge bosons under which the SM fermions are charged. In this way the heavy ``composite" gauge bosons couplings to SM fermions are suppressed by the small mixing angle and direct searches for $Z'$ resonances are much less constraining.
The UV gauge structure of such models is $SU(3)\times SU(2)_{weak}\times U(1) \times SU(N)$ where the $SU(N)$ is the strongly coupled new gauge group. The SM fermions are all charged under the $SU(3)\times SU(2) \times U(1)$ as usual and carry no $SU(N)$ charge. The $SU(N)$ is chosen such that a part of the SM gauge group containing color and hypercharge fits inside $SU(N)$. The minimal choice is $SU(4)$ but larger groups can also be used. For example, a particularly interesting choice is $SU(5)$ in which case all the SM gauge bosons (including $SU(2)_{weak}$) can mix with the heavy gauge bosons. Such a non-minimal structure is also necessary to obtain vector leptoquarks which carry $SU(2)_{weak}$ quantum  numbers. For definiteness we describe a particular $SU(4)$ model
in which the vector leptoquark $V^\mu$ couples to the SM fermion bilinear $Q^{\dagger} \bar \sigma_\mu L$. Here $Q\equiv Q_i$ and $L\equiv L_j$ can be chosen to be from any generation $i,j=1, 2, 3$ to give 9 different possible leptoquark couplings. For ease of notation we suppress generation indices in the following.
The fields and representations are shown in \tabref{tab:model}.
\begin{table}[!htbp]
   \caption{Fields and representations. There are three generations of each of the SM fermion fields but only one copy of the heavy fermion $\Psi + \bar \Psi$. $H$ is the SM Higgs doublet and $\Omega$ is a ``bi-doublet" scalar field which is in the fundamental representation of $SU(3)\times U(1)$ (thought of as a subgroup of a larger $SU(4)$) and the anti-fundamental of $SU(4)$. A TeV scale vev for $\Omega$ is responsible for breaking $[SU(3)\times U(1)] \times SU(4) \rightarrow SU(3)\times U(1)|_{diagonal}$.}
   \vspace{0.1in}
   \label{tab:model}
   \centering
   \begin{tabular}{@{} lcccc @{}} 
      \hline
   Field        & $SU(3)$ &  $SU(2)$ & $U(1)$ & $SU(4)$  \\
               \hline
$ Q$ & $\bm{3}$ & $\bm{2}$ & $1/6$ & -   \\
$ U^c$ & $\bar{\bm{ 3}}$ & - & $-2/3$ & -   \\
$ D^c$ & $\bar {\bm 3}$ & - & $1/3$ & -   \\
$ L$ & - & $\bm{2}$ & $-1/2$ & -   \\
$ E^c $ & - & - & $1$ & -   \\
$H$ & - & $\bm{2}$ & $-1/2$ & - \\
$ \Psi $ & - & $\bm{2}$ & $-$ & $\bm{4}$   \\
$ \Psi^c $ & - & $\bm{2}$ & $-$ & $\bar {\bm{4}}$   \\
$\Omega$ &$(\bm{3},\bm{1})$& - & $(-1/6,1/2)$ & $\bar {\bm{4}}$ \\
     \hline
   \end{tabular}

\end{table}

The gauge group spontaneously breaks at the TeV scale when a vev for the bifundamental field $\Omega$ develops. It is useful to think of the $SU(3)\times U(1)$ gauge groups as a subgroup of a large global $SU(4)$ symmetry. Then the reducible field $\Omega$ transforms as a $\bm{4}\otimes \bm{\bar {4}}$ representation under this $SU(4)$ times the gauged $SU(4)$, and the $\Omega$ vacuum expectation value (vev) breaks the two $SU(4)$ symmetries to the diagonal. This leaves an $SU(3)\times U(1)|_{diagonal}=SU(3)_{color}\times U(1)_Y$ gauge group unbroken.
The vev can be written as a diagonal square matrix 
\beq 
\begin{bmatrix}
   f_3 & & & \\
   & f_3 & & \\
  & & f_3 & \\
   & & & f_1
\end{bmatrix}
\eeq
The Higgs mechanism generates masses for a full $SU(4)$'s worth of gauge bosons. The massive gauge bosons are a color octet of massive ``gluons" $g'$, a $Z'$, and the complex color triplet leptoquark $V$. We are interested in the limit of large $g_4$, there the vector masses become approximately
\begin{eqnarray}
m_{g'} & = & \sqrt{g_4^2+g_3^2}\, f_3  \stackrel{g_4 \gg g_3}{\simeq} g_4 f_3\ , \\ 
m_{Z'} & = & \frac12 \sqrt{g_4^2+2g_1^2/3} \sqrt{f_3^2+3f_1^2} \stackrel{g_4 \gg g_1}{\simeq}  \frac{g_4}{2} \sqrt{f_3^2+3f_1^2}\ , \\ 
m_V & = & \frac{g_4}{\sqrt{2}}  \sqrt{f_3^2+f_1^2}\ .
\end{eqnarray}
These formulas show that by varying $f_3$ and $f_4$ one can obtain arbitrary leptoquark masses with
$m_{g'}/m_V \in [0,\sqrt{2}]$. The gauge bosons which remain massless after this symmetry breaking are the gluons with gauge coupling $g_s=g_3 g_4 /\sqrt{g_4^2+g_3^2}\simeq g_3$ and the hypercharge boson with coupling $g_Y=g_1 g_4/\sqrt{g_4^2+2g_1^2/3}\simeq g_1$. 

For pair production of the vector leptoquark we need to know the couplings of the leptoquark to gluons. These are given in part by the covariant derivatives in the kinetic term of $V$. But there is another renormalizable coupling between gluons and the leptoquark that is allowed by gauge invariance
\beq
(1+\kappa_s)\, g_s\, G_{\mu\nu}^a (V^\mu)^\dagger T^a V^\nu \ ,
\eeq  
where $G_{\mu\nu}^a$ is the gluon field strength and $T^a$ is a Gell-Mann color matrix. The parameter $\kappa_s$ is not fixed by gauge invariance of the low energy theory, but it  can be derived by expanding out the gauge kinetic terms of the unbroken 
$SU(3)\times SU(4)$ gauge theory in terms of the massive and massless vectors. One finds that $\kappa_s=0$. The vanishing of  $\kappa_s$ can also be derived without looking at the full theory by considering unitarity of leptoquark pair production. Only for $\kappa_s=0$ does the pair production amplitude $g g \rightarrow V \bar V$ have good UV behavior. 

In addition, there is a contribution to $q \bar q \rightarrow V \bar V$ from a diagram with the heavy gluon $g'$ as an $s$-channel resonance. The masses of the vectors are such that this resonance is off-shell, but it is required to unitarize the
$q \bar q \rightarrow V \bar V$ pair production amplitude. Therefore we need to determine the couplings of the heavy vectors to fermions.  $G'$ and $Z'$ are the orthogonal linear combinations to the gluons and hypercharge. 
The $G'$ coupling to colored fermions is
\beq
\frac{g_3^2}{\sqrt{g_3^2+g_4^2}} \simeq \frac{g_s^2}{g_4}\ll g_s\ ,
\eeq
and the $Z'$ couplings are proportional to hypercharges times
\beq
\frac{g_1^2}{\sqrt{g_1^2+3g_4^2/2}} \simeq \sqrt{\frac23} \frac{g_Y^2}{g_4} \ll g_Y \ .
\eeq
We see that the light fermions decouple from the heavy gauge bosons in the limit $g_4 \gg g_3, g_1$, making the model safe from precision electroweak constraints and $g'$ or $Z'$ resonance searches at the LHC. 

The $s$-channel diagram for pair production of leptoquarks with intermediate $g'$ coupling to leptoquarks
is not suppressed in the large $g_4$ limit.
This is because the coupling of the heavy gluon to the leptoquarks is proportional to $g_4$.
Then the $s$-channel is proportional to couplings  $g_s^2/g_4 \times g_4 = g_s^2$
just like the QCD diagrams. We included this contribution in \figref{fig:crossec}
and varied the mass of $g'$ within its allowed limits. In the end, this contribution turns out to not be very
important because the large $g g$ parton luminosity enhances the gluon initiated diagrams over the quark initiated ones.

Finally, we consider leptoquark decay through its coupling to light fermions. In the model described so described so far
the coupling vanishes because leptoquark is an $SU(4)$ gauge boson and the fermions are not charged under $SU(4)$. There is also no mixing of the leptoquark with any of the light gauge bosons. However, it does couple to the heavy fermions $\Psi$.
Thus we introduce mixing between $\Psi$ and $Q$ and $L$ of the SM to generate a coupling of $V$ to the light fermions.
The relevant mass terms are 
\beq
\mathcal{L}_\text{mix} =\lambda_\Omega (Q,L) \Omega \Psi^c + M \Psi \Psi^c + \text{h.c.}
\eeq
where - for simplicity - we have imposed a global $SU(4)$ symmetry on the couplings of $Q$ and $L$. In the limit
$\lambda_\Omega \langle\Omega\rangle \ll M$ the light fermions are still approximately $Q$ and $L$ but with a
small $\sin(\theta_\Psi) = \lambda_\Omega \langle\Omega\rangle/ \sqrt{M^2+(\lambda_\Omega \langle\Omega\rangle)^2} \simeq \lambda_\Omega \langle\Omega\rangle/ M$ admixture of $\Psi$. Therefore the effective coupling of the
vector leptoquark to light fermions becomes 
\beq
\lambda = g_4 (\sin\theta_\Psi)^2 \ .
\eeq
Assuming that $M$ is large enough that the leptoquark does not decay to the heavy fermions this is the only decay mode of the leptoquark, and as long as $\lambda > 10^{-7}$ the decay is prompt on detector time scales.

\bibliography{leptoquark.bib}

\begin{thebibliography}{64}%
\makeatletter
\providecommand \@ifxundefined [1]{%
 \@ifx{#1\undefined}
}%
\providecommand \@ifnum [1]{%
 \ifnum #1\expandafter \@firstoftwo
 \else \expandafter \@secondoftwo
 \fi
}%
\providecommand \@ifx [1]{%
 \ifx #1\expandafter \@firstoftwo
 \else \expandafter \@secondoftwo
 \fi
}%
\providecommand \natexlab [1]{#1}%
\providecommand \enquote  [1]{``#1''}%
\providecommand \bibnamefont  [1]{#1}%
\providecommand \bibfnamefont [1]{#1}%
\providecommand \citenamefont [1]{#1}%
\providecommand \href@noop [0]{\@secondoftwo}%
\providecommand \href [0]{\begingroup \@sanitize@url \@href}%
\providecommand \@href[1]{\@@startlink{#1}\@@href}%
\providecommand \@@href[1]{\endgroup#1\@@endlink}%
\providecommand \@sanitize@url [0]{\catcode `\\12\catcode `\$12\catcode
  `\&12\catcode `\#12\catcode `\^12\catcode `\_12\catcode `\%12\relax}%
\providecommand \@@startlink[1]{}%
\providecommand \@@endlink[0]{}%
\providecommand \url  [0]{\begingroup\@sanitize@url \@url }%
\providecommand \@url [1]{\endgroup\@href {#1}{\urlprefix }}%
\providecommand \urlprefix  [0]{URL }%
\providecommand \Eprint [0]{\href }%
\providecommand \doibase [0]{http://dx.doi.org/}%
\providecommand \selectlanguage [0]{\@gobble}%
\providecommand \bibinfo  [0]{\@secondoftwo}%
\providecommand \bibfield  [0]{\@secondoftwo}%
\providecommand \translation [1]{[#1]}%
\providecommand \BibitemOpen [0]{}%
\providecommand \bibitemStop [0]{}%
\providecommand \bibitemNoStop [0]{.\EOS\space}%
\providecommand \EOS [0]{\spacefactor3000\relax}%
\providecommand \BibitemShut  [1]{\csname bibitem#1\endcsname}%
\let\auto@bib@innerbib\@empty
\bibitem [{\citenamefont {Foster}\ \emph {et~al.}(to appear.)\citenamefont
  {Foster}, \citenamefont {Schmaltz},\ and\ \citenamefont
  {Zhong}}]{inprogress}%
  \BibitemOpen
  \bibfield  {author} {\bibinfo {author} {\bibfnamefont {Sean}\ \bibnamefont
  {Foster}}, \bibinfo {author} {\bibfnamefont {Martin}\ \bibnamefont
  {Schmaltz}}, \ and\ \bibinfo {author} {\bibfnamefont {Yi-Ming}\ \bibnamefont
  {Zhong}},\ }\bibfield  {title} {\enquote {\bibinfo {title} {{The Leptoquark
  Hunter's Guide: Single Production}},}\ }\href@noop {} {\  (\bibinfo {year}
  {to appear.})}\BibitemShut {NoStop}%
\bibitem [{\citenamefont {Aaij}\ \emph {et~al.}(2014)\citenamefont {Aaij} \emph
  {et~al.}}]{Aaij:2014ora}%
  \BibitemOpen
  \bibfield  {author} {\bibinfo {author} {\bibfnamefont {Roel}\ \bibnamefont
  {Aaij}} \emph {et~al.} (\bibinfo {collaboration} {LHCb}),\ }\bibfield
  {title} {\enquote {\bibinfo {title} {{Test of lepton universality using
  $B^{+}\rightarrow K^{+}\ell^{+}\ell^{-}$ decays}},}\ }\href {\doibase
  10.1103/PhysRevLett.113.151601} {\bibfield  {journal} {\bibinfo  {journal}
  {Phys. Rev. Lett.}\ }\textbf {\bibinfo {volume} {113}},\ \bibinfo {pages}
  {151601} (\bibinfo {year} {2014})},\ \Eprint {http://arxiv.org/abs/1406.6482}
  {arXiv:1406.6482 [hep-ex]} \BibitemShut {NoStop}%
\bibitem [{\citenamefont {Aaij}\ \emph {et~al.}(2017)\citenamefont {Aaij} \emph
  {et~al.}}]{Aaij:2017vbb}%
  \BibitemOpen
  \bibfield  {author} {\bibinfo {author} {\bibfnamefont {R.}~\bibnamefont
  {Aaij}} \emph {et~al.} (\bibinfo {collaboration} {LHCb}),\ }\bibfield
  {title} {\enquote {\bibinfo {title} {{Test of lepton universality with $B^{0}
  \rightarrow K^{*0}\ell^{+}\ell^{-}$ decays}},}\ }\href@noop {} {\  (\bibinfo
  {year} {2017})},\ \Eprint {http://arxiv.org/abs/1705.05802} {arXiv:1705.05802
  [hep-ex]} \BibitemShut {NoStop}%
\bibitem [{\citenamefont {Bennett}\ \emph {et~al.}(2006)\citenamefont {Bennett}
  \emph {et~al.}}]{Bennett:2006fi}%
  \BibitemOpen
  \bibfield  {author} {\bibinfo {author} {\bibfnamefont {G.~W.}\ \bibnamefont
  {Bennett}} \emph {et~al.} (\bibinfo {collaboration} {Muon g-2}),\ }\bibfield
  {title} {\enquote {\bibinfo {title} {{Final Report of the Muon E821 Anomalous
  Magnetic Moment Measurement at BNL}},}\ }\href {\doibase
  10.1103/PhysRevD.73.072003} {\bibfield  {journal} {\bibinfo  {journal} {Phys.
  Rev.}\ }\textbf {\bibinfo {volume} {D73}},\ \bibinfo {pages} {072003}
  (\bibinfo {year} {2006})},\ \Eprint {http://arxiv.org/abs/hep-ex/0602035}
  {arXiv:hep-ex/0602035 [hep-ex]} \BibitemShut {NoStop}%
\bibitem [{\citenamefont {Aaboud}\ \emph
  {et~al.}(2016{\natexlab{a}})\citenamefont {Aaboud} \emph
  {et~al.}}]{Aaboud:2016qeg}%
  \BibitemOpen
  \bibfield  {author} {\bibinfo {author} {\bibfnamefont {Morad}\ \bibnamefont
  {Aaboud}} \emph {et~al.} (\bibinfo {collaboration} {ATLAS}),\ }\bibfield
  {title} {\enquote {\bibinfo {title} {{Search for scalar leptoquarks in pp
  collisions at $\sqrt{s}$ = 13 TeV with the ATLAS experiment}},}\ }\href
  {\doibase 10.1088/1367-2630/18/9/093016} {\bibfield  {journal} {\bibinfo
  {journal} {New J. Phys.}\ }\textbf {\bibinfo {volume} {18}},\ \bibinfo
  {pages} {093016} (\bibinfo {year} {2016}{\natexlab{a}})},\ \Eprint
  {http://arxiv.org/abs/1605.06035} {arXiv:1605.06035 [hep-ex]} \BibitemShut
  {NoStop}%
\bibitem [{CMS(2016{\natexlab{a}})}]{CMS-PAS-EXO-16-007}%
  \BibitemOpen
  \href {http://cds.cern.ch/record/2139349} {\emph {\bibinfo {title} {{Search
  for pair-production of second-generation scalar leptoquarks in pp collisions
  at $\sqrt{s}=13~\mathrm{TeV}$ with the CMS detector}}}},\ \bibinfo {type}
  {Tech. Rep.}\ \bibinfo {number} {CMS-PAS-EXO-16-007}\ (\bibinfo
  {institution} {CERN},\ \bibinfo {address} {Geneva},\ \bibinfo {year}
  {2016})\BibitemShut {NoStop}%
\bibitem [{\citenamefont {Raj}(2017)}]{Raj:2016aky}%
  \BibitemOpen
  \bibfield  {author} {\bibinfo {author} {\bibfnamefont {Nirmal}\ \bibnamefont
  {Raj}},\ }\bibfield  {title} {\enquote {\bibinfo {title} {{Anticipating
  nonresonant new physics in dilepton angular spectra at the LHC}},}\ }\href
  {\doibase 10.1103/PhysRevD.95.015011} {\bibfield  {journal} {\bibinfo
  {journal} {Phys. Rev.}\ }\textbf {\bibinfo {volume} {D95}},\ \bibinfo {pages}
  {015011} (\bibinfo {year} {2017})},\ \Eprint
  {http://arxiv.org/abs/1610.03795} {arXiv:1610.03795 [hep-ph]} \BibitemShut
  {NoStop}%
\bibitem [{\citenamefont {Blumlein}\ \emph {et~al.}(1998)\citenamefont
  {Blumlein}, \citenamefont {Boos},\ and\ \citenamefont
  {Kryukov}}]{Blumlein:1998ym}%
  \BibitemOpen
  \bibfield  {author} {\bibinfo {author} {\bibfnamefont {Johannes}\
  \bibnamefont {Blumlein}}, \bibinfo {author} {\bibfnamefont {Edward}\
  \bibnamefont {Boos}}, \ and\ \bibinfo {author} {\bibfnamefont {Alexander}\
  \bibnamefont {Kryukov}},\ }\bibfield  {title} {\enquote {\bibinfo {title}
  {{Leptoquark pair production cross-sections at hadron colliders}},}\
  }\href@noop {} {\  (\bibinfo {year} {1998})},\ \Eprint
  {http://arxiv.org/abs/hep-ph/9811271} {arXiv:hep-ph/9811271 [hep-ph]}
  \BibitemShut {NoStop}%
\bibitem [{\citenamefont {Blumlein}\ \emph {et~al.}(1997)\citenamefont
  {Blumlein}, \citenamefont {Boos},\ and\ \citenamefont
  {Kryukov}}]{Blumlein:1996qp}%
  \BibitemOpen
  \bibfield  {author} {\bibinfo {author} {\bibfnamefont {Johannes}\
  \bibnamefont {Blumlein}}, \bibinfo {author} {\bibfnamefont {Edward}\
  \bibnamefont {Boos}}, \ and\ \bibinfo {author} {\bibfnamefont {Alexander}\
  \bibnamefont {Kryukov}},\ }\bibfield  {title} {\enquote {\bibinfo {title}
  {{Leptoquark pair production in hadronic interactions}},}\ }\href {\doibase
  10.1007/s002880050538} {\bibfield  {journal} {\bibinfo  {journal} {Z. Phys.}\
  }\textbf {\bibinfo {volume} {C76}},\ \bibinfo {pages} {137--153} (\bibinfo
  {year} {1997})},\ \Eprint {http://arxiv.org/abs/hep-ph/9610408}
  {arXiv:hep-ph/9610408 [hep-ph]} \BibitemShut {NoStop}%
\bibitem [{\citenamefont {Kramer}\ \emph {et~al.}(1997)\citenamefont {Kramer},
  \citenamefont {Plehn}, \citenamefont {Spira},\ and\ \citenamefont
  {Zerwas}}]{Kramer:1997hh}%
  \BibitemOpen
  \bibfield  {author} {\bibinfo {author} {\bibfnamefont {M.}~\bibnamefont
  {Kramer}}, \bibinfo {author} {\bibfnamefont {T.}~\bibnamefont {Plehn}},
  \bibinfo {author} {\bibfnamefont {M.}~\bibnamefont {Spira}}, \ and\ \bibinfo
  {author} {\bibfnamefont {P.~M.}\ \bibnamefont {Zerwas}},\ }\bibfield  {title}
  {\enquote {\bibinfo {title} {{Pair production of scalar leptoquarks at the
  Tevatron}},}\ }\href {\doibase 10.1103/PhysRevLett.79.341} {\bibfield
  {journal} {\bibinfo  {journal} {Phys. Rev. Lett.}\ }\textbf {\bibinfo
  {volume} {79}},\ \bibinfo {pages} {341--344} (\bibinfo {year} {1997})},\
  \Eprint {http://arxiv.org/abs/hep-ph/9704322} {arXiv:hep-ph/9704322 [hep-ph]}
  \BibitemShut {NoStop}%
\bibitem [{\citenamefont {Kramer}\ \emph {et~al.}(2005)\citenamefont {Kramer},
  \citenamefont {Plehn}, \citenamefont {Spira},\ and\ \citenamefont
  {Zerwas}}]{Kramer:2004df}%
  \BibitemOpen
  \bibfield  {author} {\bibinfo {author} {\bibfnamefont {M.}~\bibnamefont
  {Kramer}}, \bibinfo {author} {\bibfnamefont {T.}~\bibnamefont {Plehn}},
  \bibinfo {author} {\bibfnamefont {M.}~\bibnamefont {Spira}}, \ and\ \bibinfo
  {author} {\bibfnamefont {P.~M.}\ \bibnamefont {Zerwas}},\ }\bibfield  {title}
  {\enquote {\bibinfo {title} {{Pair production of scalar leptoquarks at the
  CERN LHC}},}\ }\href {\doibase 10.1103/PhysRevD.71.057503} {\bibfield
  {journal} {\bibinfo  {journal} {Phys. Rev.}\ }\textbf {\bibinfo {volume}
  {D71}},\ \bibinfo {pages} {057503} (\bibinfo {year} {2005})},\ \Eprint
  {http://arxiv.org/abs/hep-ph/0411038} {arXiv:hep-ph/0411038 [hep-ph]}
  \BibitemShut {NoStop}%
\bibitem [{\citenamefont {Mandal}\ \emph {et~al.}(2016)\citenamefont {Mandal},
  \citenamefont {Mitra},\ and\ \citenamefont {Seth}}]{Mandal:2015lca}%
  \BibitemOpen
  \bibfield  {author} {\bibinfo {author} {\bibfnamefont {Tanumoy}\ \bibnamefont
  {Mandal}}, \bibinfo {author} {\bibfnamefont {Subhadip}\ \bibnamefont
  {Mitra}}, \ and\ \bibinfo {author} {\bibfnamefont {Satyajit}\ \bibnamefont
  {Seth}},\ }\bibfield  {title} {\enquote {\bibinfo {title} {{Pair Production
  of Scalar Leptoquarks at the LHC to NLO Parton Shower Accuracy}},}\ }\href
  {\doibase 10.1103/PhysRevD.93.035018} {\bibfield  {journal} {\bibinfo
  {journal} {Phys. Rev.}\ }\textbf {\bibinfo {volume} {D93}},\ \bibinfo {pages}
  {035018} (\bibinfo {year} {2016})},\ \Eprint
  {http://arxiv.org/abs/1506.07369} {arXiv:1506.07369 [hep-ph]} \BibitemShut
  {NoStop}%
\bibitem [{ATL(2017{\natexlab{a}})}]{ATLAS-CONF-2017-036}%
  \BibitemOpen
  \href {http://cds.cern.ch/record/2265808} {\emph {\bibinfo {title} {{A search
  for B-L R-parity-violating scalar tops in $\sqrt{s}$ = 13 TeV $pp$ collisions
  with the ATLAS experiment}}}},\ \bibinfo {type} {Tech. Rep.}\ \bibinfo
  {number} {ATLAS-CONF-2017-036}\ (\bibinfo  {institution} {CERN},\ \bibinfo
  {address} {Geneva},\ \bibinfo {year} {2017})\BibitemShut {NoStop}%
\bibitem [{\citenamefont {Chatrchyan}\ \emph
  {et~al.}(2013{\natexlab{a}})\citenamefont {Chatrchyan} \emph
  {et~al.}}]{Chatrchyan:2013mys}%
  \BibitemOpen
  \bibfield  {author} {\bibinfo {author} {\bibfnamefont {Serguei}\ \bibnamefont
  {Chatrchyan}} \emph {et~al.} (\bibinfo {collaboration} {CMS}),\ }\bibfield
  {title} {\enquote {\bibinfo {title} {{Search for supersymmetry in hadronic
  final states with missing transverse energy using the variables $\alpha_T$
  and b-quark multiplicity in pp collisions at $\sqrt s=8$ TeV}},}\ }\href
  {\doibase 10.1140/epjc/s10052-013-2568-6} {\bibfield  {journal} {\bibinfo
  {journal} {Eur. Phys. J.}\ }\textbf {\bibinfo {volume} {C73}},\ \bibinfo
  {pages} {2568} (\bibinfo {year} {2013}{\natexlab{a}})},\ \Eprint
  {http://arxiv.org/abs/1303.2985} {arXiv:1303.2985 [hep-ex]} \BibitemShut
  {NoStop}%
\bibitem [{\citenamefont {Khachatryan}\ \emph
  {et~al.}(2015{\natexlab{a}})\citenamefont {Khachatryan} \emph
  {et~al.}}]{Khachatryan:2015vra}%
  \BibitemOpen
  \bibfield  {author} {\bibinfo {author} {\bibfnamefont {Vardan}\ \bibnamefont
  {Khachatryan}} \emph {et~al.} (\bibinfo {collaboration} {CMS}),\ }\bibfield
  {title} {\enquote {\bibinfo {title} {{Searches for Supersymmetry using the
  M$_{T2}$ Variable in Hadronic Events Produced in pp Collisions at 8 TeV}},}\
  }\href {\doibase 10.1007/JHEP05(2015)078} {\bibfield  {journal} {\bibinfo
  {journal} {JHEP}\ }\textbf {\bibinfo {volume} {05}},\ \bibinfo {pages} {078}
  (\bibinfo {year} {2015}{\natexlab{a}})},\ \Eprint
  {http://arxiv.org/abs/1502.04358} {arXiv:1502.04358 [hep-ex]} \BibitemShut
  {NoStop}%
\bibitem [{CMS(2017{\natexlab{a}})}]{CMS-PAS-SUS-16-033}%
  \BibitemOpen
  \href {http://cds.cern.ch/record/2256850} {\emph {\bibinfo {title} {{ Search
  for supersymmetry in multijet events with missing transverse momentum in
  proton-proton collisions at $13~\mathrm{TeV}$ }}}},\ \bibinfo {type} {Tech.
  Rep.}\ \bibinfo {number} {CMS-PAS-SUS-16-033}\ (\bibinfo  {institution}
  {CERN},\ \bibinfo {address} {Geneva},\ \bibinfo {year} {2017})\BibitemShut
  {NoStop}%
\bibitem [{CMS(2017{\natexlab{b}})}]{CMS-PAS-SUS-16-036}%
  \BibitemOpen
  \href {http://cds.cern.ch/record/2256872} {\emph {\bibinfo {title} {{Search
  for new physics in the all-hadronic final state with the MT2 variable}}}},\
  \bibinfo {type} {Tech. Rep.}\ \bibinfo {number} {CMS-PAS-SUS-16-036}\
  (\bibinfo  {institution} {CERN},\ \bibinfo {address} {Geneva},\ \bibinfo
  {year} {2017})\BibitemShut {NoStop}%
\bibitem [{\citenamefont {Sirunyan}\ \emph
  {et~al.}(2017{\natexlab{a}})\citenamefont {Sirunyan} \emph
  {et~al.}}]{Sirunyan:2017cwe}%
  \BibitemOpen
  \bibfield  {author} {\bibinfo {author} {\bibfnamefont {Albert~M}\
  \bibnamefont {Sirunyan}} \emph {et~al.} (\bibinfo {collaboration} {CMS}),\
  }\bibfield  {title} {\enquote {\bibinfo {title} {{Search for supersymmetry in
  multijet events with missing transverse momentum in proton-proton collisions
  at 13 TeV}},}\ }\href@noop {} {\  (\bibinfo {year} {2017}{\natexlab{a}})},\
  \Eprint {http://arxiv.org/abs/1704.07781} {arXiv:1704.07781 [hep-ex]}
  \BibitemShut {NoStop}%
\bibitem [{\citenamefont {Aad}\ \emph {et~al.}(2015{\natexlab{a}})\citenamefont
  {Aad} \emph {et~al.}}]{Aad:2015gna}%
  \BibitemOpen
  \bibfield  {author} {\bibinfo {author} {\bibfnamefont {Georges}\ \bibnamefont
  {Aad}} \emph {et~al.} (\bibinfo {collaboration} {ATLAS}),\ }\bibfield
  {title} {\enquote {\bibinfo {title} {{Search for Scalar Charm Quark Pair
  Production in $pp$ Collisions at $\sqrt{s}=$ 8  TeV with the ATLAS
  Detector}},}\ }\href {\doibase 10.1103/PhysRevLett.114.161801} {\bibfield
  {journal} {\bibinfo  {journal} {Phys. Rev. Lett.}\ }\textbf {\bibinfo
  {volume} {114}},\ \bibinfo {pages} {161801} (\bibinfo {year}
  {2015}{\natexlab{a}})},\ \Eprint {http://arxiv.org/abs/1501.01325}
  {arXiv:1501.01325 [hep-ex]} \BibitemShut {NoStop}%
\bibitem [{ATL(2017{\natexlab{b}})}]{ATLAS-CONF-2017-022}%
  \BibitemOpen
  \href {http://cds.cern.ch/record/2258145} {\emph {\bibinfo {title} {{Search
  for squarks and gluinos in final states with jets and missing transverse
  momentum using 36 fb$^{−1}$ of $\sqrt{s} =13$ TeV pp collision data with
  the ATLAS detector}}}},\ \bibinfo {type} {Tech. Rep.}\ \bibinfo {number}
  {ATLAS-CONF-2017-022}\ (\bibinfo  {institution} {CERN},\ \bibinfo {address}
  {Geneva},\ \bibinfo {year} {2017})\BibitemShut {NoStop}%
\bibitem [{\citenamefont {Aad}\ \emph {et~al.}(2013{\natexlab{a}})\citenamefont
  {Aad} \emph {et~al.}}]{Aad:2013ija}%
  \BibitemOpen
  \bibfield  {author} {\bibinfo {author} {\bibfnamefont {Georges}\ \bibnamefont
  {Aad}} \emph {et~al.} (\bibinfo {collaboration} {ATLAS}),\ }\bibfield
  {title} {\enquote {\bibinfo {title} {{Search for direct third-generation
  squark pair production in final states with missing transverse momentum and
  two $b$-jets in $\sqrt{s} =$ 8 TeV $pp$ collisions with the ATLAS
  detector}},}\ }\href {\doibase 10.1007/JHEP10(2013)189} {\bibfield  {journal}
  {\bibinfo  {journal} {JHEP}\ }\textbf {\bibinfo {volume} {10}},\ \bibinfo
  {pages} {189} (\bibinfo {year} {2013}{\natexlab{a}})},\ \Eprint
  {http://arxiv.org/abs/1308.2631} {arXiv:1308.2631 [hep-ex]} \BibitemShut
  {NoStop}%
\bibitem [{\citenamefont {Aad}\ \emph {et~al.}(2016)\citenamefont {Aad} \emph
  {et~al.}}]{Aad:2015caa}%
  \BibitemOpen
  \bibfield  {author} {\bibinfo {author} {\bibfnamefont {Georges}\ \bibnamefont
  {Aad}} \emph {et~al.} (\bibinfo {collaboration} {ATLAS}),\ }\bibfield
  {title} {\enquote {\bibinfo {title} {{Searches for scalar leptoquarks in pp
  collisions at ${\sqrt{s}}$ = 8 TeV with the ATLAS detector}},}\ }\href
  {\doibase 10.1140/epjc/s10052-015-3823-9} {\bibfield  {journal} {\bibinfo
  {journal} {Eur. Phys. J.}\ }\textbf {\bibinfo {volume} {C76}},\ \bibinfo
  {pages} {5} (\bibinfo {year} {2016})},\ \Eprint
  {http://arxiv.org/abs/1508.04735} {arXiv:1508.04735 [hep-ex]} \BibitemShut
  {NoStop}%
\bibitem [{\citenamefont {Aad}\ \emph {et~al.}(2015{\natexlab{b}})\citenamefont
  {Aad} \emph {et~al.}}]{Aad:2015pfx}%
  \BibitemOpen
  \bibfield  {author} {\bibinfo {author} {\bibfnamefont {Georges}\ \bibnamefont
  {Aad}} \emph {et~al.} (\bibinfo {collaboration} {ATLAS}),\ }\bibfield
  {title} {\enquote {\bibinfo {title} {{ATLAS Run 1 searches for direct pair
  production of third-generation squarks at the Large Hadron Collider}},}\
  }\href {\doibase 10.1140/epjc/s10052-015-3726-9,
  10.1140/epjc/s10052-016-3935-x} {\bibfield  {journal} {\bibinfo  {journal}
  {Eur. Phys. J.}\ }\textbf {\bibinfo {volume} {C75}},\ \bibinfo {pages} {510}
  (\bibinfo {year} {2015}{\natexlab{b}})},\ \bibinfo {note} {[Erratum: Eur.
  Phys. J.C76,no.3,153(2016)]},\ \Eprint {http://arxiv.org/abs/1506.08616}
  {arXiv:1506.08616 [hep-ex]} \BibitemShut {NoStop}%
\bibitem [{\citenamefont {Aaboud}\ \emph
  {et~al.}(2016{\natexlab{b}})\citenamefont {Aaboud} \emph
  {et~al.}}]{Aaboud:2016nwl}%
  \BibitemOpen
  \bibfield  {author} {\bibinfo {author} {\bibfnamefont {Morad}\ \bibnamefont
  {Aaboud}} \emph {et~al.} (\bibinfo {collaboration} {ATLAS}),\ }\bibfield
  {title} {\enquote {\bibinfo {title} {{Search for bottom squark pair
  production in proton--proton collisions at $\sqrt{s}=13$ TeV with the ATLAS
  detector}},}\ }\href {\doibase 10.1140/epjc/s10052-016-4382-4} {\bibfield
  {journal} {\bibinfo  {journal} {Eur. Phys. J.}\ }\textbf {\bibinfo {volume}
  {C76}},\ \bibinfo {pages} {547} (\bibinfo {year} {2016}{\natexlab{b}})},\
  \Eprint {http://arxiv.org/abs/1606.08772} {arXiv:1606.08772 [hep-ex]}
  \BibitemShut {NoStop}%
\bibitem [{\citenamefont {Sirunyan}\ \emph {et~al.}(2016)\citenamefont
  {Sirunyan} \emph {et~al.}}]{Sirunyan:2016jpr}%
  \BibitemOpen
  \bibfield  {author} {\bibinfo {author} {\bibfnamefont {Albert~M}\
  \bibnamefont {Sirunyan}} \emph {et~al.} (\bibinfo {collaboration} {CMS}),\
  }\bibfield  {title} {\enquote {\bibinfo {title} {{Searches for pair
  production for third-generation squarks in sqrt(s)=13 TeV pp collisions}},}\
  }\href@noop {} {\  (\bibinfo {year} {2016})},\ \Eprint
  {http://arxiv.org/abs/1612.03877} {arXiv:1612.03877 [hep-ex]} \BibitemShut
  {NoStop}%
\bibitem [{ATL(2017{\natexlab{c}})}]{ATLAS-CONF-2017-38}%
  \BibitemOpen
  \href@noop {} {\emph {\bibinfo {title} {{Search for supersymmetry in events
  with b-tagged jets and missing transverse momentum in pp collisions at $\sqrt
  s=13$ TeV with the ATLAS detector}}}},\ \bibinfo {type} {Tech. Rep.}\
  \bibinfo {number} {ATLAS-CONF-2017-038}\ (\bibinfo  {institution} {CERN},\
  \bibinfo {address} {Geneva},\ \bibinfo {year} {2017})\BibitemShut {NoStop}%
\bibitem [{\citenamefont {Chatrchyan}\ \emph
  {et~al.}(2013{\natexlab{b}})\citenamefont {Chatrchyan} \emph
  {et~al.}}]{Chatrchyan:2013xna}%
  \BibitemOpen
  \bibfield  {author} {\bibinfo {author} {\bibfnamefont {Serguei}\ \bibnamefont
  {Chatrchyan}} \emph {et~al.} (\bibinfo {collaboration} {CMS}),\ }\bibfield
  {title} {\enquote {\bibinfo {title} {{Search for top-squark pair production
  in the single-lepton final state in pp collisions at $\sqrt{s}$ = 8 TeV}},}\
  }\href {\doibase 10.1140/epjc/s10052-013-2677-2} {\bibfield  {journal}
  {\bibinfo  {journal} {Eur. Phys. J.}\ }\textbf {\bibinfo {volume} {C73}},\
  \bibinfo {pages} {2677} (\bibinfo {year} {2013}{\natexlab{b}})},\ \Eprint
  {http://arxiv.org/abs/1308.1586} {arXiv:1308.1586 [hep-ex]} \BibitemShut
  {NoStop}%
\bibitem [{\citenamefont {Aad}\ \emph {et~al.}(2014)\citenamefont {Aad} \emph
  {et~al.}}]{Aad:2014kra}%
  \BibitemOpen
  \bibfield  {author} {\bibinfo {author} {\bibfnamefont {Georges}\ \bibnamefont
  {Aad}} \emph {et~al.} (\bibinfo {collaboration} {ATLAS}),\ }\bibfield
  {title} {\enquote {\bibinfo {title} {{Search for top squark pair production
  in final states with one isolated lepton, jets, and missing transverse
  momentum in $\sqrt s =$8 TeV $pp$ collisions with the ATLAS detector}},}\
  }\href {\doibase 10.1007/JHEP11(2014)118} {\bibfield  {journal} {\bibinfo
  {journal} {JHEP}\ }\textbf {\bibinfo {volume} {11}},\ \bibinfo {pages} {118}
  (\bibinfo {year} {2014})},\ \Eprint {http://arxiv.org/abs/1407.0583}
  {arXiv:1407.0583 [hep-ex]} \BibitemShut {NoStop}%
\bibitem [{ATL(2016)}]{ATLAS-CONF-2016-077}%
  \BibitemOpen
  \href {http://cds.cern.ch/record/2206250} {\emph {\bibinfo {title} {{Search
  for the Supersymmetric Partner of the Top Quark in the Jets+Emiss Final State
  at sqrt(s) = 13 TeV}}}},\ \bibinfo {type} {Tech. Rep.}\ \bibinfo {number}
  {ATLAS-CONF-2016-077}\ (\bibinfo  {institution} {CERN},\ \bibinfo {address}
  {Geneva},\ \bibinfo {year} {2016})\BibitemShut {NoStop}%
\bibitem [{\citenamefont {Khachatryan}\ \emph
  {et~al.}(2017{\natexlab{a}})\citenamefont {Khachatryan} \emph
  {et~al.}}]{Khachatryan:2016pxa}%
  \BibitemOpen
  \bibfield  {author} {\bibinfo {author} {\bibfnamefont {Vardan}\ \bibnamefont
  {Khachatryan}} \emph {et~al.} (\bibinfo {collaboration} {CMS}),\ }\bibfield
  {title} {\enquote {\bibinfo {title} {{Search for top squark pair production
  in compressed-mass-spectrum scenarios in proton-proton collisions at
  $\sqrt{s}$ = 8 TeV using the $\alpha_T$ variable}},}\ }\href {\doibase
  10.1016/j.physletb.2017.02.007} {\bibfield  {journal} {\bibinfo  {journal}
  {Phys. Lett.}\ }\textbf {\bibinfo {volume} {B767}},\ \bibinfo {pages}
  {403--430} (\bibinfo {year} {2017}{\natexlab{a}})},\ \Eprint
  {http://arxiv.org/abs/1605.08993} {arXiv:1605.08993 [hep-ex]} \BibitemShut
  {NoStop}%
\bibitem [{\citenamefont {Khachatryan}\ \emph
  {et~al.}(2016{\natexlab{a}})\citenamefont {Khachatryan} \emph
  {et~al.}}]{Khachatryan:2016oia}%
  \BibitemOpen
  \bibfield  {author} {\bibinfo {author} {\bibfnamefont {Vardan}\ \bibnamefont
  {Khachatryan}} \emph {et~al.} (\bibinfo {collaboration} {CMS}),\ }\bibfield
  {title} {\enquote {\bibinfo {title} {{Search for direct pair production of
  supersymmetric top quarks decaying to all-hadronic final states in pp
  collisions at $\sqrt{s} = 8\;\text {TeV} $}},}\ }\href {\doibase
  10.1140/epjc/s10052-016-4292-5} {\bibfield  {journal} {\bibinfo  {journal}
  {Eur. Phys. J.}\ }\textbf {\bibinfo {volume} {C76}},\ \bibinfo {pages} {460}
  (\bibinfo {year} {2016}{\natexlab{a}})},\ \Eprint
  {http://arxiv.org/abs/1603.00765} {arXiv:1603.00765 [hep-ex]} \BibitemShut
  {NoStop}%
\bibitem [{\citenamefont {Khachatryan}\ \emph
  {et~al.}(2016{\natexlab{b}})\citenamefont {Khachatryan} \emph
  {et~al.}}]{Khachatryan:2016pup}%
  \BibitemOpen
  \bibfield  {author} {\bibinfo {author} {\bibfnamefont {Vardan}\ \bibnamefont
  {Khachatryan}} \emph {et~al.} (\bibinfo {collaboration} {CMS}),\ }\bibfield
  {title} {\enquote {\bibinfo {title} {{Search for direct pair production of
  scalar top quarks in the single- and dilepton channels in proton-proton
  collisions at $ \sqrt{s}=8 $ TeV}},}\ }\href {\doibase
  10.1007/JHEP07(2016)027, 10.1007/JHEP09(2016)056} {\bibfield  {journal}
  {\bibinfo  {journal} {JHEP}\ }\textbf {\bibinfo {volume} {07}},\ \bibinfo
  {pages} {027} (\bibinfo {year} {2016}{\natexlab{b}})},\ \bibinfo {note}
  {[Erratum: JHEP09,056(2016)]},\ \Eprint {http://arxiv.org/abs/1602.03169}
  {arXiv:1602.03169 [hep-ex]} \BibitemShut {NoStop}%
\bibitem [{ATL(2017{\natexlab{d}})}]{ATLAS-CONF-2017-020}%
  \BibitemOpen
  \href {http://cds.cern.ch/record/2258142} {\emph {\bibinfo {title} {{Search
  for a Scalar Partner of the Top Quark in the Jets+ETmiss Final State at
  $\sqrt{s}$ = 13 TeV with the ATLAS detector}}}},\ \bibinfo {type} {Tech.
  Rep.}\ \bibinfo {number} {ATLAS-CONF-2017-020}\ (\bibinfo  {institution}
  {CERN},\ \bibinfo {address} {Geneva},\ \bibinfo {year} {2017})\BibitemShut
  {NoStop}%
\bibitem [{ATL(2017{\natexlab{e}})}]{ATLAS-CONF-2017-024}%
  \BibitemOpen
  \href {https://cds.cern.ch/record/2258147} {\emph {\bibinfo {title} {{Search
  for new phenomena in events with missing transverse momentum and a Higgs
  boson decaying into two photons at $\sqrt{s}$ = 13 TeV with the ATLAS
  detector}}}},\ \bibinfo {type} {Tech. Rep.}\ \bibinfo {number}
  {ATLAS-CONF-2017-024}\ (\bibinfo  {institution} {CERN},\ \bibinfo {address}
  {Geneva},\ \bibinfo {year} {2017})\BibitemShut {NoStop}%
\bibitem [{CMS(2017{\natexlab{c}})}]{CMS-PAS-SUS-17-001}%
  \BibitemOpen
  \href {http://cds.cern.ch/record/2256753} {\emph {\bibinfo {title} {{Search
  for direct stop pair production in the dilepton final state at $\sqrt{s}=$13
  TeV}}}},\ \bibinfo {type} {Tech. Rep.}\ \bibinfo {number}
  {CMS-PAS-SUS-17-001}\ (\bibinfo  {institution} {CERN},\ \bibinfo {address}
  {Geneva},\ \bibinfo {year} {2017})\BibitemShut {NoStop}%
\bibitem [{CMS(2017{\natexlab{d}})}]{CMS-PAS-SUS-16-049}%
  \BibitemOpen
  \href {http://cds.cern.ch/record/2256439} {\emph {\bibinfo {title} {{Search
  for direct top squark pair production in the all-hadronic final state in
  proton-proton collisions at sqrt(s) = 13 TeV}}}},\ \bibinfo {type} {Tech.
  Rep.}\ \bibinfo {number} {CMS-PAS-SUS-16-049}\ (\bibinfo  {institution}
  {CERN},\ \bibinfo {address} {Geneva},\ \bibinfo {year} {2017})\BibitemShut
  {NoStop}%
\bibitem [{\citenamefont {Sirunyan}\ \emph
  {et~al.}(2017{\natexlab{b}})\citenamefont {Sirunyan} \emph
  {et~al.}}]{Sirunyan:2017wif}%
  \BibitemOpen
  \bibfield  {author} {\bibinfo {author} {\bibfnamefont {Albert~M}\
  \bibnamefont {Sirunyan}} \emph {et~al.} (\bibinfo {collaboration} {CMS}),\
  }\bibfield  {title} {\enquote {\bibinfo {title} {{Search for direct
  production of supersymmetric partners of the top quark in the all-jets final
  state in proton-proton collisions at sqrt(s) = 13 TeV}},}\ }\href@noop {} {\
  (\bibinfo {year} {2017}{\natexlab{b}})},\ \Eprint
  {http://arxiv.org/abs/1707.03316} {arXiv:1707.03316 [hep-ex]} \BibitemShut
  {NoStop}%
\bibitem [{\citenamefont {Khachatryan}\ \emph
  {et~al.}(2011{\natexlab{a}})\citenamefont {Khachatryan} \emph
  {et~al.}}]{Khachatryan:2010mq}%
  \BibitemOpen
  \bibfield  {author} {\bibinfo {author} {\bibfnamefont {Vardan}\ \bibnamefont
  {Khachatryan}} \emph {et~al.} (\bibinfo {collaboration} {CMS}),\ }\bibfield
  {title} {\enquote {\bibinfo {title} {{Search for Pair Production of
  Second-Generation Scalar Leptoquarks in $pp$ Collisions at $\sqrt{s}=7$
  TeV}},}\ }\href {\doibase 10.1103/PhysRevLett.106.201803} {\bibfield
  {journal} {\bibinfo  {journal} {Phys. Rev. Lett.}\ }\textbf {\bibinfo
  {volume} {106}},\ \bibinfo {pages} {201803} (\bibinfo {year}
  {2011}{\natexlab{a}})},\ \Eprint {http://arxiv.org/abs/1012.4033}
  {arXiv:1012.4033 [hep-ex]} \BibitemShut {NoStop}%
\bibitem [{\citenamefont {Khachatryan}\ \emph
  {et~al.}(2011{\natexlab{b}})\citenamefont {Khachatryan} \emph
  {et~al.}}]{Khachatryan:2010mp}%
  \BibitemOpen
  \bibfield  {author} {\bibinfo {author} {\bibfnamefont {Vardan}\ \bibnamefont
  {Khachatryan}} \emph {et~al.} (\bibinfo {collaboration} {CMS}),\ }\bibfield
  {title} {\enquote {\bibinfo {title} {{Search for Pair Production of
  First-Generation Scalar Leptoquarks in $pp$ Collisions at $\sqrt{s}=7$
  TeV}},}\ }\href {\doibase 10.1103/PhysRevLett.106.201802} {\bibfield
  {journal} {\bibinfo  {journal} {Phys. Rev. Lett.}\ }\textbf {\bibinfo
  {volume} {106}},\ \bibinfo {pages} {201802} (\bibinfo {year}
  {2011}{\natexlab{b}})},\ \Eprint {http://arxiv.org/abs/1012.4031}
  {arXiv:1012.4031 [hep-ex]} \BibitemShut {NoStop}%
\bibitem [{\citenamefont {Aad}\ \emph {et~al.}(2012{\natexlab{a}})\citenamefont
  {Aad} \emph {et~al.}}]{Aad:2011ch}%
  \BibitemOpen
  \bibfield  {author} {\bibinfo {author} {\bibfnamefont {Georges}\ \bibnamefont
  {Aad}} \emph {et~al.} (\bibinfo {collaboration} {ATLAS}),\ }\bibfield
  {title} {\enquote {\bibinfo {title} {{Search for first generation scalar
  leptoquarks in $pp$ collisions at $\sqrt{s}=7$ TeV with the ATLAS
  detector}},}\ }\href {\doibase 10.1016/j.physletb.2012.03.023,
  10.1016/j.physletb.2012.02.004} {\bibfield  {journal} {\bibinfo  {journal}
  {Phys. Lett.}\ }\textbf {\bibinfo {volume} {B709}},\ \bibinfo {pages}
  {158--176} (\bibinfo {year} {2012}{\natexlab{a}})},\ \bibinfo {note}
  {[Erratum: Phys. Lett.B711,442(2012)]},\ \Eprint
  {http://arxiv.org/abs/1112.4828} {arXiv:1112.4828 [hep-ex]} \BibitemShut
  {NoStop}%
\bibitem [{\citenamefont {Aad}\ \emph {et~al.}(2011)\citenamefont {Aad} \emph
  {et~al.}}]{Aad:2011uv}%
  \BibitemOpen
  \bibfield  {author} {\bibinfo {author} {\bibfnamefont {Georges}\ \bibnamefont
  {Aad}} \emph {et~al.} (\bibinfo {collaboration} {ATLAS}),\ }\bibfield
  {title} {\enquote {\bibinfo {title} {{Search for pair production of first or
  second generation leptoquarks in proton-proton collisions at $\sqrt{s}=7$ TeV
  using the ATLAS detector at the LHC}},}\ }\href {\doibase
  10.1103/PhysRevD.83.112006} {\bibfield  {journal} {\bibinfo  {journal} {Phys.
  Rev.}\ }\textbf {\bibinfo {volume} {D83}},\ \bibinfo {pages} {112006}
  (\bibinfo {year} {2011})},\ \Eprint {http://arxiv.org/abs/1104.4481}
  {arXiv:1104.4481 [hep-ex]} \BibitemShut {NoStop}%
\bibitem [{\citenamefont {Chatrchyan}\ \emph {et~al.}(2011)\citenamefont
  {Chatrchyan} \emph {et~al.}}]{Chatrchyan:2011ar}%
  \BibitemOpen
  \bibfield  {author} {\bibinfo {author} {\bibfnamefont {Serguei}\ \bibnamefont
  {Chatrchyan}} \emph {et~al.} (\bibinfo {collaboration} {CMS}),\ }\bibfield
  {title} {\enquote {\bibinfo {title} {{Search for First Generation Scalar
  Leptoquarks in the e$\nu$jj channel in $pp$ collisions at $\sqrt{s}=$ 7
  TeV}},}\ }\href {\doibase 10.1016/j.physletb.2011.07.089} {\bibfield
  {journal} {\bibinfo  {journal} {Phys. Lett.}\ }\textbf {\bibinfo {volume}
  {B703}},\ \bibinfo {pages} {246--266} (\bibinfo {year} {2011})},\ \Eprint
  {http://arxiv.org/abs/1105.5237} {arXiv:1105.5237 [hep-ex]} \BibitemShut
  {NoStop}%
\bibitem [{\citenamefont {Aad}\ \emph {et~al.}(2012{\natexlab{b}})\citenamefont
  {Aad} \emph {et~al.}}]{ATLAS:2012aq}%
  \BibitemOpen
  \bibfield  {author} {\bibinfo {author} {\bibfnamefont {Georges}\ \bibnamefont
  {Aad}} \emph {et~al.} (\bibinfo {collaboration} {ATLAS}),\ }\bibfield
  {title} {\enquote {\bibinfo {title} {{Search for second generation scalar
  leptoquarks in $pp$ collisions at $\sqrt{s}=7$ TeV with the ATLAS
  detector}},}\ }\href {\doibase 10.1140/epjc/s10052-012-2151-6} {\bibfield
  {journal} {\bibinfo  {journal} {Eur. Phys. J.}\ }\textbf {\bibinfo {volume}
  {C72}},\ \bibinfo {pages} {2151} (\bibinfo {year} {2012}{\natexlab{b}})},\
  \Eprint {http://arxiv.org/abs/1203.3172} {arXiv:1203.3172 [hep-ex]}
  \BibitemShut {NoStop}%
\bibitem [{\citenamefont {Chatrchyan}\ \emph
  {et~al.}(2012{\natexlab{a}})\citenamefont {Chatrchyan} \emph
  {et~al.}}]{Chatrchyan:2012vza}%
  \BibitemOpen
  \bibfield  {author} {\bibinfo {author} {\bibfnamefont {Serguei}\ \bibnamefont
  {Chatrchyan}} \emph {et~al.} (\bibinfo {collaboration} {CMS}),\ }\bibfield
  {title} {\enquote {\bibinfo {title} {{Search for pair production of first-
  and second-generation scalar leptoquarks in $pp$ collisions at $\sqrt{s}= 7$
  TeV}},}\ }\href {\doibase 10.1103/PhysRevD.86.052013} {\bibfield  {journal}
  {\bibinfo  {journal} {Phys. Rev.}\ }\textbf {\bibinfo {volume} {D86}},\
  \bibinfo {pages} {052013} (\bibinfo {year} {2012}{\natexlab{a}})},\ \Eprint
  {http://arxiv.org/abs/1207.5406} {arXiv:1207.5406 [hep-ex]} \BibitemShut
  {NoStop}%
\bibitem [{CMS(2014)}]{CMS-PAS-EXO-12-041}%
  \BibitemOpen
  \href {http://cds.cern.ch/record/1742179} {\emph {\bibinfo {title} {{Search
  for Pair-production of First Generation Scalar Leptoquarks in pp Collisions
  at sqrt s = 8 TeV}}}},\ \bibinfo {type} {Tech. Rep.}\ \bibinfo {number}
  {CMS-PAS-EXO-12-041}\ (\bibinfo  {institution} {CERN},\ \bibinfo {address}
  {Geneva},\ \bibinfo {year} {2014})\BibitemShut {NoStop}%
\bibitem [{CMS(2013)}]{CMS-PAS-EXO-12-042}%
  \BibitemOpen
  \href {http://cds.cern.ch/record/1542374} {\emph {\bibinfo {title} {{Search
  for Pair-production of Second generation Leptoquarks in 8 TeV proton-proton
  collisions.}}}},\ \bibinfo {type} {Tech. Rep.}\ \bibinfo {number}
  {CMS-PAS-EXO-12-042}\ (\bibinfo  {institution} {CERN},\ \bibinfo {address}
  {Geneva},\ \bibinfo {year} {2013})\BibitemShut {NoStop}%
\bibitem [{\citenamefont {Khachatryan}\ \emph
  {et~al.}(2016{\natexlab{c}})\citenamefont {Khachatryan} \emph
  {et~al.}}]{Khachatryan:2015vaa}%
  \BibitemOpen
  \bibfield  {author} {\bibinfo {author} {\bibfnamefont {Vardan}\ \bibnamefont
  {Khachatryan}} \emph {et~al.} (\bibinfo {collaboration} {CMS}),\ }\bibfield
  {title} {\enquote {\bibinfo {title} {{Search for pair production of first and
  second generation leptoquarks in proton-proton collisions at sqrt(s) = 8
  TeV}},}\ }\href {\doibase 10.1103/PhysRevD.93.032004} {\bibfield  {journal}
  {\bibinfo  {journal} {Phys. Rev.}\ }\textbf {\bibinfo {volume} {D93}},\
  \bibinfo {pages} {032004} (\bibinfo {year} {2016}{\natexlab{c}})},\ \Eprint
  {http://arxiv.org/abs/1509.03744} {arXiv:1509.03744 [hep-ex]} \BibitemShut
  {NoStop}%
\bibitem [{CMS(2016{\natexlab{b}})}]{CMS-PAS-EXO-16-043}%
  \BibitemOpen
  \href {http://cds.cern.ch/record/2205285} {\emph {\bibinfo {title} {{Search
  for pair-production of first generation scalar leptoquarks in pp collisions
  at $\sqrt{s}=13~\mathrm{TeV}$ with $2.6~\mathrm{fb}^{-1}$}}}},\ \bibinfo
  {type} {Tech. Rep.}\ \bibinfo {number} {CMS-PAS-EXO-16-043}\ (\bibinfo
  {institution} {CERN},\ \bibinfo {address} {Geneva},\ \bibinfo {year}
  {2016})\BibitemShut {NoStop}%
\bibitem [{ATL(2015)}]{ATLAS-CONF-2015-015}%
  \BibitemOpen
  \href {http://cds.cern.ch/record/2002885} {\emph {\bibinfo {title} {{A search
  for $B-L$ $R$-Parity violating scalar top decays in $\sqrt{s} = 8$ TeV $pp$
  collisions with the ATLAS experiment}}}},\ \bibinfo {type} {Tech. Rep.}\
  \bibinfo {number} {ATLAS-CONF-2015-015}\ (\bibinfo  {institution} {CERN},\
  \bibinfo {address} {Geneva},\ \bibinfo {year} {2015})\BibitemShut {NoStop}%
\bibitem [{\citenamefont {Chatrchyan}\ \emph
  {et~al.}(2013{\natexlab{c}})\citenamefont {Chatrchyan} \emph
  {et~al.}}]{Chatrchyan:2012sv}%
  \BibitemOpen
  \bibfield  {author} {\bibinfo {author} {\bibfnamefont {Serguei}\ \bibnamefont
  {Chatrchyan}} \emph {et~al.} (\bibinfo {collaboration} {CMS}),\ }\bibfield
  {title} {\enquote {\bibinfo {title} {{Search for pair production of
  third-generation leptoquarks and top squarks in $pp$ collisions at
  $\sqrt{s}=7$ TeV}},}\ }\href {\doibase 10.1103/PhysRevLett.110.081801}
  {\bibfield  {journal} {\bibinfo  {journal} {Phys. Rev. Lett.}\ }\textbf
  {\bibinfo {volume} {110}},\ \bibinfo {pages} {081801} (\bibinfo {year}
  {2013}{\natexlab{c}})},\ \Eprint {http://arxiv.org/abs/1210.5629}
  {arXiv:1210.5629 [hep-ex]} \BibitemShut {NoStop}%
\bibitem [{\citenamefont {Chatrchyan}\ \emph
  {et~al.}(2012{\natexlab{b}})\citenamefont {Chatrchyan} \emph
  {et~al.}}]{Chatrchyan:2012st}%
  \BibitemOpen
  \bibfield  {author} {\bibinfo {author} {\bibfnamefont {Serguei}\ \bibnamefont
  {Chatrchyan}} \emph {et~al.} (\bibinfo {collaboration} {CMS}),\ }\bibfield
  {title} {\enquote {\bibinfo {title} {{Search for third-generation leptoquarks
  and scalar bottom quarks in $pp$ collisions at $\sqrt{s}=7$ TeV}},}\ }\href
  {\doibase 10.1007/JHEP12(2012)055} {\bibfield  {journal} {\bibinfo  {journal}
  {JHEP}\ }\textbf {\bibinfo {volume} {12}},\ \bibinfo {pages} {055} (\bibinfo
  {year} {2012}{\natexlab{b}})},\ \Eprint {http://arxiv.org/abs/1210.5627}
  {arXiv:1210.5627 [hep-ex]} \BibitemShut {NoStop}%
\bibitem [{\citenamefont {Aad}\ \emph {et~al.}(2013{\natexlab{b}})\citenamefont
  {Aad} \emph {et~al.}}]{ATLAS:2013oea}%
  \BibitemOpen
  \bibfield  {author} {\bibinfo {author} {\bibfnamefont {Georges}\ \bibnamefont
  {Aad}} \emph {et~al.} (\bibinfo {collaboration} {ATLAS}),\ }\bibfield
  {title} {\enquote {\bibinfo {title} {{Search for third generation scalar
  leptoquarks in pp collisions at $\sqrt{s}$ = 7 TeV with the ATLAS
  detector}},}\ }\href {\doibase 10.1007/JHEP06(2013)033} {\bibfield  {journal}
  {\bibinfo  {journal} {JHEP}\ }\textbf {\bibinfo {volume} {06}},\ \bibinfo
  {pages} {033} (\bibinfo {year} {2013}{\natexlab{b}})},\ \Eprint
  {http://arxiv.org/abs/1303.0526} {arXiv:1303.0526 [hep-ex]} \BibitemShut
  {NoStop}%
\bibitem [{\citenamefont {Khachatryan}\ \emph {et~al.}(2014)\citenamefont
  {Khachatryan} \emph {et~al.}}]{Khachatryan:2014ura}%
  \BibitemOpen
  \bibfield  {author} {\bibinfo {author} {\bibfnamefont {Vardan}\ \bibnamefont
  {Khachatryan}} \emph {et~al.} (\bibinfo {collaboration} {CMS}),\ }\bibfield
  {title} {\enquote {\bibinfo {title} {{Search for pair production of
  third-generation scalar leptoquarks and top squarks in proton--proton
  collisions at $\sqrt{s}$=8 TeV}},}\ }\href {\doibase
  10.1016/j.physletb.2014.10.063} {\bibfield  {journal} {\bibinfo  {journal}
  {Phys. Lett.}\ }\textbf {\bibinfo {volume} {B739}},\ \bibinfo {pages}
  {229--249} (\bibinfo {year} {2014})},\ \Eprint
  {http://arxiv.org/abs/1408.0806} {arXiv:1408.0806 [hep-ex]} \BibitemShut
  {NoStop}%
\bibitem [{\citenamefont {Khachatryan}\ \emph
  {et~al.}(2017{\natexlab{b}})\citenamefont {Khachatryan} \emph
  {et~al.}}]{Khachatryan:2016jqo}%
  \BibitemOpen
  \bibfield  {author} {\bibinfo {author} {\bibfnamefont {Vardan}\ \bibnamefont
  {Khachatryan}} \emph {et~al.} (\bibinfo {collaboration} {CMS}),\ }\bibfield
  {title} {\enquote {\bibinfo {title} {{Search for heavy neutrinos or
  third-generation leptoquarks in final states with two hadronically decaying
  $\tau$ leptons and two jets in proton-proton collisions at $ \sqrt{s}=13 $
  TeV}},}\ }\href {\doibase 10.1007/JHEP03(2017)077} {\bibfield  {journal}
  {\bibinfo  {journal} {JHEP}\ }\textbf {\bibinfo {volume} {03}},\ \bibinfo
  {pages} {077} (\bibinfo {year} {2017}{\natexlab{b}})},\ \Eprint
  {http://arxiv.org/abs/1612.01190} {arXiv:1612.01190 [hep-ex]} \BibitemShut
  {NoStop}%
\bibitem [{\citenamefont {Sirunyan}\ \emph
  {et~al.}(2017{\natexlab{c}})\citenamefont {Sirunyan} \emph
  {et~al.}}]{Sirunyan:2017yrk}%
  \BibitemOpen
  \bibfield  {author} {\bibinfo {author} {\bibfnamefont {Albert~M}\
  \bibnamefont {Sirunyan}} \emph {et~al.} (\bibinfo {collaboration} {CMS}),\
  }\bibfield  {title} {\enquote {\bibinfo {title} {{Search for the
  third-generation scalar leptoquarks and heavy right-handed neutrinos in final
  states with two tau leptons and two jets in proton-proton collisions at
  $\sqrt{s}$ = 13 TeV}},}\ }\href@noop {} {\  (\bibinfo {year}
  {2017}{\natexlab{c}})},\ \Eprint {http://arxiv.org/abs/1703.03995}
  {arXiv:1703.03995 [hep-ex]} \BibitemShut {NoStop}%
\bibitem [{\citenamefont {Khachatryan}\ \emph
  {et~al.}(2015{\natexlab{b}})\citenamefont {Khachatryan} \emph
  {et~al.}}]{Khachatryan:2015bsa}%
  \BibitemOpen
  \bibfield  {author} {\bibinfo {author} {\bibfnamefont {Vardan}\ \bibnamefont
  {Khachatryan}} \emph {et~al.} (\bibinfo {collaboration} {CMS}),\ }\bibfield
  {title} {\enquote {\bibinfo {title} {{Search for Third-Generation Scalar
  Leptoquarks in the t$\tau$ Channel in Proton-Proton Collisions at $\sqrt{s}$
  = 8 TeV}},}\ }\href {\doibase 10.1007/JHEP11(2016)056,
  10.1007/JHEP07(2015)042} {\bibfield  {journal} {\bibinfo  {journal} {JHEP}\
  }\textbf {\bibinfo {volume} {07}},\ \bibinfo {pages} {042} (\bibinfo {year}
  {2015}{\natexlab{b}})},\ \bibinfo {note} {[Erratum: JHEP11,056(2016)]},\
  \Eprint {http://arxiv.org/abs/1503.09049} {arXiv:1503.09049 [hep-ex]}
  \BibitemShut {NoStop}%
\bibitem [{\citenamefont {Gripaios}\ \emph {et~al.}(2011)\citenamefont
  {Gripaios}, \citenamefont {Papaefstathiou}, \citenamefont {Sakurai},\ and\
  \citenamefont {Webber}}]{Gripaios:2010hv}%
  \BibitemOpen
  \bibfield  {author} {\bibinfo {author} {\bibfnamefont {Ben}\ \bibnamefont
  {Gripaios}}, \bibinfo {author} {\bibfnamefont {Andreas}\ \bibnamefont
  {Papaefstathiou}}, \bibinfo {author} {\bibfnamefont {Kazuki}\ \bibnamefont
  {Sakurai}}, \ and\ \bibinfo {author} {\bibfnamefont {Bryan}\ \bibnamefont
  {Webber}},\ }\bibfield  {title} {\enquote {\bibinfo {title} {{Searching for
  third-generation composite leptoquarks at the LHC}},}\ }\href {\doibase
  10.1007/JHEP01(2011)156} {\bibfield  {journal} {\bibinfo  {journal} {JHEP}\
  }\textbf {\bibinfo {volume} {01}},\ \bibinfo {pages} {156} (\bibinfo {year}
  {2011})},\ \Eprint {http://arxiv.org/abs/1010.3962} {arXiv:1010.3962
  [hep-ph]} \BibitemShut {NoStop}%
\bibitem [{CMS(2016{\natexlab{c}})}]{CMS-PAS-BTV-16-001}%
  \BibitemOpen
  \href {http://cds.cern.ch/record/2205149} {\emph {\bibinfo {title}
  {{Identification of c-quark jets at the CMS experiment}}}},\ \bibinfo {type}
  {Tech. Rep.}\ \bibinfo {number} {CMS-PAS-BTV-16-001}\ (\bibinfo
  {institution} {CERN},\ \bibinfo {address} {Geneva},\ \bibinfo {year}
  {2016})\BibitemShut {NoStop}%
\bibitem [{\citenamefont {Khachatryan}\ \emph
  {et~al.}(2016{\natexlab{d}})\citenamefont {Khachatryan} \emph
  {et~al.}}]{Khachatryan:2015qda}%
  \BibitemOpen
  \bibfield  {author} {\bibinfo {author} {\bibfnamefont {Vardan}\ \bibnamefont
  {Khachatryan}} \emph {et~al.} (\bibinfo {collaboration} {CMS}),\ }\bibfield
  {title} {\enquote {\bibinfo {title} {{Search for single production of scalar
  leptoquarks in proton-proton collisions at $\sqrt{s} = 8$ $TeV$}},}\ }\href
  {\doibase 10.1103/PhysRevD.95.039906, 10.1103/PhysRevD.93.032005} {\bibfield
  {journal} {\bibinfo  {journal} {Phys. Rev.}\ }\textbf {\bibinfo {volume}
  {D93}},\ \bibinfo {pages} {032005} (\bibinfo {year} {2016}{\natexlab{d}})},\
  \bibinfo {note} {[Erratum: Phys. Rev.D95,no.3,039906(2017)]},\ \Eprint
  {http://arxiv.org/abs/1509.03750} {arXiv:1509.03750 [hep-ex]} \BibitemShut
  {NoStop}%
\bibitem [{CMS(2017{\natexlab{e}})}]{CMS-PAS-SUS-16-041}%
  \BibitemOpen
  \href {http://cds.cern.ch/record/2256435} {\emph {\bibinfo {title} {{Search
  for new physics with multileptons and jets in $35.9~\mathrm{fb}^{-1}$ of pp
  collision data at $\sqrt{s}=13~\mathrm{TeV}$}}}},\ \bibinfo {type} {Tech.
  Rep.}\ \bibinfo {number} {CMS-PAS-SUS-16-041}\ (\bibinfo  {institution}
  {CERN},\ \bibinfo {address} {Geneva},\ \bibinfo {year} {2017})\BibitemShut
  {NoStop}%
\bibitem [{\citenamefont {Alwall}\ \emph {et~al.}(2014)\citenamefont {Alwall},
  \citenamefont {Frederix}, \citenamefont {Frixione}, \citenamefont {Hirschi},
  \citenamefont {Maltoni}, \citenamefont {Mattelaer}, \citenamefont {Shao},
  \citenamefont {Stelzer}, \citenamefont {Torrielli},\ and\ \citenamefont
  {Zaro}}]{Alwall:2014hca}%
  \BibitemOpen
  \bibfield  {author} {\bibinfo {author} {\bibfnamefont {J.}~\bibnamefont
  {Alwall}}, \bibinfo {author} {\bibfnamefont {R.}~\bibnamefont {Frederix}},
  \bibinfo {author} {\bibfnamefont {S.}~\bibnamefont {Frixione}}, \bibinfo
  {author} {\bibfnamefont {V.}~\bibnamefont {Hirschi}}, \bibinfo {author}
  {\bibfnamefont {F.}~\bibnamefont {Maltoni}}, \bibinfo {author} {\bibfnamefont
  {O.}~\bibnamefont {Mattelaer}}, \bibinfo {author} {\bibfnamefont {H.~S.}\
  \bibnamefont {Shao}}, \bibinfo {author} {\bibfnamefont {T.}~\bibnamefont
  {Stelzer}}, \bibinfo {author} {\bibfnamefont {P.}~\bibnamefont {Torrielli}},
  \ and\ \bibinfo {author} {\bibfnamefont {M.}~\bibnamefont {Zaro}},\
  }\bibfield  {title} {\enquote {\bibinfo {title} {{The automated computation
  of tree-level and next-to-leading order differential cross sections, and
  their matching to parton shower simulations}},}\ }\href {\doibase
  10.1007/JHEP07(2014)079} {\bibfield  {journal} {\bibinfo  {journal} {JHEP}\
  }\textbf {\bibinfo {volume} {07}},\ \bibinfo {pages} {079} (\bibinfo {year}
  {2014})},\ \Eprint {http://arxiv.org/abs/1405.0301} {arXiv:1405.0301
  [hep-ph]} \BibitemShut {NoStop}%
\bibitem [{\citenamefont {Patrignani}\ \emph {et~al.}(2016)\citenamefont
  {Patrignani} \emph {et~al.}}]{Olive:2016xmw}%
  \BibitemOpen
  \bibfield  {author} {\bibinfo {author} {\bibfnamefont {C.}~\bibnamefont
  {Patrignani}} \emph {et~al.} (\bibinfo {collaboration} {Particle Data
  Group}),\ }\bibfield  {title} {\enquote {\bibinfo {title} {{Review of
  Particle Physics}},}\ }\href {\doibase 10.1088/1674-1137/40/10/100001}
  {\bibfield  {journal} {\bibinfo  {journal} {Chin. Phys.}\ }\textbf {\bibinfo
  {volume} {C40}},\ \bibinfo {pages} {100001} (\bibinfo {year}
  {2016})}\BibitemShut {NoStop}%
\bibitem [{\citenamefont {Buchmuller}\ \emph {et~al.}(1987)\citenamefont
  {Buchmuller}, \citenamefont {Ruckl},\ and\ \citenamefont
  {Wyler}}]{Buchmuller:1986zs}%
  \BibitemOpen
  \bibfield  {author} {\bibinfo {author} {\bibfnamefont {W.}~\bibnamefont
  {Buchmuller}}, \bibinfo {author} {\bibfnamefont {R.}~\bibnamefont {Ruckl}}, \
  and\ \bibinfo {author} {\bibfnamefont {D.}~\bibnamefont {Wyler}},\ }\bibfield
   {title} {\enquote {\bibinfo {title} {{Leptoquarks in Lepton - Quark
  Collisions}},}\ }\href {\doibase 10.1016/S0370-2693(99)00014-3,
  10.1016/0370-2693(87)90637-X} {\bibfield  {journal} {\bibinfo  {journal}
  {Phys. Lett.}\ }\textbf {\bibinfo {volume} {B191}},\ \bibinfo {pages}
  {442--448} (\bibinfo {year} {1987})},\ \bibinfo {note} {[Erratum: Phys.
  Lett.B448,320(1999)]}\BibitemShut {NoStop}%
\bibitem [{\citenamefont {Weinberg}\ and\ \citenamefont
  {Witten}(1980)}]{Weinberg:1980kq}%
  \BibitemOpen
  \bibfield  {author} {\bibinfo {author} {\bibfnamefont {Steven}\ \bibnamefont
  {Weinberg}}\ and\ \bibinfo {author} {\bibfnamefont {Edward}\ \bibnamefont
  {Witten}},\ }\bibfield  {title} {\enquote {\bibinfo {title} {{Limits on
  Massless Particles}},}\ }\href {\doibase 10.1016/0370-2693(80)90212-9}
  {\bibfield  {journal} {\bibinfo  {journal} {Phys. Lett.}\ }\textbf {\bibinfo
  {volume} {B96}},\ \bibinfo {pages} {59--62} (\bibinfo {year}
  {1980})}\BibitemShut {NoStop}%
\end{thebibliography}%

\end{document}